\documentclass[11pt,a4paper]{article}

\usepackage{jheppub}
\usepackage{graphicx}
\usepackage{dcolumn}
\usepackage{bm}
\usepackage{amsmath}
\usepackage{braket}
\usepackage{slashed}    
\usepackage{epstopdf}
\usepackage{placeins}
\usepackage{multirow}
\usepackage{makecell}
\usepackage[shortlabels]{enumitem}
\usepackage{cleveref}
\usepackage{xcolor}
\usepackage[normalem]{ulem}
\usepackage[font=small, labelfont=bf, textfont={small,it}]{caption}
\usepackage{orcidlink}

\graphicspath{{figs/}}

\newcommand{\beq}{\begin{eqnarray}}
\newcommand{\eeq}{\end{eqnarray}}
\newcommand{\beqnn}{\begin{eqnarray*}}
\newcommand{\eeqnn}{\end{eqnarray*}}

\newcommand{\Nf}{N_{\scriptscriptstyle{\rm f}}}
\newcommand{\DW}{D_{\scriptscriptstyle{\rm W}}}
\newcommand{\DDW}{D_{\scriptscriptstyle{\rm DW}}}
\newcommand{\DT}{D_{\scriptscriptstyle{\rm T}}}
\newcommand{\DTp}{\widetilde{D}_{\scriptscriptstyle{\rm T}}}

\newcommand{\Nfive}{N_{\scriptscriptstyle{5}}}
\newcommand{\HW}{\mathcal{H}_{\scriptscriptstyle{\rm W}}}

\newcommand{\RMT}{{\scriptscriptstyle{\rm RMT}}}

\newcommand{\R}{{\scriptscriptstyle{\rm R}}}
\newcommand{\ZA}{Z_{\scriptscriptstyle{\rm A}}}
\newcommand{\ZS}{Z_{\scriptscriptstyle{\rm S}}}
\newcommand{\ZP}{Z_{\scriptscriptstyle{\rm P}}}
\newcommand{\Zm}{Z_{\scriptscriptstyle{\rm m}}}

\newcommand{\PCAC}{{\scriptscriptstyle{\rm PCAC}}}
\newcommand{\TEK}{{\scriptscriptstyle{\rm TEK}}}
\newcommand{\sstrm}[1]{{\scriptscriptstyle{\mathrm{#1}}}}
\newcommand{\dd}{\mathrm{d}}
\newcommand{\ee}{\mathrm{e}}
\newcommand{\ii}{\mathrm{i}}
\newcommand{\Tr}{\mathrm{Tr}}
\newcommand{\sign}{\mathrm{sign}}

\newcommand{\SW}{S_{\scriptscriptstyle{\rm W}}}

\newcommand{\A}{{\scriptscriptstyle{\rm A}}}
\newcommand{\s}{{\scriptscriptstyle{\rm s}}}

\newcommand{\chir}{{\scriptscriptstyle{(\chi)}}}
\newcommand{\chirW}{{\scriptscriptstyle{(\chi,\mathrm{W})}}}

\newcommand{\SU}{\mathrm{SU}}

\begin{document}

\title{\Huge Large-$N$ chiral dynamics from twisted volume reduction with chiral quarks}

\author[a]{Claudio Bonanno\,\orcidlink{0000-0002-1421-6510},}
\author[b]{Margarita Garc\'ia P\'erez\,\orcidlink{0000-0001-6566-2687},}
\author[b,c]{Antonio Gonz\'alez-Arroyo\,\orcidlink{0000-0002-5263-1088},}
\author[d,e]{\\Ken-Ichi Ishikawa\,\orcidlink{0000-0001-6425-1894},}
\author[e]{Masanori Okawa\,\orcidlink{0009-0007-7983-4364},}
\author[f]{Dario Panfalone\,\orcidlink{0009-0007-6651-7490}.}

\affiliation[a]{Albert Einstein Center for Fundamental Physics, Institute for Theoretical Physics, University of Bern, Sidlerstra{\ss}e 5, CH-3012 Bern, Switzerland}

\affiliation[b]{Instituto de F\'isica Te\'orica UAM-CSIC, Calle Nicol\'as Cabrera 13-15,\\Universidad Aut\'onoma de Madrid, Cantoblanco, E-28049 Madrid, Spain}

\affiliation[c]{Departamento de F\'isica Te\'orica, Universidad Aut\'onoma de Madrid,\\M\'odulo 15, Cantoblanco, E-28049 Madrid, Spain}

\affiliation[d]{Core of Research for the Energetic Universe,\\Graduate School of Advanced Science and Engineering,\\Hiroshima University, Higashi-Hiroshima, Hiroshima 739-8526, Japan}

\affiliation[e]{Graduate School of Advanced Science and Engineering, Hiroshima University,\\Higashi-Hiroshima, Hiroshima 739-8526, Japan}

\affiliation[f]{Department of Physics, University of Turin \& INFN, Turin,\\Via Pietro Giuria 1, I-10125 Turin, Italy}

\emailAdd{claudio.bonanno@unibe.ch}
\emailAdd{margarita.garcia@csic.es}
\emailAdd{A.Gonzalez-Arroyo@pm.me}
\emailAdd{ishikawa@theo.phys.sci.hiroshima-u.ac.jp}
\emailAdd{okawa@hiroshima-u.ac.jp}
\emailAdd{dario.panfalone@unito.it}

\abstract{We investigate chiral dynamics in large-$N$ QCD on the lattice adopting a chiral quark discretization, and exploiting the Twisted Eguchi--Kawai (TEK) formulation of the theory. We present results in the continuum limit for the chiral condensate, extracted from the low-lying Dirac spectrum according to Random Matrix Theory predictions, and for the quark mass dependence of several quantities down to the chiral limit: the pion mass, the pion decay constant, and the rho meson mass. These determinations are then compared with recent large-$N$ TEK results obtained with a non-chiral Wilson discretization.}


\maketitle

\section{Introduction}

One of the most fundamental non-perturbative features of Quantum Chromo-Dynamics (QCD) is the spontaneous breaking of chiral symmetry, $\SU(\Nf)_{\scriptscriptstyle{\rm L}} \times \SU(\Nf)_{\scriptscriptstyle{\rm R}} \to \SU(\Nf)_{\scriptscriptstyle{\rm V}}$, in the chiral limit, implying the existence of $\Nf^2 - 1$ Nambu--Goldstone bosons in the spectrum (identified, e.g., with the three pions for $\Nf=2$). These properties allow a low-energy description of QCD dynamics using Chiral Perturbation Theory ($\chi$PT)~\cite{Gasser:1983yg,Gasser:1984gg}, an effective theory written in terms of the pseudo-Goldstone fields, where interaction terms in the effective Lagrangian are organized into a systematic expansion in powers of the squared momenta and of the quark masses.

Another fundamental theoretical tool to investigate the non-perturbative domain of QCD is the expansion in the inverse number of colors $N$. In the large-$N$ limit $N\to\infty$~\cite{tHooft:1973alw}, QCD enjoys several simplifications, while at the same time retaining many of its key theoretical features, including spontaneous chiral symmetry breaking. Including a further light flavor-singlet in the chiral Lagrangian, the $1/N$ expansion can be embedded into $\chi$PT, allowing one to define a low-energy chiral effective theory for large-$N$ QCD.

The combination of these two frameworks has proven to be capable of providing valuable phenomenological predictions for strong interactions, such as the resolution of the $\mathrm{U}(1)_{\scriptscriptstyle{\rm A}}$    puzzle~\cite{tHooft:1976rip,tHooft:1986nc,Witten:1979vv,Veneziano:1979ec,Witten:1980sp,DiVecchia:1980yfw}, and has also been employed as a tool to obtain important insights into the nature of tetraquarks and non-ordinary mesons, see, e.g., Refs.~\cite{Cirigliano:2003yq,Weinberg:2013cfa,Cohen:2014vta,Pelaez:2015qba}. Furthermore, large-$N$ chiral dynamics is an intriguing theoretical topic in the context of understanding non-perturbative gauge dynamics from first principles. As an example, large-$N$ QCD in the chiral limit is one of the targets of the S-matrix bootstrap research program~\cite{Albert:2022oes,Fernandez:2022kzi,Ma:2023vgc,Albert:2023jtd,Albert:2023seb,Li:2023qzs,Guerrieri:2024jkn,Albert:2026xyz,Cheung:2026lpv,Bocchia:2026kew,Bocchia:2026lty,Cordoba:2026wsy}.

At leading order (LO), only two parameters appear in the chiral Lagrangian, namely, the chiral condensate and the pion decay constant. Beyond LO, a growing set of further low-energy constants (LECs) appears at each higher order of the chiral expansion. On general grounds, all $\chi$PT LECs possess a well-defined and non-trivial scaling with $N$~\cite{Kaiser:2000gs}, and first-principles determinations of the coefficients of their $1/N$ expansion constitute important theoretical predictions. In this respect, lattice QCD is a natural systematic non-perturbative approach to compute QCD LECs and their $1/N$ expansion. As of today, many determinations of LO and NLO $\chi$PT LECs are available for $N=3$~\cite{Giusti:2008vb,Frezzotti:2008dr,Feng:2009ij,ETM:2009ztk,MILC:2010hzw,Bazavov:2010yq,ETM:2010cqp,Beane:2011zm,Borsanyi:2012zv,Budapest-Marseille-Wuppertal:2013vij,Cichy:2013gja,Brandt:2013dua,Engel:2014eea,ETM:2015bzg,Boyle:2015exm,Gulpers:2015bba,Cossu:2016eqs,Aoki:2017paw,Alexandrou:2017bzk,FlavourLatticeAveragingGroupFLAG:2021npn,Bonanno:2023xkg,Bonanno:2025cmh,vonHippel:2025uhr,Ottnad:2025hkw}, and some studies for $N>3$ have appeared in the literature too~\cite{DelDebbio:2007wk,Bali:2013kia,Donini:2016lwz,DeGrand:2016pur,Hernandez:2019qed,Hernandez:2020tbc,Donini:2020qfu,Baeza-Ballesteros:2022azb,DeGrand:2023hzz,Baeza-Ballesteros:2025iee}.

The standard strategy to explore the large-$N$ limit on the lattice consists in simulating the theory with increasing gauge-group ranks and in extrapolating finite-$N$ data towards $N\to \infty$, see~\cite{Lucini:2012gg,Hernandez:2020tbc} for reviews. Typically employed values of the number of colors are $N\lesssim\mathcal{O}(10)$. An alternative approach to obtain physical results in the large-$N$ limit is to make use of volume independence~\cite{PhysRevLett.48.1063,BHANOT198247,Gross:1982at,Gonzalez-Arroyo:1982hwr,PhysRevD.27.2397,Aldazabal:1983ec,Kiskis:2002gr,Narayanan:2003fc,Kovtun:2007py,Unsal:2008ch,Gonzalez-Arroyo:2010omx,Neuberger:2020wpx,Yaffe:2026ouz}. In the presence of an unbroken center symmetry, a $d$-dimensional $\mathrm{SU}(N)$ gauge theory is equivalent, at leading order in $1/N$, to the same theory discretized on a one-point lattice. This approach is computationally advantageous over the traditional one for the purpose of investigating the large-$N$ limit, as the space-time volume reduction allows the simulation of much larger values of $N$ while keeping the computational effort feasible. The Twisted Eguchi--Kawai (TEK)~\cite{Gonzalez-Arroyo:1982hwr,PhysRevD.27.2397} model implements volume reduction by imposing twisted boundary conditions on the one-point box, as they prevent the spontaneous breaking of center symmetry (other prescriptions to rescue center symmetry on a one-site lattice have been put forward in the literature, see, e.g., Refs.~\cite{BHANOT198247,Narayanan:2003fc,Kovtun:2007py,Unsal:2008ch}). The model maps to a lattice field theory defined on an effective physical volume $V \propto N^2$, allowing one to access large-physical-volume and large-$N$ dynamics simultaneously while simulating a single-site box. Indeed, one can show that the TEK model is equivalent to a lattice-discretized non-commutative field theory formulated on a non-commutative torus of linear size $\sqrt{N}$~\cite{Gonzalez-Arroyo:1983zev,Eguchi:1982ta,Douglas:2001ba,Ambjorn:1999ts,Ambjorn:2000nb,Ambjorn:2000cs}. This implies that planar finite-$N$ corrections show up as finite effective-volume effects. Additional non-planar finite-$N$ corrections can be monitored and reduced by appropriate choices of the twist 't~Hooft flux. In any case, these corrections are of a different nature compared to those in standard approaches. Thus, determining the physical $N$-dependence requires the combination of the TEK and the standard approaches.

The methodology based on volume reduction at large $N$ has been used extensively over the years to obtain numerical results on the physical observables of large-$N$ QCD. Many of these results~\cite{Gonzalez-Arroyo:1983cyv,Gonzalez-Arroyo:2012euf,Gonzalez-Arroyo:2013bta,GarciaPerez:2014azn,Gonzalez-Arroyo:2014dua,Gonzalez-Arroyo:2015bya,GarciaPerez:2015rda,Perez:2015ssa,Perez:2020vbn,Bonanno:2023ypf,Butti:2023hfp,Bonanno:2025hzr,Bonanno:2026gay,Zhu:2026zvb} have been obtained using the TEK formulation that we will follow in this paper.\footnote{Other formulations have also been employed, see, e.g., Refs.~\cite{Narayanan:2004cp,Gonzalez-Arroyo:2005dgf,Kiskis:2009rf,Hietanen:2009ex,Hietanen:2009tu,Hietanen:2010fx,Bringoltz:2011by,Hietanen:2012ma,Lohmayer:2013spa,Karthik:2022fdb}.} The large majority of these works have used the Wilson discretization of lattice fermions. Explicit chiral invariance of the theory is lost in this formulation, leading to an additive renormalization of the quark mass. As is well known, the chiral limit is then obtained by monitoring the vanishing of the PCAC (Partially Conserved Axial-Vector Current) mass or of the pion mass. Recently~\cite{Bonanno:2025hzr}, this formulation has been used successfully to determine important chiral observables, such as the pion decay constant, the chiral condensate and other $\chi$PT low-energy constants (see also Refs.~\cite{Perez:2020vbn,Bonanno:2023ypf} for earlier results).

Naturally, as in standard lattice QCD, other lattice discretizations are also possible in the TEK approach. Therefore, a natural extension of the work done in~\cite{Bonanno:2025hzr} is to carry out a TEK investigation of large-$N$ chiral dynamics in QCD adopting \emph{chiral quarks}. While the use of a chirally-symmetric fermion lattice discretization requires a larger computational effort than a non-chiral one, it also offers several conceptual and practical advantages. With chiral fermions, the Ginsparg--Wilson relation protects the quark mass from additive renormalizations, making the massless point coincide with the vanishing bare quark mass limit. Moreover, the lattice artifacts induced by the overlap and the Wilson discretizations are different (and vanish at different rates as the continuum limit is approached), so comparing results obtained from the two methods provides a stringent test of universality, as well as an \emph{a posteriori} confirmation that continuum extrapolations are under control.

A first TEK investigation of large-$N$ QCD with a chiral (overlap) lattice Dirac operator has appeared in~\cite{Bonanno:2025bla}, where it was shown that Random Matrix Theory predictions~\cite{Shuryak:1992pi,Verbaarschot:1993pm,Verbaarschot:1994qf} could be successfully reproduced in the large-effective-volume limit. This study was however limited to a single lattice spacing, and the chiral condensate was the only quantity considered there. In the present work we will substantially extend the research program initiated in~\cite{Bonanno:2025bla}. Using three values of the lattice spacing, four values of the quark mass for each lattice spacing, and several values of the number of colors up to $1156$, we will show that the pion becomes massless in the chiral limit, and we will report results for the chiral condensate and for the quark-mass dependence of the $\rho$ mass and the $\pi$ decay constant down to the chiral limit, thus determining also their associated NLO $\chi$PT LECs. All these results will be compared with those obtained (in the continuum limit) in Ref.~\cite{Bonanno:2025hzr} using non-chiral Wilson fermions.

The paper is organized as follows. Section~\ref{sec:setup} describes the twisted reduced model and the adopted chiral fermion formulation. Section~\ref{sec:eff} summarizes the effective-theory predictions used to compare with lattice data. Section~\ref{sec:res} presents the numerical results and their chiral and continuum extrapolations. Finally, we outline our conclusions in Section~\ref{sec:conclusions}. The raw lattice data analyzed in this work are collected in Appendix~\ref{app:data}, while in Appendix~\ref{app:mass} we discuss the normalization of the bare quark mass.

\section{Twisted reduced model for \texorpdfstring{large-$N$}{large-N} QCD with chiral quarks}\label{sec:setup}

\subsection{Lattice formulation}

We investigate QCD in the 't Hooft large-$N$ limit, i.e., $N\to\infty$ at fixed $\lambda=g^2N$ and fixed number of quark flavors $\Nf$ (so that $\Nf/N\to 0$). In this limit, quark-loop contributions are suppressed with respect to the gluonic sector by a factor of $\Nf/N$. The leading large-$N$ dynamics can therefore be studied using quenched quarks: gauge configurations are generated with the pure Yang--Mills path integral measure without including the fermion determinant, and fermion observables are subsequently evaluated on these gauge backgrounds.

To simulate the pure-glue dynamics, we use the TEK formulation of the pure Yang--Mills lattice action. In particular, we adopt the Wilson plaquette action, defined on a single-site lattice, and we assume twisted boundary conditions in all four directions. The TEK partition function of the $d = 4$ single-site model describing the gluonic sector is:
\beq\label{eq:part_func}
Z_\TEK \equiv \int [\dd U]\, \exp\left\{-\SW[U]\right\}, \qquad [\dd U] \equiv \prod_{\mu\,=\,1}^{d}\left[\dd U_\mu\right],
\eeq
where $[\dd U]$ denotes the $\SU(N)$ invariant Haar measure and
\beq\label{eq:TEK_Wilson_action}
\SW[U] = -N b \sum_{\mu\,=\,1}^{d}\sum_{\nu \,\ne\, \mu} z_{\nu\mu} \Tr\left\{U_\mu U_\nu U_\mu^\dagger U_\nu^\dagger\right\},
\eeq
is the TEK Wilson action. The Monte Carlo procedure used to generate gauge configurations according to the probability distribution $\mathcal{P}[U]=\exp\left\{-\SW[U]\right\}/Z_{\TEK}$ is described in Ref.~\cite{Perez:2015ssa}. In Eq.~\eqref{eq:TEK_Wilson_action} $b=1/(Ng^2)$ is the inverse bare 't Hooft coupling, $U_\mu$ are the $d=4$ $\SU(N)$ gauge link matrices describing lattice gluon fields, and $z_{\nu \mu}$ is the twist factor used to implement twisted boundary conditions. The twist factor is an element of $\mathbb{Z}_N$:
\beq\label{eq:twist_def}
z_{\nu\mu} = z_{\mu\nu}^* = \exp\left\{\frac{2\pi \ii}{N} n_{\nu\mu} \right\}.
\eeq
We choose $n_{\nu\mu}$ according to the \emph{symmetric twist} definition, for which the number of colors is a perfect square, $N=L^2$, and the antisymmetric integer twist tensor is specified by:
\beq
n_{\nu\mu}=-n_{\mu\nu}=k(L)L,
\qquad \nu>\mu,
\eeq
with $k(L)$ coprime to $L$. The corresponding twist phases are therefore:
\beq
z_{\nu\mu}
=\exp\left\{2\pi\ii\frac{k(L)}{L}\varepsilon_{\nu\mu}\right\},
\qquad
\varepsilon_{\nu\mu}=-\varepsilon_{\mu\nu},
\qquad
\varepsilon_{\nu\mu}=1 \quad (\nu>\mu).
\eeq
The flux $k(L)$ and its modular inverse $\bar{k}(L)$, defined by
$k\bar{k}=1\pmod L$, are scaled to grow with $L$ in order to maintain both $k/L$ and $\bar{k}/L$ above approximately $0.1$~\cite{Gonzalez-Arroyo:2010omx,Chamizo:2016msz,GarciaPerez:2018fkj}.
This prescription is used to prevent spontaneous breaking of center symmetry that could occur by increasing $N$ at fixed $k$~\cite{Ishikawa:2003,Bietenholz:2006cz,Teper:2006sp,Azeyanagi:2007su}, whose preservation is required for large-$N$ volume reduction to work. Perturbative studies also show that an appropriate choice of these flux parameters can help in suppressing non-planar finite-$N$ corrections~\cite{Perez:2017jyq,Bribian:2019ybc}. Despite its single-site formulation, the TEK model describes fields propagating in an effective volume whose size grows with $N$. For the symmetric twist, the associated physical effective torus size is:
\beq
\ell=aL=a\sqrt{N},
\eeq
where $a=a(b)$ is the lattice spacing. Thus, increasing $N$ at fixed $b$ enlarges the effective volume, and the limit $N\to\infty$ coincides with the infinite-volume limit $\ell\to\infty$.

Concerning the valence fermion sector, we discretize the Dirac operator using a particular formulation of the chiral overlap operator~\cite{Neuberger:1997fp,Neuberger:1998wv,Luscher:1998pqa} known as \emph{truncated overlap}~\cite{Kaplan:1992bt,Shamir:1993zy,Furman:1994ky,Borici:1999zw,Edwards:2000qv,Ishikawa:2013rxa}, identical to the one already adopted in Ref.~\cite{Bonanno:2025bla}. Its construction starts from the TEK Dirac operator in the Wilson formulation~\cite{Gonzalez-Arroyo:2015bya},
\beq\label{eq:Wilson_kernel}
\DW(-M)
=(4-M)\mathbb{I}
-\frac{1}{2}\sum_{\mu\,=\,1}^{4}
\left[
(1-\gamma_\mu)\otimes\mathcal W_\mu
+(1+\gamma_\mu)\otimes\mathcal W_\mu^\dagger
\right],
\eeq
evaluated at a negative kernel mass $-M$. The matrices entering
the hopping terms are
\beq\label{eq:fermion_links}
\mathcal W_\mu=U^{\sstrm{(stout)}}_\mu \otimes\Gamma_\mu^*,
\qquad
\Gamma_\mu\Gamma_\nu=z_{\nu\mu}\Gamma_\nu\Gamma_\mu,
\eeq
with $U^{\sstrm{(stout)}}_\mu$ representing stout-smeared~\cite{Morningstar:2003gk} gauge
links, and with the twist eaters
$\Gamma_\mu\in\mathrm{SU}(N)$ satisfying a commutation relation defined by the same twist appearing in the gauge action~\cite{Gonzalez-Arroyo:1997ugn}. Each $\mathcal W_\mu$ acts on an $N^2$-dimensional space, with the Dirac indices included separately in Eq.~\eqref{eq:Wilson_kernel}.
This construction, derived in~\cite{Gonzalez-Arroyo:2015bya}, describes quarks on an extended periodic lattice
of linear size $L=\sqrt{N}$, interacting with a gauge background
obtained by replicating the single-site links $\sqrt{N}$ times.

The truncated overlap operator can be derived from a
five-dimensional domain-wall formulation
\cite{Borici:1999zw,Edwards:2000qv,Kikukawa:2000ac,Ishikawa:2013rxa}.
The additional auxiliary direction, whose extent is denoted with $\Nfive$, controls the approximation to the overlap operator. In the M\"obius formulation~\cite{Brower:2005qw}, the five-dimensional operator can be written, in components along the fifth dimension, as:
\beq\label{eq:domain_wall_operator}
\left[\DDW(\hat{m},M,\Nfive)\right]_{st}
=
\DW(-M)\left(b_s\,\delta_{st}+c_s\left[M_5(\hat{m})\right]_{st}\right)
+\delta_{st}-\left[M_5(\hat{m})\right]_{st},
\eeq
where $s,t=1,\dots,\Nfive$ label the fifth-dimension coordinates, $\hat{m}$ is (proportional to) the bare quark mass in lattice units, $\DW(-M)$ acts on color, Dirac and (effective) space-time indices, $b_s$ and $c_s$ are real coefficients, and the matrix $M_5(\hat m)$ acts on fifth-dimension and Dirac indices:
\beq
[M_5(\hat m)]_{st}
= P_+\delta_{s+1,t}+P_-\delta_{s-1,t} -\hat m\left(
P_+\delta_{s,\Nfive}\delta_{t,1}
+P_-\delta_{s,1}\delta_{t,\Nfive}
\right),
\eeq
with $P_\pm=(\mathbb{I}\pm\gamma_5)/2$ the chiral projectors. Several choices for the coefficients $b_s$ and $c_s$ are possible, each one corresponding to a different discretization of the same continuum operator. Here we choose $b_s=1$ and $c_s=0$, corresponding to the Shamir formulation. The associated four-dimensional truncated overlap operator $\DT$ is defined as
\beq
\DT(\hat{m},M,\Nfive) = \left[\mathcal{P}^{\dagger}\DDW^{-1}(\hat{m}=1,M,\Nfive)\DDW(\hat{m},M,\Nfive)\mathcal{P}\right]_{1,1},
\eeq
with $\mathcal{P} = P_+\delta_{s,t} + P_-(\delta_{s+1,t}+\delta_{s,\Nfive}\delta_{t,1})$, and reads~\cite{Borici:1999zw,Kikukawa:2000ac,Edwards:2000qv,Brower:2005qw}:
\beq\label{eq:truncated_overlap}
\DT(\hat m,M,\Nfive)
=\frac{1+\hat m}{2}\mathbb{I}
+\frac{1-\hat m}{2}\gamma_5 \, s(M,\Nfive).
\eeq
Here, $s(M,\Nfive)$ is an approximation of the sign function that becomes exact when the size of the $5^{\text{th}}$ dimension goes to infinity:
\beq
s(M,\Nfive)
=\frac{[\mathbb{I}+\HW(M)]^{\Nfive}-[\mathbb{I}-\HW(M)]^{\Nfive}}
{[\mathbb{I}+\HW(M)]^{\Nfive}+[\mathbb{I}-\HW(M)]^{\Nfive}}
\xrightarrow{\Nfive\to\infty}\operatorname{sign}[\HW(M)],
\eeq
with $\HW(M)$ the (Hermitian) kernel
\beq
\HW(M)
=\gamma_5 \DW(-M)
\left[2\mathbb{I}+\DW(-M)\right]^{-1}.
\eeq
In the limit $\Nfive\to \infty$, we thus recover the overlap operator $D$, $\DT\underset{\Nfive\to\infty}{\to} D$. The overlap operator takes the following form:
\beq\label{eq:overlap}
D(\hat{m},M) = \frac{1+\hat{m}}{2} \, 1 + \frac{1-\hat{m}}{2} \, V(M) = (1-\hat{m})D_0(M) + \hat{m},
\eeq
with $V(M) = \gamma_5 \, \sign\left[\HW(M)\right]$, and $D_0(M)$ the massless overlap operator,
\beq\label{eq:overlap_massless}
D_0(M) = \frac{1+V(M)}{2}.
\eeq

At finite $\Nfive$, the approximation to the sign function
introduces chirality violations, whose size must be kept small enough to be negligible. As discussed in Ref.~\cite{Bonanno:2025bla}, the adoption of smeared links in the Wilson kernel allows one to have negligibly small chirality violations even for moderate values of $\Nfive$ (the $\Nfive$ dependence of the $\pi$ and $\rho$ masses with the truncated overlap operator has also been investigated in standard quenched lattice QCD in Ref.~\cite{Murakami:2021kdk}). In particular, in Ref.~\cite{Bonanno:2025bla} we have shown that, adopting 5 stout steps with $\rho_{\sstrm{stout}}=0.1$ (corresponding to a smoothing radius of 2 lattice spacings) and $\Nfive=24$, violations of the Ginsparg--Wilson relation are below $10^{-8}$, and the first few tens of eigenvalues lie on the overlap circle for all analyzed gauge configurations.
We verified that this is the case also for all the ensembles employed in the present work: on all gauge configurations, the first ten eigenvalues of the massless truncated overlap operator lie on the overlap circle, $|\hat\lambda-1/2|=1/2$, with deviations below $5\times10^{-10}$ (see also Fig.~\ref{fig:overlap-circle} in Sec.~\ref{sec:res-condensate}).

The matching between the lattice operator and the physical one depends on the chosen kernel. For the kernel chosen in this work, matching the free lattice operator in momentum space to $\ii \slashed{p}$ gives a factor of $M(2-M)/a$, see~\cite{Aoki:2001su,Bonanno:2025bla}. The dimensionful massless lattice operator and bare quark mass are consequently given by:
\beq\label{eq:overlap_normalization}
\slashed{D}=\frac{M(2-M)}{a}D_0(M),
\qquad
m=\frac{M(2-M)}{a}\,\hat m.
\eeq
Note that, although in Eq.~\eqref{eq:overlap} the mass term appears multiplied by a factor $1/(1-\hat m)$ relative to $D_0$, this factor cancels in the quark propagator, so that  the formulas in Eq.~\eqref{eq:overlap_normalization} hold up to $O(a^2)$ corrections, see Appendix~\ref{app:mass}.
The same normalization applies to the eigenvalues of the massless
overlap operator
\beq
D_0(M) u_{\scriptscriptstyle{\hat{\lambda}}} = \hat{\lambda}(M) u_{\scriptscriptstyle{\hat{\lambda}}}.
\eeq
For the exact overlap operator, these eigenvalues lie on a circle
of radius $1/2$ centered at $(1/2,0)$ in the complex plane. They can
therefore be parametrized as
\beq\label{eq:overlap_spectrum}
\hat{\lambda}(M)=\frac{1}{2}
\left[1+\ee^{\ii\theta(M)}\right].
\eeq
Their dimensionful counterparts, used in the spectral analysis
to extract the chiral condensate, are
\beq\label{eq:scaled_overlap_eigenvalue}
\lambda(M)
= \frac{M(2-M)}{a} \hat{\lambda}(M).
\eeq
After this rescaling, any residual dependence of physical
observables on the kernel parameter $M$ is expected to reflect
discretization effects that vanish in the continuum
limit~\cite{DelDebbio:2004ns}. In Ref.~\cite{Bonanno:2025bla}, we verified that $M=1.2$ and $M=1.5$ gave perfectly compatible results for the eigenvalue spectrum, thus we chose $M=1.2$ in all cases. Recall also that, in the continuum limit, Dirac eigenvalues are purely imaginary, while on the lattice overlap eigenvalues have a real part that vanishes only in the continuum limit. Therefore, following what has been done in Ref.~\cite{Bonanno:2025bla}, in the following we will identify the imaginary part of the physical Dirac eigenvalue with the absolute value of the overlap one.

\subsection{Calculation of meson masses and matrix elements}

Meson masses have been extracted from the large Euclidean-time decay of appropriate zero-momentum correlation functions (recall that at $N=\infty$ all resonances become stable):
\beq\label{eq:meson_correlator}
C_{\A}(\tau)=
\left\langle\mathcal O_{\A}^\dagger(\tau)\mathcal O_{\A}(0)\right\rangle,
\qquad
\mathcal O_{\A}(\tau)=
\sum_{\vec{x}}
\overline{\psi}(\vec{x},\tau)\gamma_{\A}\psi(\vec{x},\tau),
\eeq
where flavor indices are left implicit, and where $\mathrm{A}$ denotes the desired channel. We consider the pseudoscalar $\mathrm{A}=\pi$ and vector $\mathrm{A}=\rho$ channels, corresponding to $\gamma_{\pi}=\gamma_5$ and $\gamma_{\rho}=\gamma_i$, respectively. Following the approach put forward in Ref.~\cite{Gonzalez-Arroyo:2015bya}, the correlators in coordinate space are reconstructed from the Fourier transform of the correlators in momentum space, where momenta are discretized in units of the inverse effective lattice size given by $1/\sqrt{N}$ (see below). 

To improve the projection onto the ground state, we also constructed a basis of interpolating operators and solved a Generalized EigenValue Problem (GEVP):
\beq\label{eq:setup-GEVP}
C_{\A}^{\ell_\s\ell_\s^{\prime}}(\tau)v_{\ell_\s^\prime}
=\bar{\lambda}(\tau,\tau_0)C_{\A}^{\ell_\s\ell_\s^\prime}(\tau_0)v_{\ell_\s^\prime}.
\eeq
In this expression $C_{\A}^{\ell_\s\ell_\s^\prime}(\tau)$ is the inverse Fourier transform of
\beq\label{eq:setup-CA-momentum}
\widetilde{C}_{\A}^{\ell_\s\ell_\s^\prime}(p_0)=
\left\langle\Tr\left[\gamma_{\A}^{\ell_\s}\,\widetilde{S}(0,\vec{0})\,\gamma_{\A}^{\ell_\s^\prime}\,\widetilde{S}(p_0,\vec{0})\right]\right\rangle,
\eeq
where $\widetilde{S}(p)$ represents the fermion propagator in momentum space, while $\gamma_{\A}^{\ell_\s}$, defined as
\beq
\gamma_{\A}^{\ell_\s} \equiv D_{\s}^{\ell_\s} \gamma_{\A}, \qquad D_{\s} \equiv \frac{1}{1+6c}\left[1+c \sum_{i\,=\,1}^{3}\left(\overline{U}_i\otimes\Gamma_i^{*}+\overline{U}^{\dagger}_i\otimes\Gamma_i^{\rm t}\right)\right],
\eeq
denotes the extended $\gamma_{\A}$ operator with a spin-diagonal APE--Wuppertal smearing operator $D_\s$ at smearing level $\ell_\s$. The links $\overline{U}_i \equiv U_i(n_{\sstrm{APE}})$ entering the definition of $D_\s$ are APE-smeared spatial gauge links obtained after $n_{\sstrm{APE}}$ steps. The smeared link is defined iteratively according to:
\begin{equation}
U_i(n+1) = \mathrm{Proj}_{\scriptscriptstyle{\SU(N)}}\left\{(1-f)U_i(n)+\frac{f}{4}\sum_{j \, \ne \, i}\left[z_{ij} U_j(n) U_i(n) U_j^\dagger(n) + z_{ji} U_j^\dagger(n) U_i(n) U_j(n) \right]\right\}
\end{equation}
where $n=0,1,\dots,n_{\sstrm{APE}}$ counts the number of three-dimensional APE-smearing steps ($n=0$ is the original link). In this study we employed the same definition of $D_\s$ and the operator basis used for Wilson fermions in Ref.~\cite{Bonanno:2025hzr}: we used 10 steps of three-dimensional APE smearing with $f=0.081$, and the operator basis was constructed by defining $D_\s$ with $c=0.4$ and considering $\ell_\s=16,36,64,100$ levels of Wuppertal smearing. The trace in Eq.~\eqref{eq:setup-CA-momentum} has been evaluated with a stochastic noise method using $n_{\sstrm{src}}=5$ sources. Finally, to extract the PCAC mass and the pion decay constant, we computed suitable pseudoscalar correlation functions obtained by replacing one of the $\gamma_{5}^{\ell_\s}$ factors in Eq.~\eqref{eq:setup-CA-momentum} with $\gamma_0\gamma_5$ (i.e., without the meson smearing operator). More details are given in Ref.~\cite{Bonanno:2025hzr}.

In Eq.~\eqref{eq:setup-CA-momentum} we employ the subtracted propagator defined by
\beq\label{eq:setup-subtracted-propagator}
\widetilde{S}(p)=\frac{1}{1-\hat{m}}\left[\DTp^{-1}(p)-\mathbb{I}\right],
\eeq
where $\DTp(p)$ is the truncated overlap operator of Eq.~\eqref{eq:truncated_overlap} constructed with a momentum-dependent Wilson kernel instead of the one in Eq.~\eqref{eq:Wilson_kernel}. The Wilson kernel carrying momentum $p$ is defined by:
\beq
\widetilde{D}_{\scriptscriptstyle{\rm W}}(p,-M)=(4-M)\mathbb{I}
-\frac{1}{2}\sum_{\mu=1}^{4}
\left[
(1-\gamma_\mu)\otimes\widetilde{\mathcal{W}}_{\mu}
+(1+\gamma_\mu)\otimes\widetilde{\mathcal{W}}_{\mu}^{\dagger}
\right],
\eeq
with $\widetilde{\mathcal{W}}_{\mu}=U^{\sstrm{(stout)}}_\mu\otimes e^{\ii a p_\mu}\Gamma_{\mu}^*$, see Ref.~\cite{Gonzalez-Arroyo:2015bya}. The momenta are discretized as $ap_i = 2\pi m_i / L$ and $a p_0 = 2\pi m_0 / T$, where $L=\sqrt{N}$ and $T=\ell_0 L= \ell_0\sqrt{N}$ (we chose $\ell_0=2$), and where the integers $m_i$ and $m_0$ run over $m_i=0,1,\dots,L-1$ and $m_0=0,1,\dots,T-1$. In the $\Nfive\to\infty$ limit, the propagator satisfies the Ward identity
\beq
\left\{\widetilde{S},\gamma_5\right\}=2\hat{m}\,\widetilde{S}\gamma_5\widetilde{S}
\eeq
thanks to the Ginsparg--Wilson relation.

From Eq.~\eqref{eq:setup-GEVP}, the ground-state $\pi$ and $\rho$ masses are determined from the optimized correlators in two ways: either via a cosh fit of the best-overlapping operator (obtained by solving the GEVP at fixed times $\tau_0$ and $\tau=\tau_0 + \delta$, $\delta > 0$, and by contracting $C_{\A}^{\ell_\s\ell_\s^\prime}$ with the eigenvector relative to the largest GEVP eigenvalue), or from a constant fit of the effective mass built from the largest GEVP eigenvalue, see~\cite{Bonanno:2025hzr} for more details. The two methods give compatible results in all cases.

\section{Effective theory predictions} \label{sec:eff}
In this study, we will match our calculations to effective theory predictions to extract physical quantities. We briefly summarize these predictions in this section.

\subsection{Chiral Perturbation Theory}
At leading order in the momentum and quark-mass expansion, the chiral effective theory describing QCD with two degenerate light flavors is characterized by two low-energy constants, the scheme- and scale-dependent quark condensate and the pion decay constant:
\beq
\Sigma \equiv -\lim_{m\to0}\langle\bar\psi\psi\rangle, \qquad
 \sqrt{2}F=\lim_{m\to0}\frac{1}{m_\pi}
 \langle0|\bar{\psi}\gamma_0 \gamma_5 \psi|\pi,\mathbf{p}=\mathbf{0}\rangle.
\eeq
The adopted normalization for $F$ corresponds to $F_\pi\simeq92$~MeV at the physical point for $N=3$. The large-$N$ behavior of these quantities is:
\beq
 \Sigma\sim O(N),\qquad F\sim O(\sqrt{N}),
 \label{eq:chpt-largeN}
\eeq
so that $\Sigma/N$ and $F/\sqrt{N}$ remain finite in the large-$N$ limit. These low-energy couplings enter the celebrated Gell-Mann--Oakes--Renner (GMOR) relation, expressing the quark-mass dependence of the pion mass. At leading order of $\chi$PT, this relation reads:
\beq
\label{eq:GMOR_LO}
m_\pi^2 = M_\pi^2 [1 + O(M_\pi^2)], \qquad M_\pi^2= 2 B_\R m_\R,
\eeq
where the subscript “R" denotes renormalized quantities and
\beq
B_\R\equiv\frac{\Sigma_\R}{F^2}\sim O(N^0).
\eeq
Although $\Sigma_\R$ and $B_\R$ depend on the renormalization scheme and scale, the products $m_\R\Sigma_\R$ and $m_\R B_\R$ are renormalization-group invariant.

At NLO of the chiral expansion, one has to introduce further LECs~\cite{Gasser:1983yg,Gasser:1984gg,Colangelo:2001df} (more precisely, seven more in the two-flavor effective theory). In this paper we will be interested in the two NLO couplings parameterizing the sub-leading quark mass dependence of the pion mass and of the pion decay constant, namely, $\bar{\ell}_3$ and $\bar{\ell}_4$ (recall that the bar denotes renormalization-group-invariant quantities). At NLO, neglecting chiral logs which are sub-leading in the large-$N$ limit, the GMOR relation becomes:
\beq
\label{eq:GMOR_NLO}
m_\pi^2=M_\pi^2\left[1 - \frac{1}{2}\bar{\ell}_3 x + O(x^2)\right], \qquad x = \left(\frac{M_\pi}{4\pi F}\right)^2\sim O\left(\frac{1}{N}\right), \qquad \bar{\ell}_3\sim O(N).
\eeq
Recall that chiral logs in this formula would show up in the form $\mathcal{L}(M_\pi) x$ --- where $\mathcal{L}(M_\pi)=\log(M_\pi^2/M^2_{\pi,\rm phys})$ and $M_{\pi,\rm phys}=135$ MeV --- with $x\sim O(1/N)$, thus they vanish at infinite $N$. Using the expression for $M_\pi^2$ in terms of the quark mass, one gets:
\beq
\label{eq:chpt-pion-NLO}
m_\pi^2=2B_\R m_\R\left[1-\frac{\bar\ell_3}{N}\frac{B_\R}{16\pi^2F^2/N} m_\R+O(m_\R^2)\right], \qquad \frac{\bar\ell_3}{N}\sim O(N^0).
\eeq
Concerning the quark-mass dependence of the pion decay constant at NLO in $\chi$PT, neglecting again sub-leading chiral logs, one has:
\beq
F_\pi = F[1 + \bar{\ell}_4 x + O(x^2)],
\eeq
where now $F_\pi$ denotes the pion decay constant away from the chiral limit. Since $m_\pi^2=M_\pi^2[1+O(M_\pi^2)]$, this formula can be equivalently rewritten (at this order of the chiral expansion) as:
\beq
\label{eq:chpt-F-NLO}
F_\pi=F\left[1+\frac{\bar\ell_4}{N} \frac{1}{16\pi^2F^2/N} m_\pi^2+O(m_\pi^4)\right], \qquad \frac{\bar\ell_4}{N}\sim O(N^0).
\eeq

Finally, in this paper we also consider the vector meson $\rho$, which is not part of the pseudo-Goldstone bosons arising from spontaneous chiral symmetry breaking. Thus, building a low-energy effective theory for vector mesons requires further assumptions. A proposal for a large-$N$ low-energy effective theory for vector mesons has been put forward in Ref.~\cite{Jenkins:1995vb}, where the authors build the vector effective Lagrangian in terms of heavy static matter fields. The prediction for the large-$N$ quark-mass dependence of the $\rho$ mass is:
\beq
m_{\rho} = m_{\rho}^{\chir} + C_\rho m_{\pi}^2 + O(m_\pi^4),
\eeq
where $m_{\rho}^{\chir}\sim\mathcal{O}(N^0)$ denotes the non-zero chiral limit of the $\rho$ meson mass, and $C_\rho\sim\mathcal{O}(N^0)$. It is worth recalling that, in the presence of quenched quarks and at finite $N$, the prediction is that the quark mass dependence of $m_{\rho}$ is actually given by a power series in $\sqrt{m}\sim\mathcal{O}(m_\pi)$; however, the coefficients of all odd powers of $m_\pi$ are $O(1/N)$ suppressed~\cite{Jenkins:1995vb,Booth:1996hk} (see also~\cite{Bali:2013kia} for a numerical verification), thus these terms do not appear at large $N$.

\subsection{Random Matrix Theory}

We will determine the condensate directly from the low-lying spectrum of the massless overlap Dirac operator, following the same Random Matrix Theory (RMT) strategy already adopted in Ref.~\cite{Bonanno:2025bla} as well as in other lattice studies of the Dirac spectrum~\cite{Edwards:1999ra,Damgaard:1999tk,Hernandez:1999cu,Bietenholz:2003mi,Giusti:2003gf,Narayanan:2004cp,Gonzalez-Arroyo:2005dgf,Wennekers:2005wa,JLQCD:2007ppn,TWQCD:2007eit,Bernardoni:2008ei,Hasenfratz:2008ce,Bernardoni:2010nf,Deuzeman:2011dh,Catillo:2017qbz}, which we will briefly summarize in the following. The appropriate universality class to describe large-$N$ QCD is the chiral unitary RMT~\cite{Leutwyler:1992yt,Shuryak:1992pi,Verbaarschot:1993pm,Verbaarschot:1994qf,Nishigaki:1998is}. In a fixed topological sector, RMT predicts universal probability densities $p_k(z)$ for the $k^{\rm th}$ eigenvalue $z_k$ of the random matrix ensemble. As an example, for the $\nu=0$ topological sector, the distributions of the two lowest eigenvalues and of their ratio $r=z_1/z_2$ are~\cite{Narayanan:2004cp,Damgaard:2000ah}
\begin{align}
 p_1(z_1)&=\frac{z_1}{2}\exp\left(-\frac{z_1^2}{4}\right),
 \label{eq:setup-RMT-p1}\\
 p_2(z_2)&=\frac{z_2}{4}\exp\left(-\frac{z_2^2}{4}\right)
 \int_0^{z_2} \dd u\,u\left[I_2(u)^2-I_1(u)I_3(u)\right],
 \label{eq:setup-RMT-densities}\\
 p(r)&=\frac{1}{4}\,\frac{r}{(1-r^2)^2}
 \int_0^{\infty} \dd u\,\exp\left(-\frac{u^2}{4(1-r^2)}\right)u^3\left[I_2(u)^2-I_1(u)I_3(u)\right],
 \label{eq:setup-RMT-ratio}
\end{align}
where $I_n(x) = \int_0^{2 \pi} \frac{\dd\theta}{2 \pi} \, \ee^{\ii \theta n+x\cos(\theta)}$ is the $n^{\rm th}$ modified Bessel function of the first kind.

In the infinite-volume limit, $V \to \infty$, the following identification is expected to hold in each topological sector for the eigenvalues of the massless Dirac operator:
\beq \label{eq:QCD_RMT_matching}
z_k = \Sigma V \lambda_k, \qquad \slashed{D}v_{\lambda_k} = \ii \lambda_k v_{\lambda_k}, \qquad \lambda_k \in \mathbb{R}.
\eeq
For the reduced model, following Ref.~\cite{Bonanno:2025bla}, we assume that the identification between the microscopic RMT variables and Dirac eigenvalues is modified as:
\beq\label{eq:RMT_QCD_eigval_matching_largeN_TEK}
z_k = \frac{\Sigma}{N} V \lambda_k, \qquad V=a^4 N^2,
\eeq
where $\lambda_k$ is computed from the absolute value of the $k^{\rm th}$ overlap eigenvalue, cf.~Eq.~\eqref{eq:scaled_overlap_eigenvalue}. This assumption was successfully tested in Ref.~\cite{Bonanno:2025bla}, and it will be further checked below in Sec.~\ref{sec:res-condensate}.
However, on the lattice we compute the spectrum of the massless chiral Dirac operator at finite volume, and deviations from the predictions of chiral RMT are therefore expected. As $V \to \infty$, these deviations should become progressively smaller and occur at increasingly higher values of $k$. Assuming Eq.~\eqref{eq:RMT_QCD_eigval_matching_largeN_TEK}, such deviations manifest themselves as a dependence of the extracted value of $\Sigma / N$ on the choice of $k$, which should disappear as the effective volume is increased.

The applicability of the effective RMT description to the lattice can be first checked in a parameter-free and scale-independent way by studying, as a function of the volume, the deviation from unity of the ratios:
\beq \label{eq:ratio_rmt}
R_{k_1,k_2} \equiv \frac{\left\langle \lambda_{k_1}\right\rangle}{\left\langle\lambda_{k_2}\right\rangle} \times \left(\frac{\left\langle z_{k_1}\right\rangle_{\RMT}}{\left\langle z_{k_2}\right\rangle_{\RMT}}\right)^{-1}.
\eeq
Here $\braket{\mathcal{O}}$ and $\braket{\mathcal{O}}_{\RMT}$ denote, respectively, the expectation values taken in the full theory and in the RMT effective model. Once the correct volume regime has been established, the matching between lattice data and analytic results allows us to extract the bare quark condensate, assuming the relation~\eqref{eq:RMT_QCD_eigval_matching_largeN_TEK}, in two different ways. The first consists in performing a one-parameter best fit of the probability distributions of the lowest lattice Dirac eigenvalues to the corresponding RMT predictions. The second one relies on the use of the analytic RMT predictions for the expectation values of the Dirac eigenvalues:
\beq\label{eq:RMT_condensate}
\frac{\Sigma}{N} = \frac{1}{V}\frac{\left\langle z_k\right\rangle_{\RMT}}{\left\langle\lambda_k\right\rangle}.
\eeq
These two approaches, as already shown in~\cite{Bonanno:2025bla}, lead to perfectly compatible estimates of the bare quark condensate; we explicitly verify this also for the present ensembles in Sec.~\ref{sec:res-condensate}. In this work, we therefore restrict ourselves to the second approach for the final determination of the condensate, and determine $\Sigma/N$ from Eq.~\eqref{eq:RMT_condensate}.

\section{Numerical results}\label{sec:res}
In this section we present the chiral and continuum extrapolation of the $\pi$ and $\rho$ mass and of the $\pi$ decay constant, and the continuum limit of the chiral condensate extracted from the low-lying overlap Dirac spectrum using RMT.

\subsection{Simulation details}\label{sec:simulation_details}
In this paper we will consider two different sets of simulations, whose parameters are summarized in Tabs.~\ref{tab:ensembles} and~\ref{tab:ensembles-mesons}. All raw data can be found in Appendix~\ref{app:data}.
\begin{itemize}
\item A set of simulations, performed in the chiral limit $\hat{m}=0$, will be used to compute the first four low-lying eigenvalues of the massless overlap operator, and extract the chiral condensate using RMT predictions. We will consider 3 different lattice spacings, corresponding to bare couplings $b=0.355,0.360,0.365$. For each lattice spacing, we have considered 4 different values of the number of colors, ranging from $N=289$ up to $N=1156$, corresponding to effective sizes $\ell\sqrt{\sigma}=aL\sqrt{\sigma}=a \sqrt{N}\sqrt{\sigma}$ ranging from $\sim 3.5$ up to $\sim 7$. These will be instrumental to monitor the approach of the computed eigenvalues to the RMT predictions as $\ell\to\infty$.
\item Another set of simulations will be employed for the computation of the quark-mass dependence of $m_\pi$, $m_\rho$ and $F_\pi$. In this case we consider again 3 bare couplings $b=0.355,0.360,0.365$, and for each of them, 4 non-zero values of the bare quark mass $\hat{m}$. In this case, due to the much larger computational cost compared to the computation of the first few low-lying eigenvalues, we have employed $N=361$ for $b=0.355$, $N=529$ for $b=0.365$, and for $b=0.360$ $N=361$ for the two heaviest masses and $N=529$ for the two lightest ones. These choices are motivated by the necessity of ensuring that $\ell\sqrt{\sigma} \gtrsim 4$ and $m_\pi \ell = a m_\pi L = a m_\pi \sqrt{N}\gtrsim 4$, values that were observed to be sufficient to keep finite size effects below the typical statistical errors in the Wilson meson paper~\cite{Bonanno:2025hzr}. At $b=0.360$, the two lightest masses have also been simulated at $N=361$. These data, reported in Appendix~\ref{app:data} for completeness, have not been included in the analysis, and have only been employed to check finite-size effects.
\end{itemize}

\begin{table}[!t]
\centering
\renewcommand{\arraystretch}{1.15}
\begin{tabular}{ccccccccc}
\hline\hline
$b$ & $N$ & $L$ & $k$ & $\bar k$ & $a\sqrt{\sigma}$ & $a/\sqrt{8t_1}$ & $\ell\sqrt{\sigma}$ & $n_{\rm conf}$ \\
\hline
\multirow{4}{*}{$0.355$} & $289$ & $17$ & $5$ & $7$  & \multirow{4}{*}{$0.2410(30)$} & \multirow{4}{*}{$0.37707(57)$} & $4.097(51)$ & $400$\\
                         & $361$ & $19$ & $7$ & $11$ & & & $4.579(57)$ & $300$\\
                         & $529$ & $23$ & $7$ & $10$ & & & $5.543(69)$ & $420$\\
                         & $841$ & $29$ & $9$ & $13$ & & & $6.989(87)$ & $240$\\
\hline
\multirow{4}{*}{$0.360$} & $289$ & $17$ & $5$ & $7$  & \multirow{4}{*}{$0.2058(25)$} & \multirow{4}{*}{$0.31776(50)$} & $3.499(42)$ & $326$\\
                         & $361$ & $19$ & $7$ & $11$ & & & $3.910(48)$ & $200$\\
                         & $529$ & $23$ & $7$ & $10$ & & & $4.733(58)$ & $220$\\
                         & $841$ & $29$ & $9$ & $13$ & & & $5.968(72)$ & $210$\\
\hline
\multirow{4}{*}{$0.365$} & $361$  & $19$ & $7$  & $11$ & \multirow{4}{*}{$0.1783(17)$} & \multirow{4}{*}{$0.27181(74)$} & $3.388(32)$ & $400$\\
                         & $529$  & $23$ & $7$  & $10$ & & & $4.101(39)$ & $345$\\
                         & $841$  & $29$ & $9$  & $13$ & & & $5.171(49)$ & $220$\\
                         & $1156$ & $34$ & $13$ & $21$ & & & $6.062(58)$ & $200$\\
\hline\hline
\end{tabular}
\caption{Summary of the simulation parameters for the massless Dirac spectrum ($\hat{m}=0$). The column $n_{\mathrm{conf}}$ stands for the number of employed gauge configurations. The string tension $\sigma$ was determined in Ref.~\cite{GonzalezArroyo:2012bh}, the gradient flow time $t_1$ in Refs.~\cite{Butti:2023hfp,Bonanno:2025hzr}.}
\label{tab:ensembles}
\end{table}
 \begin{table}[!t]
\centering
\renewcommand{\arraystretch}{1.15}
\begin{tabular}{cccccccc}
\hline\hline
$b$ & $N$ & $L$ & $k$ & $\bar k$ & $n_{\rm conf}$ & $\hat{m}$ & $am$ \\
\hline
\multirow{4}{*}{$0.355$} & \multirow{4}{*}{$361$} & \multirow{4}{*}{$19$} & \multirow{4}{*}{$7$} & \multirow{4}{*}{$11$} & $396$ & $0.025$ & $0.02400$\\
 & & & & & $385$ & $0.042$ & $0.04032$\\
 & & & & & $400$ & $0.063$ & $0.06048$\\
 & & & & & $400$ & $0.088$ & $0.08448$\\
\hline
\multirow{4}{*}{$0.360$} & \multirow{2}{*}{$529$} & \multirow{2}{*}{$23$} & \multirow{2}{*}{$7$} & \multirow{2}{*}{$10$} & $301$ & $0.020$ & $0.01920$\\
 & & & & & $244$ & $0.034$ & $0.03264$\\
\cline{2-8}
 & \multirow{2}{*}{$361$} & \multirow{2}{*}{$19$} & \multirow{2}{*}{$7$} & \multirow{2}{*}{$11$} & $400$ & $0.055$ & $0.05280$\\
 & & & & & $400$ & $0.077$ & $0.07392$\\
\hline
\multirow{4}{*}{$0.365$} & \multirow{4}{*}{$529$} & \multirow{4}{*}{$23$} & \multirow{4}{*}{$7$} & \multirow{4}{*}{$10$} & $209$ & $0.020$ & $0.01920$\\
 & & & & & $186$ & $0.025$ & $0.02400$\\
 & & & & & $200$ & $0.046$ & $0.04416$\\
 & & & & & $199$ & $0.064$ & $0.06144$\\
\hline\hline
\end{tabular}
\caption{Same as Tab.~\ref{tab:ensembles}, for the simulations used to compute meson observables ($m_\pi$, $m_\rho$, $F_\pi$). The values of $a\sqrt{\sigma}$, $a/\sqrt{8t_1}$ and $\ell\sqrt{\sigma}$ for each $(b,N)$ can be read from Tab.~\ref{tab:ensembles}. The bare mass in lattice units $am$ is related to the input mass $\hat{m}$ via $am = M(2-M)\hat{m}=0.96\, \hat{m}$ for our choice $M=1.2$.}
\label{tab:ensembles-mesons}
\end{table}
We will set the scale in units of the reference gradient flow~\cite{Narayanan:2006rf,Luscher:2010iy} time $t_1$:
\beq
\frac{1}{N} \braket{t^2 E(t)}\bigg\vert_{t \, = \, t_1} = 0.05,
\eeq
with $E(t)=\frac{1}{2}\Tr\left[G_{\mu\nu}(t)G_{\mu\nu}(t)\right]$ the action density computed at positive flow time $t>0$ (discretized on the lattice via the usual clover formulation of the field strength). The final dimensionless results will be converted to string tension units using:
\beq
 \label{eq:t1_sigma_conversion}
 \sqrt{8t_1\sigma}=0.6759(92),
\eeq
a continuum conversion factor obtained in Ref.~\cite{Bonanno:2025hzr}.

We use $t_1$ in place of the more common gradient flow reference scale $t_0$:
\beq\label{eq:def_t0}
\frac{N}{N^2-1} \braket{t^2 E(t)}\bigg\vert_{t \, = \, t_0} = 0.1125,
\eeq
where Eq.~\eqref{eq:def_t0} reflects the conventional definition introduced in Ref.~\cite{Ce:2016awn} to extend the $N=3$ scale to $N>3$, since $t_1$ is found to suffer from smaller finite-size effects (at the price of being affected by slightly larger cut-off effects).

Since determinations of $a/\sqrt{8t_1}$ are more precise than those of $a\sqrt{\sigma}$, this procedure ultimately yields smaller uncertainties on dimensionless ratios. As a consistency check, we have explicitly verified that repeating the whole analysis setting the scale directly in terms of $a\sqrt{\sigma}$, without passing through $t_1$, leads to results compatible within errors for all quantities. Errors obtained with this scale-setting procedure turn out in all cases to be larger by about $15\%$, except for the chiral condensate, whose relative error increases by about $40\%$ when scale-setting is not performed passing through $t_1$. The reason is that, being $\Sigma$ a dimension-three observable, the relative uncertainty on the chosen scale-setting quantity is amplified by a factor of 3 when propagated to the chiral condensate.

\subsection{Quark mass and renormalization constants}\label{sec:res-quark-mass}
\begin{figure}[!t]
 \centering
 \includegraphics[width=\textwidth]
 {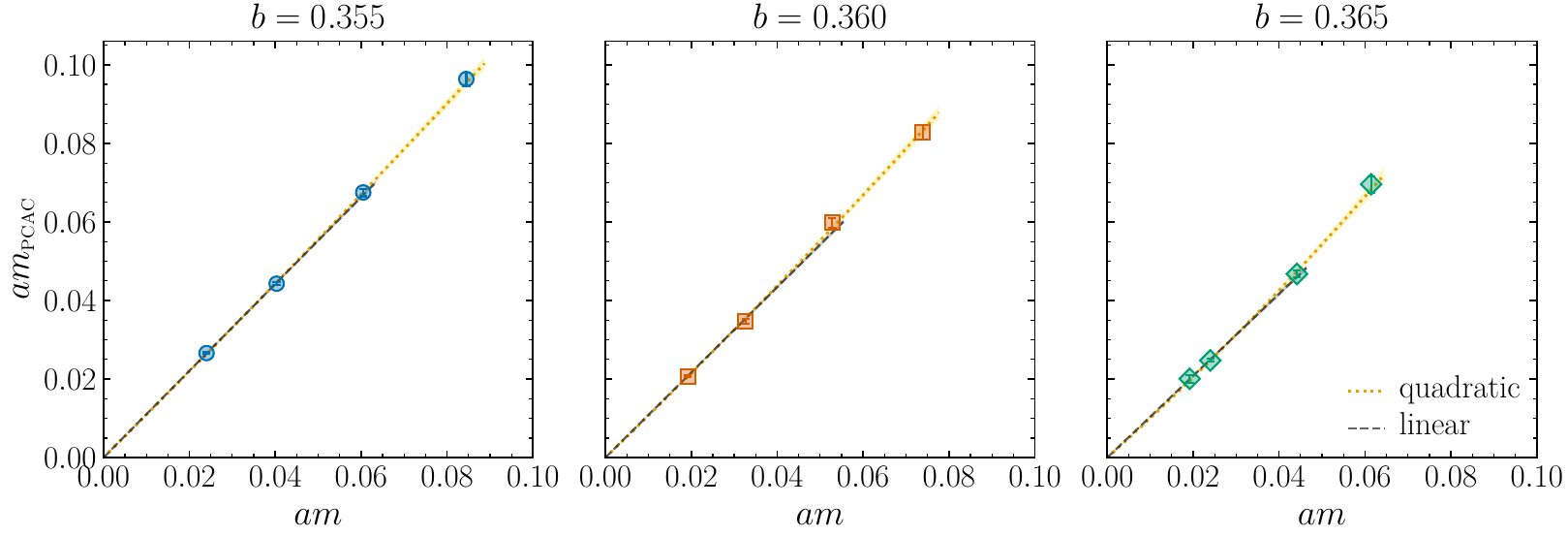}
\caption{PCAC mass as a function of the bare overlap quark mass, in lattice units. Dashed black lines show the linear fits through the origin of Eq.~\eqref{eq:Fpi-PCAC-fit} to the three lightest masses, from which the values of $\ZA$ in Tab.~\ref{tab:Fpi-ZA} are obtained, dotted orange lines show quadratic fits through the origin including all points.}
\label{fig:Fpi-PCAC}
\end{figure}

With overlap fermions, chiral symmetry protects the quark mass from additive renormalizations, unlike the Wilson case. We tested this expectation by studying the PCAC mass, obtained as usual from the axial Ward identity as in Ref.~\cite{Bonanno:2025hzr}, to which we refer for further details, as a function of the bare quark mass $am=M(2-M)\hat{m}$. The same analysis also provides the finite renormalization factor $\ZA$ needed to obtain the physical $F_\pi$. Indeed, for overlap fermions $\ZP=\ZS$ and $m_\R = m / \ZS$, thus:
\beq
m_\PCAC=\frac{\ZP}{\ZA}m_\R = \frac{\ZP}{\ZA} \frac{m}{\ZS}
 = \frac{m}{\ZA}.
 \label{eq:Fpi-Ward-identity}
\eeq
We thus fitted the PCAC mass at each bare coupling according to:
\beq
a m_\PCAC=\frac{am}{\ZA}=\frac{M(2-M)}{\ZA}\hat{m},
 \label{eq:Fpi-PCAC-fit}
\eeq
to determine $\ZA$. The results of the fit are reported in Tab.~\ref{tab:Fpi-ZA} and shown in Fig.~\ref{fig:Fpi-PCAC}. At all bare couplings the fit includes the three lightest masses, the heaviest one showing a departure from the linear behavior. Including also the heaviest points and adding a quadratic term, $am_\PCAC = am/\ZA + \beta_2(am)^2$, we obtain $\ZA=0.920(14)$, $0.948(17)$ and $1.023(34)$ at $b=0.355$, $0.360$ and $0.365$, compatible with the values in Tab.~\ref{tab:Fpi-ZA} within at most two standard deviations. A linear fit to the three lightest masses with a free intercept gives intercepts compatible with zero within about one standard deviation. The PCAC mass is thus linear in the bare mass to a very good approximation, and it vanishes at vanishing bare mass, confirming the absence of an additive renormalization of the overlap quark mass.

\begin{table}[t]
 \centering
 \begin{tabular}{ccc}
 \hline\hline
 $b$ & $\ZA$ & $\chi^2/{\rm dof}$\\
 \hline
 $0.355$ & $0.9050(55)$  & $1.05/2$\\
 $0.360$ & $0.9223(80)$  & $5.67/2$\\
 $0.365$ & $0.960(12)$ & $1.10/2$\\
 \hline\hline
 \end{tabular}
 \caption{Axial renormalization factors obtained from the linear  behavior of the PCAC mass.}
 \label{tab:Fpi-ZA}
\end{table}
    
\begin{figure}[!t]
\centering
\includegraphics[width=\textwidth]{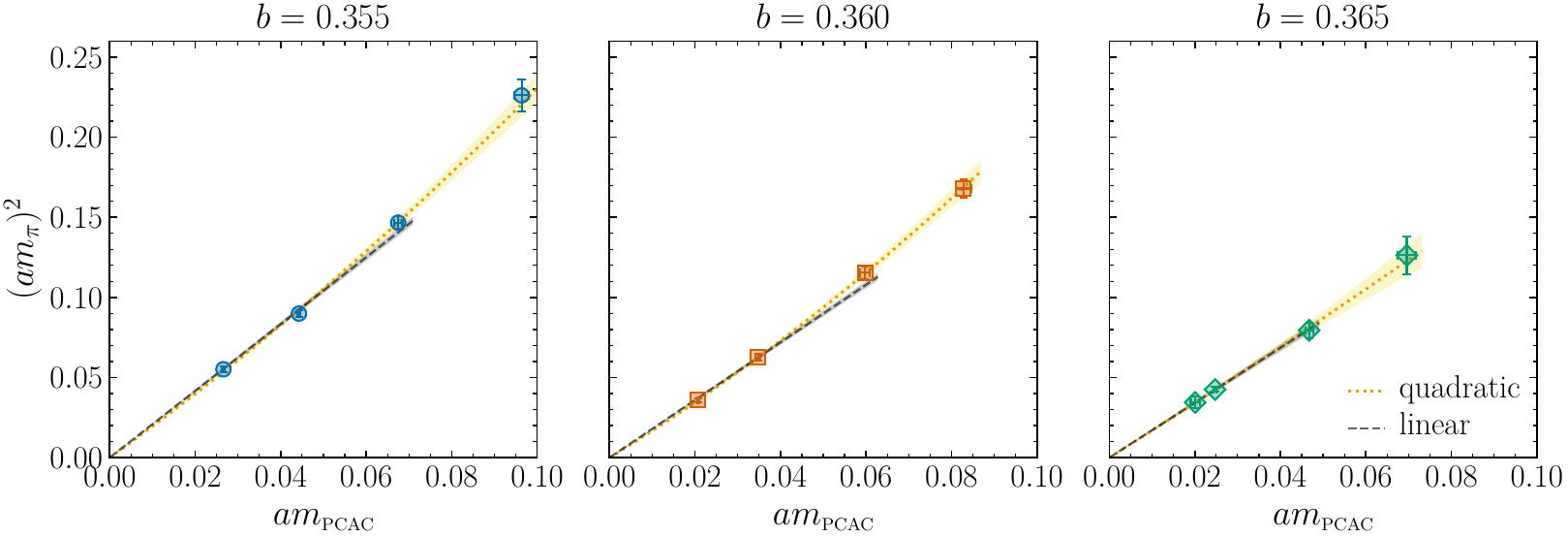}
\caption{Squared pion mass as a function of the PCAC mass, in lattice units, for the three bare couplings. Dashed black lines show linear fits through the origin to the three lightest masses, dotted orange lines show quadratic fits through the origin, Eq.~\eqref{eq:mpi-bare-fit} with $\alpha_0=0$, including all points.}
\label{fig:mpi-mpcac}
\end{figure}

As a further check, we fitted the squared pion mass as a function of the PCAC mass according to:
\beq
(am_\pi)^2 = \alpha_0 + \alpha_1\, am_\PCAC + \alpha_2\, (am_\PCAC)^2,
\label{eq:mpi-bare-fit}
\eeq
where the quadratic term, again, is required only to describe the data in the whole explored mass range. The results are shown in Fig.~\ref{fig:mpi-mpcac} and collected in Tab.~\ref{tab:mpi-bare}. The intercept $\alpha_0$ is compatible with zero within $0.1$ and $0.4$ standard deviations at $b=0.360$ and $b=0.365$, respectively, and within $1.3$ standard deviations at $b=0.355$. Imposing $\alpha_0=0$, the fits remain of good quality at all bare couplings. Our data thus confirm that the pion becomes massless when the quark mass vanishes, as expected for the Goldstone bosons of spontaneously broken chiral symmetry.

\begin{table}[t]
\centering
\begin{tabular}{cccc}
\hline\hline
$b$ & $\alpha_0$ & $\chi^2/{\rm dof}$ & $\chi^2/{\rm dof}$ ($\alpha_0=0$)\\
\hline
$0.355$ & $0.0129(96)$  & $0.12/1$ & $1.91/2$\\
$0.360$ & $-0.0006(66)$ & $0.01/1$ & $0.02/2$\\
$0.365$ & $0.007(18)$   & $0.06/1$ & $0.22/2$\\
\hline\hline
\end{tabular}
\caption{Results of the quadratic fits of Eq.~\eqref{eq:mpi-bare-fit} to the squared pion mass as a function of the PCAC mass, in lattice units. The last column refers to the fits with vanishing intercept.}
\label{tab:mpi-bare}
\end{table}

\subsubsection{Scalar renormalization constant}\label{sec:res-ZS}

To compare the overlap results at different lattice spacings with the Wilson ones, the quark mass and the chiral condensate must be renormalized. Lattice chiral symmetry implies $\Zm \ZS = 1$, therefore:
\beq
 \Sigma_\R = \ZS \Sigma,
 \qquad
 m_\R = \Zm m = \frac{m}{\ZS},
 \label{eq:overlap-renormalization}
\eeq
where $m$ is the bare quark mass, and $m_\R $ the renormalized overlap valence quark mass. The renormalization scheme and scale of $\Sigma_\R$ and $m_\R$ are inherited from those used to define $\ZS$. From now on, we will always quote renormalized quantities in the
$\overline{\mathrm{MS}}$ scheme at the scale $\mu\sqrt{8t_0}=4.30$, following Refs.~\cite{DeGrand:2023hzz,Bonanno:2025bla,Bonanno:2025hzr}. This corresponds to the customary $\mu = 2\,\mathrm{GeV}$ scale assuming $\sqrt{t_0} = 0.15\,\mathrm{fm}$.

\begin{figure}[!t]
 \centering
 \includegraphics[width=.8\textwidth]
 {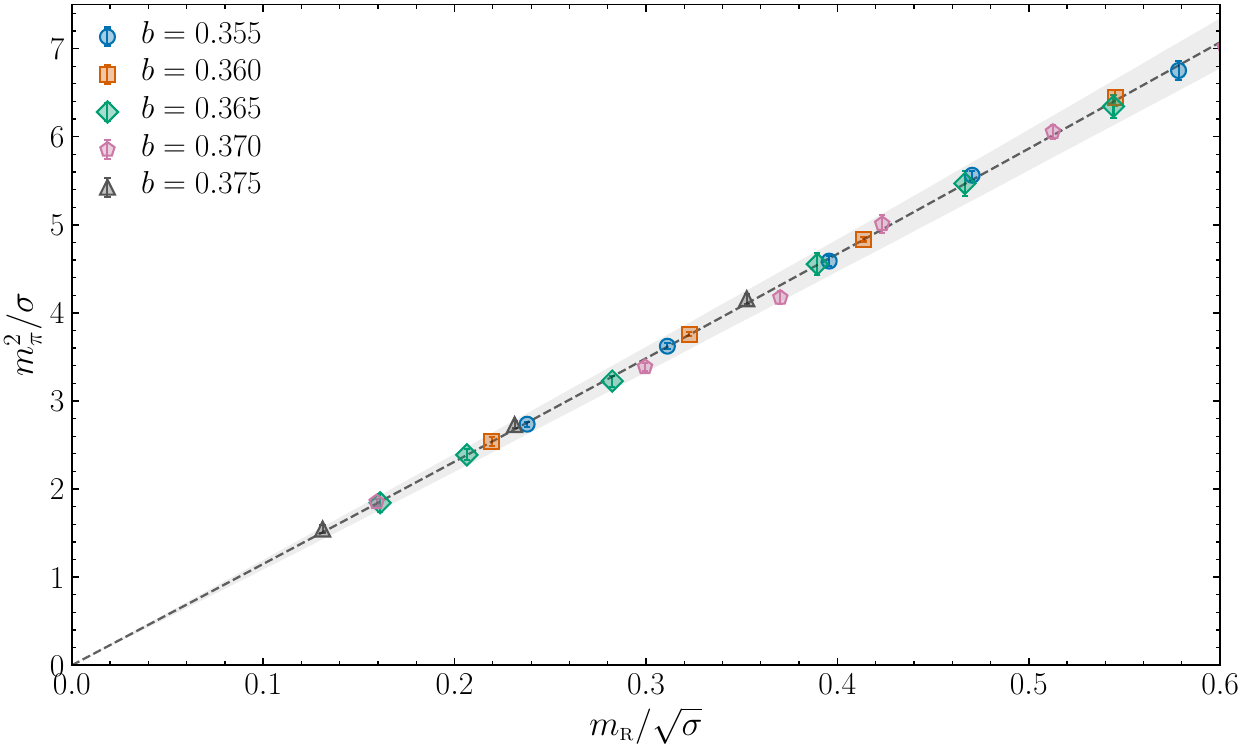}
 \caption{Continuum determination of the squared pion mass as a function of the renormalized quark mass obtained from the TEK data with Wilson quarks of Refs.~\cite{Bonanno:2025hzr,Bonanno:2026gay}. The renormalized quark mass $m_\R$ is obtained from the Wilson hopping parameter according to Eq.~\eqref{eq:Wilson-renormalized-mass}, in the $\overline{\mathrm{MS}}$ scheme at $\mu \sqrt{8t_0}=4.30$, corresponding to $\mu=2$ $\mathrm{GeV}$ for $\sqrt{t_0}=0.15$ $\mathrm{fm}$. The fitted lattice artifact terms $A_1a\sqrt\sigma+A_3a\sqrt\sigma\,m_\R/\sqrt{\sigma}$ have been subtracted from the data. The black curve and shaded band, corresponding to the continuum limit of the best fit ($\chi^2/\mathrm{dof}=15.7/26$), allow the overlap-Wilson matching of the quark mass at equal pion mass.}
 \label{fig:mR_vs_mpi}
\end{figure}

For overlap quarks, we follow Ref.~\cite{Wennekers:2005wa} and determine $\ZS$ by matching the overlap bare quark mass to the continuum renormalized quark mass at the same pion mass (computed from Wilson fermions). This procedure is repeated for each overlap ensemble, and then the chiral limit of $\ZS$ is taken. The matching condition reads:
\beq\label{eq:ZS_overlap}
\ZS = \frac{ m(m_\pi) }{m_{\R}(m_\pi)}.
\eeq
The function $m_{\R}(m_\pi)$ entering this equation is determined
(in the continuum limit) from the Wilson data of Ref.~\cite{Bonanno:2025hzr}, complemented by the new determinations of Ref.~\cite{Bonanno:2026gay}. For each value of the Wilson hopping parameter $\kappa$ and bare coupling $b$ we compute:
\beq
 m_\R(b,\kappa)
 =\frac{1}{\ZS^{\chirW}(b)} \frac{1}{a(b)}
 \left(\frac{1}{2\kappa}-\frac{1}{2\kappa_c(b)}\right),
 \label{eq:Wilson-renormalized-mass}
\eeq
with $\ZS^{\chirW}(b)$ the scalar renormalization constant computed for Wilson fermions in the chiral limit, and $\kappa_c(b)$ the critical Wilson hopping parameter. Then we perform the following global chiral-continuum fit of the squared pion mass as a function of the renormalized quark mass:
\beq
\frac{m_\pi^2(b,\kappa)}{\sigma} = A_1 \, x_1(b) + A_2 \, y(b,\kappa) + A_3 \, x_1(b)\, y(b,\kappa) + A_4\, y^2(b,\kappa),
\label{eq:Wilson-matching-fit}
\eeq
where
\beq
x_1 = a\sqrt{\sigma}, \qquad \quad y = \frac{m_\R}{\sqrt{\sigma}},
\eeq
and the $A_i$ are four fit parameters.

\begin{figure}[!t]
\centering
\includegraphics[width=.49\textwidth]{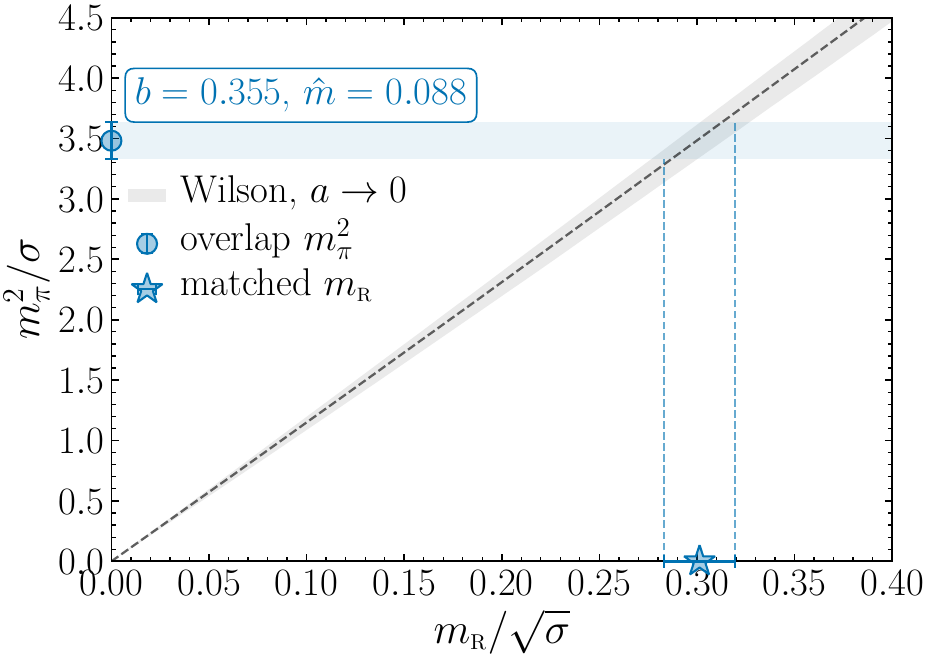}
\includegraphics[width=.49\textwidth]{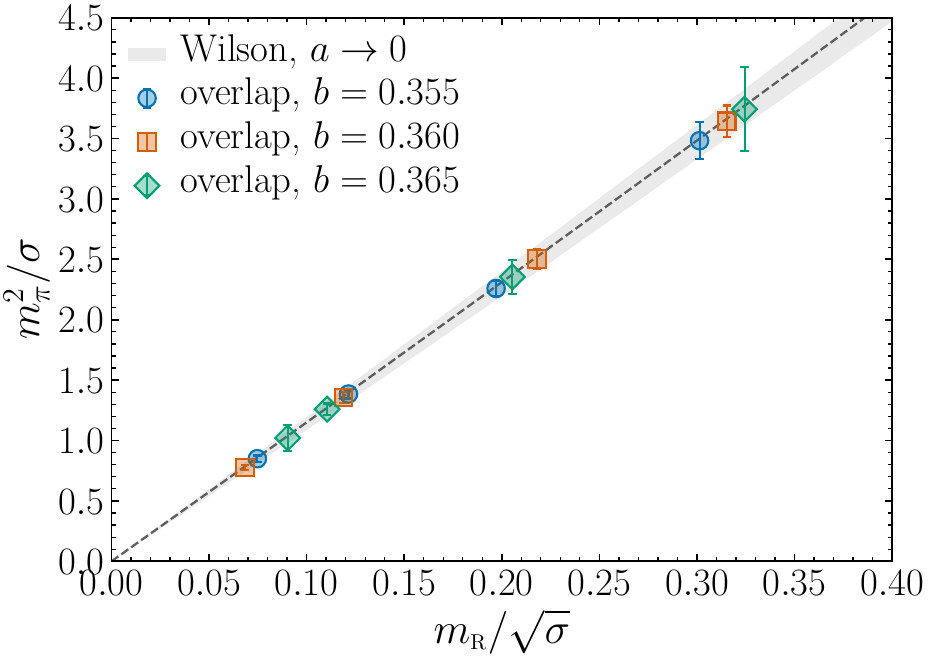}
\caption{Left: example of the matching between the overlap pion mass ($b=0.355$, $\hat m=0.088$) and the Wilson continuum curve of Eq.~\eqref{eq:Wilson-continuum-mass}; the horizontal band is the overlap $m_\pi^2/\sigma$ and the star the resulting $m_\R/\sqrt\sigma$. Right: all overlap determinations at their matched renormalized quark mass, compared with the Wilson continuum curve.}
\label{fig:Wilson_overlap_match}
\end{figure}

\noindent In the continuum limit, the squared pion mass as a function of the renormalized quark mass thus becomes:
\beq
 \frac{m_\pi^2}{\sigma}
 =A_2\frac{m_\R}{\sqrt{\sigma}}
 +A_4\left(\frac{m_\R}{\sqrt{\sigma}}\right)^2,
 \label{eq:Wilson-continuum-mass}
\eeq
which is inverted to obtain $m_\R(m_\pi)$. In Fig.~\ref{fig:mR_vs_mpi}, we show the curve defined in Eq.~\eqref{eq:Wilson-continuum-mass}, together with the fitted data points at $b=0.355, 0.360, 0.365, 0.370, 0.375$, after subtracting the lattice artifacts, i.e., the contributions proportional to $A_1$ and $A_3$. The best fit, already performed in Ref.~\cite{Bonanno:2026gay}, gives $\chi^2/\mathrm{dof}=15.7/26$. We note that $A_2/2=B_\R/\sqrt{\sigma}=5.68(30)$, in very good agreement with the value $B_\R/\sqrt{\sigma}=5.58(26)$ obtained in Ref.~\cite{Bonanno:2025hzr} from separate chiral fits, performed at each value of $b$, to the quark-mass dependence of the pion mass, followed by a continuum limit. From $A_4$ one also gets a bound on the NLO LEC $\bar{\ell}_3/N=-0.02(2)$, already reported in Ref.~\cite{Bonanno:2026gay}, which turns out to be very small (as can be seen from the linearity of the curve in Fig.~\ref{fig:mR_vs_mpi}).

The function $m_\R(m_\pi)$ obtained in this way, which is all that enters the matching, can be equivalently determined from a fit of $m_\R$ as a function of $m_\pi^2$. We checked that the resulting $\ZS$ are compatible within errors with those obtained from Eq.~\eqref{eq:Wilson-matching-fit}. We adopt Eq.~\eqref{eq:Wilson-matching-fit} as our main determination, since the pion mass is the quantity used to define the matching point.

\begin{figure}[!t]
 \centering
 \includegraphics[width=\textwidth]
 {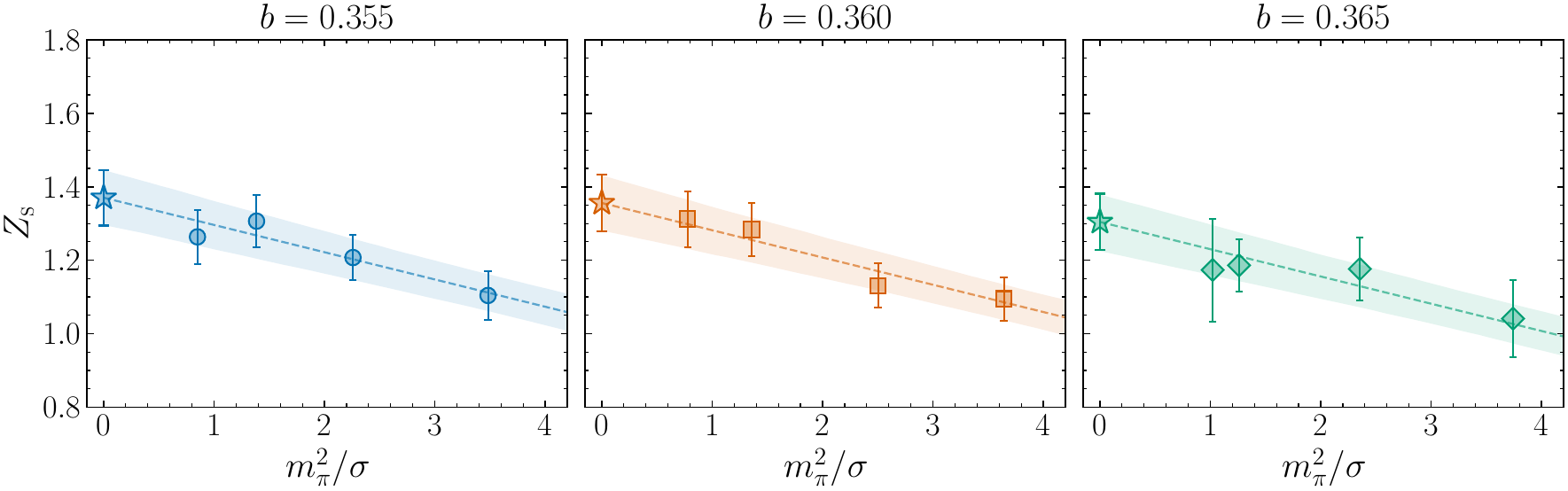}
 \caption{Chiral extrapolation of the overlap scalar renormalization factors with a common slope. The stars at $m_\pi^2=0$ denote the chiral extrapolations $\ZS^{\chir}$ used to renormalize the chiral condensate.}
 \label{fig:ZS-chiral}
\end{figure}

For each overlap measurement of the pion mass we use Eq.~\eqref{eq:Wilson-continuum-mass} to compute $\ZS$. The results are collected in Tab.~\ref{tab:ZS_overlap}. The matching procedure is shown in Fig.~\ref{fig:Wilson_overlap_match}. The matching factor has a mild dependence on the matching mass. We remove this dependence by fitting all determinations simultaneously to:
\beq
\ZS(b,m_\pi)=\ZS^{\chir}(b) + d_{\scriptscriptstyle{Z}} \frac{m_\pi^2}{\sigma},
 \label{eq:ZS-chiral-fit}
\eeq
with an independent chiral intercept for each $b$, and a common slope $d_{\scriptscriptstyle{Z}}$. This choice is supported by the compatible slopes observed by performing individual fits for each lattice spacing. The global fit, shown in Fig.~\ref{fig:ZS-chiral}, gives $d_{\scriptscriptstyle{Z}}=-0.074(15)$ and $\chi^2/\mathrm{dof}=2.47/8$. The chiral extrapolations are reported in Tab.~\ref{tab:ZS_overlap} as the $m_\pi=0$ values.
The final errors on $\ZS^{\chir}(b)$ are estimated with a parametric bootstrap. For each bootstrap sample, all observables at the same lattice spacing are resampled via $\mathcal{O}_{\rm boot}(b) = \braket{O}(b)+x(b)\times\sqrt{\mathrm{Var}_{_\mathcal{O}}(b)}$, with $x(b)$ a Gaussian variable with zero mean and unit variance. The only quantity that is not resampled is $\ZS^{\chirW}(b)$, since these were obtained from a large-$N$ extrapolation of the finite-$N$ results of~\cite{Castagnini:2015ejr} (see the previous TEK study~\cite{Bonanno:2025hzr} for more details). For each bootstrap sample, the best fit of the quark-mass dependence of $m_\pi$ is done considering only uncertainties on the pion mass. Once this is done, the matching and the chiral extrapolation of the scalar overlap renormalization constants are repeated for each bootstrap sample. Finally, central values and errors on $\ZS^{\chir}(b)$ are determined from the mean and standard deviation of the bootstrap samples.

\begin{table}[t]
 \centering
 \begin{tabular}{cc@{\hspace{0.9cm}}cc@{\hspace{0.9cm}}cc}
 \hline\hline
 \multicolumn{2}{c}{$b=0.355$} &
 \multicolumn{2}{c}{$b=0.360$} &
 \multicolumn{2}{c}{$b=0.365$}\\
 $m_\pi^2/\sigma$ & $\ZS$ &
 $m_\pi^2/\sigma$ & $\ZS$ &
 $m_\pi^2/\sigma$ & $\ZS$\\
 \hline
 3.482 & 1.104(67) & 3.643 & 1.095(59) & 3.742 & 1.04(10) \\
 2.259 & 1.207(62) & 2.502 & 1.131(61) & 2.355 & 1.176(86) \\
 1.385 & 1.306(70) & 1.356 & 1.283(72) & 1.260 & 1.186(71) \\
 0.850 & 1.264(74) & 0.777 & 1.312(76) & 1.021 & 1.17(14) \\
 0     & 1.370(76) & 0     & 1.356(76) & 0     & 1.304(77) \\
 \hline\hline
 \end{tabular}
 \caption{Scalar renormalization factor $\ZS$ for overlap quarks in the $\overline{\mathrm{MS}}$ scheme at $\mu\sqrt{8t_0}=4.30$, corresponding to $\mu=2$ $\mathrm{GeV}$ for $\sqrt{t_0}=0.15$ $\mathrm{fm}$, obtained matching the overlap bare quark mass to the continuum renormalized quark mass (computed from Wilson quarks) at equal pion masses. Errors are obtained from the parametric bootstrap described in the text. The $m_\pi=0$ results are obtained by means of a chiral extrapolation, see the text for more details.} \label{tab:ZS_overlap}
\end{table}

Since the overlap quark mass is renormalized by matching to the Wilson continuum curve at equal pion mass, the resulting quark-mass dependence of the overlap pion mass reproduces by construction the continuum Wilson one. Thus, we do not quote independent overlap determinations of $B_\R$ and $\bar\ell_3$. We only use $\ZS$ to renormalize the chiral condensate.

\subsection{Chiral condensate from the Dirac spectrum}
\label{sec:res-condensate}
\begin{figure}[!t]
 \centering
 \includegraphics[width=\textwidth]{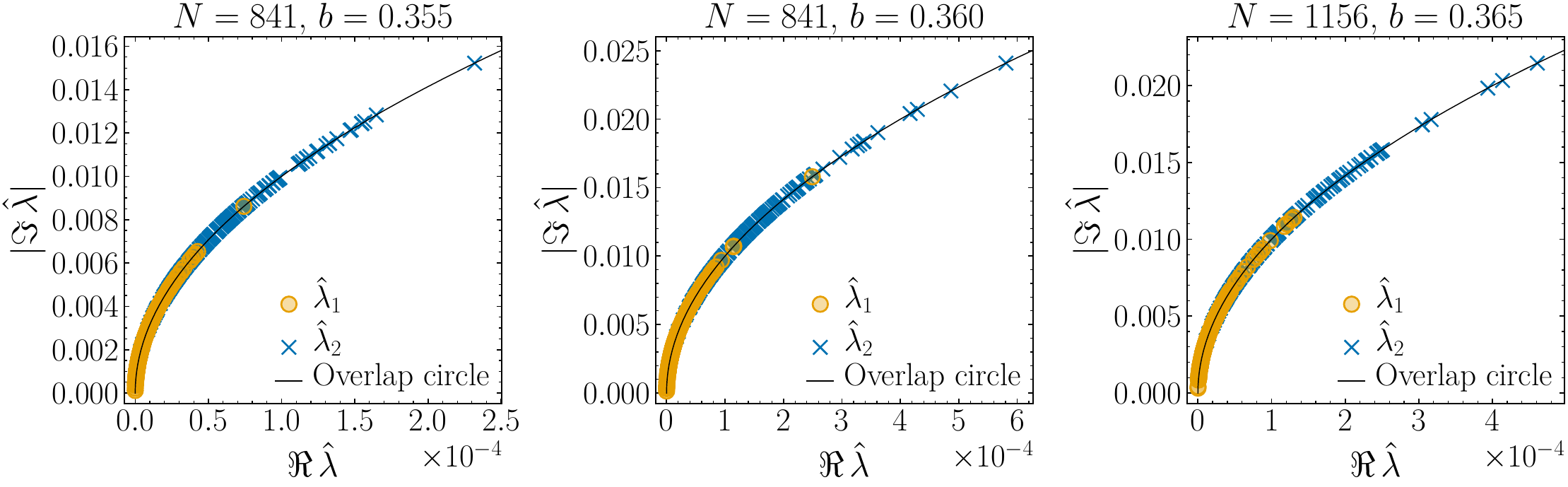}
 \caption{First two eigenvalues $\hat\lambda_1$ and $\hat\lambda_2$ (with positive imaginary part) of the massless truncated overlap Dirac operator, obtained with $M=1.2$, $\Nfive=24$ and 5 stout-smearing steps with $\rho_{\sstrm{stout}}=0.1$, on all gauge configurations at the largest value of $N$ for each bare coupling: $N=841$ at $b=0.355$ and $b=0.360$, $N=1156$ at $b=0.365$. The solid black line denotes the overlap circle, Eq.~\eqref{eq:overlap_spectrum}.}
 \label{fig:overlap-circle}
\end{figure}

As a preliminary check of the chiral properties of our lattice Dirac operator, in Fig.~\ref{fig:overlap-circle} we show the two lowest eigenvalues of the massless overlap operator, $\hat\lambda_1$ and $\hat\lambda_2$, computed on all gauge configurations at the largest value of $N$ for each bare coupling. As anticipated in Sec.~\ref{sec:setup}, all of them lie on the overlap circle of Eq.~\eqref{eq:overlap_spectrum}.
 
\begin{figure}[!t]
 \centering
 \includegraphics[width=\textwidth]
 {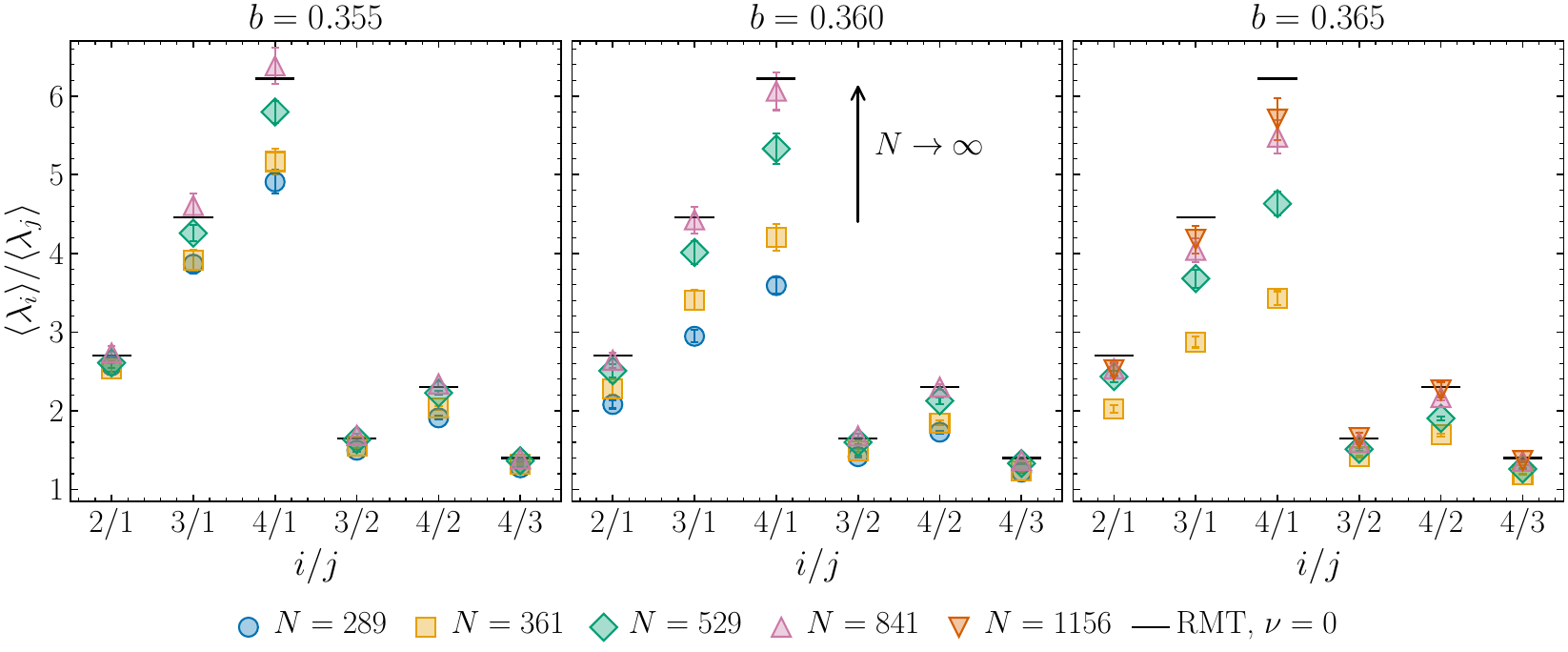}
 \caption{Ratios of expectation values of the low-lying overlap Dirac  eigenvalues for three bare couplings. All determinations of a given ratio are displayed at the same horizontal position, while colors and markers identify the number of colors $N$. Black segments denote the parameter-free RMT predictions. The arrow in the central panel indicates the direction $N\to\infty$, i.e., the infinite-volume limit.}
 \label{fig:rmt-ratios}
\end{figure}

We start by checking whether the low-lying overlap eigenvalues lie in the
universal RMT regime. To this end, we consider the scale-independent ratios
introduced in Eq.~\eqref{eq:ratio_rmt}, using all combinations of the first
four eigenvalues. The numerical values of the corresponding RMT predictions in the $\nu=0$ topological sector, taken from Tab.~VII of Ref.~\cite{Bernardoni:2008ei}, are reported in Tab.~\ref{tab:rmt-ratios}, together with the measured ratios at the largest value of $N$ for each bare coupling. Since the factor $(\Sigma/N)V$ cancels in these ratios,
this comparison contains no adjustable parameter. Exact agreement with RMT
corresponds to
\beq
\frac{\langle\lambda_{k_1}\rangle}
     {\langle\lambda_{k_2}\rangle}
=
\frac{\langle z_{k_1}\rangle_{\RMT}}
     {\langle z_{k_2}\rangle_{\RMT}}.
 \label{eq:ratio-RMT}
\eeq
Our results are shown in Fig.~\ref{fig:rmt-ratios} (given that no zero mode was found in our spectra, we employed the $\nu=0$ RMT predictions), and the values of $R_{k_1,k_2}$ for all ensembles are collected in Tab.~\ref{tab:rawdata-ratios} in Appendix~\ref{app:data}. For each bare coupling, we display six eigenvalue ratios and compare them
directly with the corresponding parameter-free RMT predictions. All determinations of a given ratio are placed at the same horizontal position, while colors and markers distinguish the values of $N$. Since the effective volume is $V=a^4(b)N^2$, increasing $N$ at fixed $b$ corresponds to approaching the infinite-volume limit, as indicated by the arrow in the
central panel.
\begin{table}[!t]
\centering
\renewcommand{\arraystretch}{1.2}
\begin{tabular}{cccccccc}
\hline\hline
 & & $2/1$ & $3/1$ & $4/1$ & $3/2$ & $4/2$ & $4/3$\\
\hline
\multicolumn{2}{c}{RMT} & $2.70$ & $4.46$ & $6.22$ & $1.65$ & $2.30$ & $1.40$\\
\hline
$b=0.355$ & $N=841$  & $2.733(89)$ & $4.61(15)$ & $6.38(23)$ & $1.686(24)$ & $2.341(38)$ & $1.388(13)$\\
$b=0.360$ & $N=841$  & $2.641(96)$ & $4.43(17)$ & $6.06(24)$ & $1.680(25)$ & $2.299(39)$ & $1.368(14)$\\
$b=0.365$ & $N=1156$ & $2.476(87)$ & $4.10(16)$ & $5.59(23)$ & $1.655(42)$ & $2.260(69)$ & $1.366(16)$\\
\hline\hline
\end{tabular}
\caption{RMT predictions for the ratios $\langle z_{k_1}\rangle_{\RMT}/\langle z_{k_2}\rangle_{\RMT}$ of the expectation values of the lowest eigenvalues in the $\nu=0$ topological sector, taken from Tab.~VII of Ref.~\cite{Bernardoni:2008ei}, compared with the measured ratios $\langle\lambda_{k_1}\rangle/\langle\lambda_{k_2}\rangle$ at the largest value of $N$ for each bare coupling. The RMT values are shown as black segments in Fig.~\ref{fig:rmt-ratios}; the results for all values of $N$ are reported in Tab.~\ref{tab:rawdata-ratios} in Appendix~\ref{app:data}.}
\label{tab:rmt-ratios}
\end{table}

\begin{figure}[!t]
 \centering
 \includegraphics[width=\textwidth]{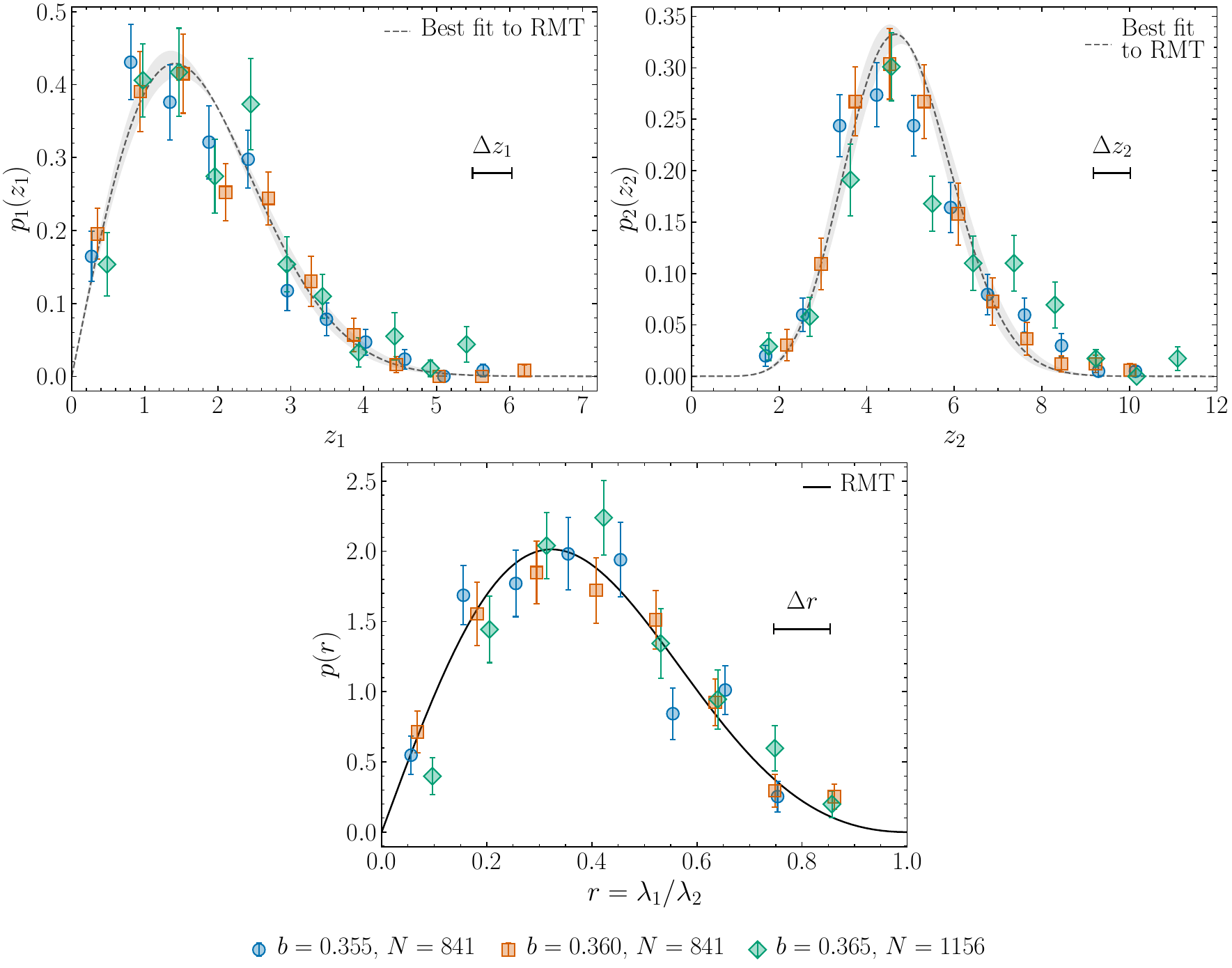}
 \caption{Distributions of the lowest (top left) and second-lowest (top right) eigenvalue in RMT units, $z_k=\lambda_k(\Sigma/N)V$ (cf.~Eq.~\eqref{eq:RMT_QCD_eigval_matching_largeN_TEK}), and of the ratio $r=\lambda_1/\lambda_2$ (bottom), for the largest value of $N$ at each bare coupling. Left and central panels: the data of each ensemble are converted to RMT units with the value of $\Sigma/N$ obtained from its own one-parameter best fit to $p_1(z_1)$ or $p_2(z_2)$; the dashed curve is the best fit to RMT, and the shaded band shows its uncertainty (envelope of the bands obtained by varying $\Sigma/N$ within its error for each ensemble). Right panel: the solid curve is the parameter-free RMT prediction of Eq.~\eqref{eq:setup-RMT-ratio}. All curves refer to the $\nu=0$ topological sector. Errors are obtained with a blocked bootstrap (blocks of five configurations). In all panels we also show the constant bin sizes $\Delta z_1$, $\Delta z_2$ and $\Delta r$ used to obtain the displayed distributions.}
 \label{fig:rmt-distributions-all}
\end{figure}

At the smallest effective volumes we observe clear deviations from RMT, which are largest for larger eigenvalue ratios. As $N$ is increased, the ratios move towards the RMT predictions. At the largest values of $N$, all measured ratios agree with the RMT predictions within at most $2.6$ standard deviations (cf.~Tab.~\ref{tab:rmt-ratios}), the largest deviations being observed at $b=0.365$ for the ratios involving the lowest eigenvalue.
The approach to the RMT regime can be tested in a more stringent way by looking at the full probability distributions of the lowest eigenvalues at the largest effective volumes. In Fig.~\ref{fig:rmt-distributions-all} we show, for the largest available value of $N$ at each bare coupling, the distributions of $z_1$ and $z_2$ and of the ratio $r=\lambda_1/\lambda_2$. For $z_1$ and $z_2$, the only free parameter, $\Sigma/N$, is determined separately for each ensemble by a one-parameter best fit of the binned distribution to $p_1(z_1)$ and $p_2(z_2)$, cf.~Eqs.~\eqref{eq:setup-RMT-p1} and~\eqref{eq:setup-RMT-densities}. Once expressed in RMT units, the data at the three bare couplings fall on the same universal curves. The distribution of $r$, instead, does not depend on $\Sigma/N$, and is compared with the parameter-free RMT prediction, Eq.~\eqref{eq:setup-RMT-ratio}.
The resulting values of $\Sigma/N$ are reported in Tab.~\ref{tab:rmt-fit}, together with those extracted from the mean values $\langle\lambda_1\rangle$ and $\langle\lambda_2\rangle$ through Eq.~\eqref{eq:RMT_condensate}, and with the reduced chi-squared $\tilde\chi^2\equiv\chi^2/\rm dof$ and $p$-values of the fits. The distribution of the lowest eigenvalue is well described by RMT at all bare couplings, and so is the parameter-free distribution of $r$. The distribution of $z_2$ is well reproduced at $b=0.360$, while at $b=0.355$ and $0.365$ the data are somewhat broader than the RMT prediction, resulting in larger values of $\tilde\chi^2$. At $b=0.365$, this is likely due to the limited statistics of the $N=1156$ ensemble, whose second eigenvalue also shows a stronger autocorrelation than in the other ensembles, so that the errors on the corresponding distribution are likely underestimated. A large $\tilde\chi^2$ in the fit of $p_2(z_2)$ was also found at $b=0.355$, $N=529$ in Ref.~\cite{Bonanno:2025bla}. In all cases, the condensates obtained from the fits agree, within at most $1.6$ standard deviations, with those obtained from the mean values, as already found in Ref.~\cite{Bonanno:2025bla}. We recall that RMT has also been used in the past to study the low-lying Dirac spectrum and to estimate the chiral condensate in large-$N$ reduced and twisted lattice models, see Refs.~\cite{Narayanan:2004cp,Gonzalez-Arroyo:2005dgf}.
 
\begin{table}[!t]
\centering
\renewcommand{\arraystretch}{1.15}
\begin{tabular}{ccccccc}
\hline\hline
 & $N$ & $b$ & $\tilde\chi^2$ & $p$-value & \makecell{$10^4\times a^3\Sigma/N$\\ $[$RMT fit$]$} & \makecell{$10^4\times a^3\Sigma/N$\\ $[$Eq.~\eqref{eq:RMT_condensate}$]$}\\
\hline
\multirow{3}{*}{$\lambda_1$} & $841$  & $0.355$ & $0.96$ & $47\%$ & $9.43(34)$ & $9.37(39)$\\
                             & $841$  & $0.360$ & $1.03$ & $41\%$ & $5.58(23)$ & $5.61(22)$\\
                             & $1156$ & $0.365$ & $1.19$ & $29\%$ & $3.47(14)$ & $3.25(19)$\\
\hline
\multirow{3}{*}{$\lambda_2$} & $841$  & $0.355$ & $2.15$ & $2\%$  & $9.49(18)$ & $9.27(23)$\\
                             & $841$  & $0.360$ & $0.57$ & $84\%$ & $5.86(11)$ & $5.75(11)$\\
                             & $1156$ & $0.365$ & $4.42$ & $<0.01\%$ & $3.93(12)$ & $3.55(20)$\\
\hline\hline
\end{tabular}
\caption{Extraction of the bare condensate from the first two low-lying eigenvalues, both from the one-parameter best fit to the RMT predictions for their probability distributions and via Eq.~\eqref{eq:RMT_condensate}, for the largest value of $N$ at each $b$ (same ensembles as in Fig.~\ref{fig:rmt-distributions-all}). We also report the reduced chi-squared, $\tilde\chi^2\equiv\chi^2/\rm dof$, and the corresponding $p$-value. Errors are obtained with a blocked bootstrap.}
\label{tab:rmt-fit}
\end{table}

This behavior shows that finite-volume effects need to be properly taken into account in the determination of the chiral condensate. We therefore extract a bare condensate $\Sigma_k(N,b)$ in lattice units from each of the first four eigenvalues through:
\beq
\frac{a^3\Sigma_k(N,b)}{N}
 =\frac{\langle z_k\rangle_{\RMT}}
 {N^2\langle a\lambda_k\rangle},
 \label{eq:condensate-from-means-results}
\eeq
and extrapolate the four determinations jointly to $V\to\infty$, equivalent to $N\to\infty$ at fixed $b$, before performing the renormalization and continuum extrapolation.

\begin{figure}[!t]
 \centering
 \includegraphics[width=0.75\textwidth]
 {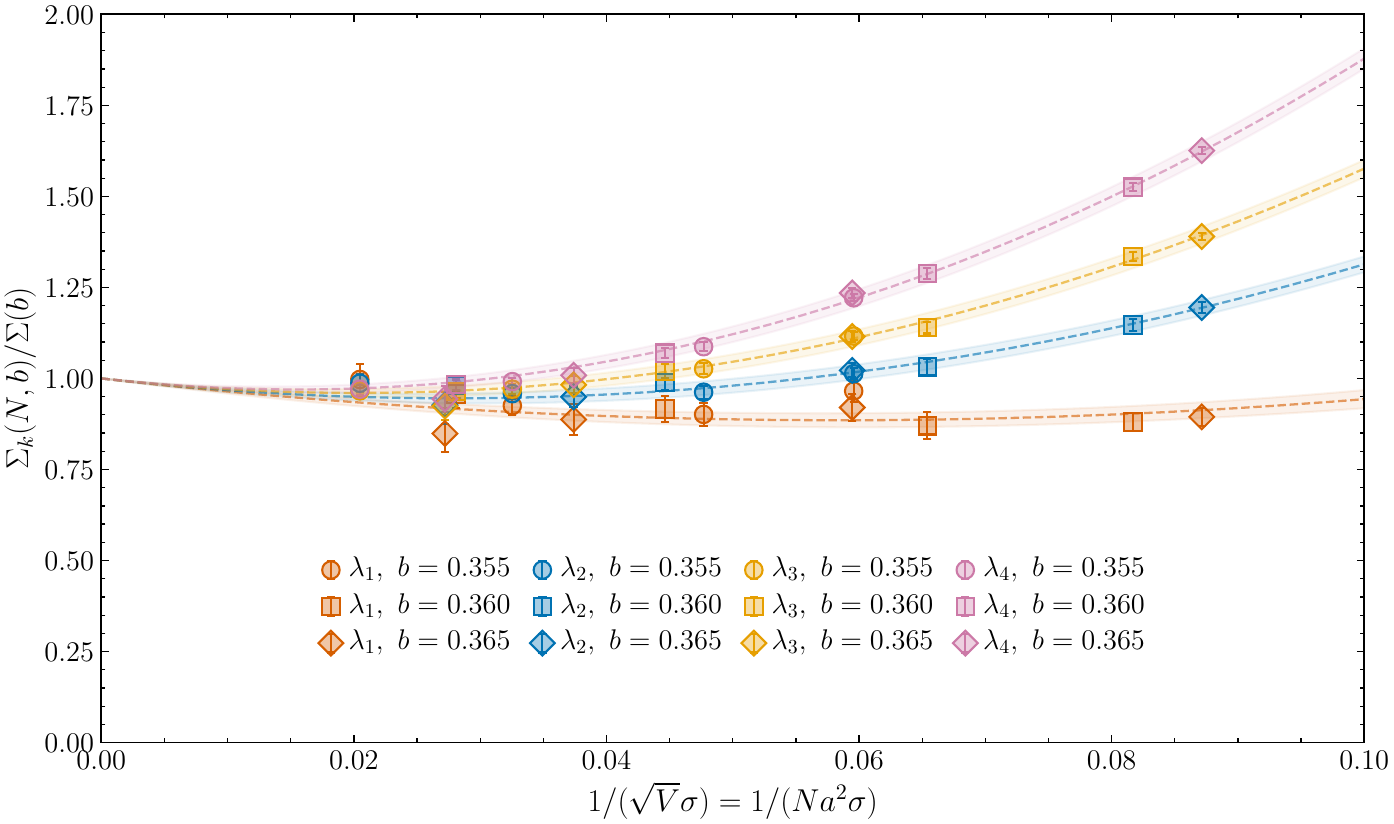}
 \caption{Infinite-volume extrapolation of the bare condensate extracted from the first four overlap eigenvalues. Results are normalized to the $N\to\infty$ condensate at each $b$. Colors distinguish the eigenvalue index and markers distinguish the value of the bare coupling.}
 \label{fig:condensate-infinite-volume}
\end{figure}

\begin{figure}[!t]
 \centering
\includegraphics[width=0.75\textwidth]{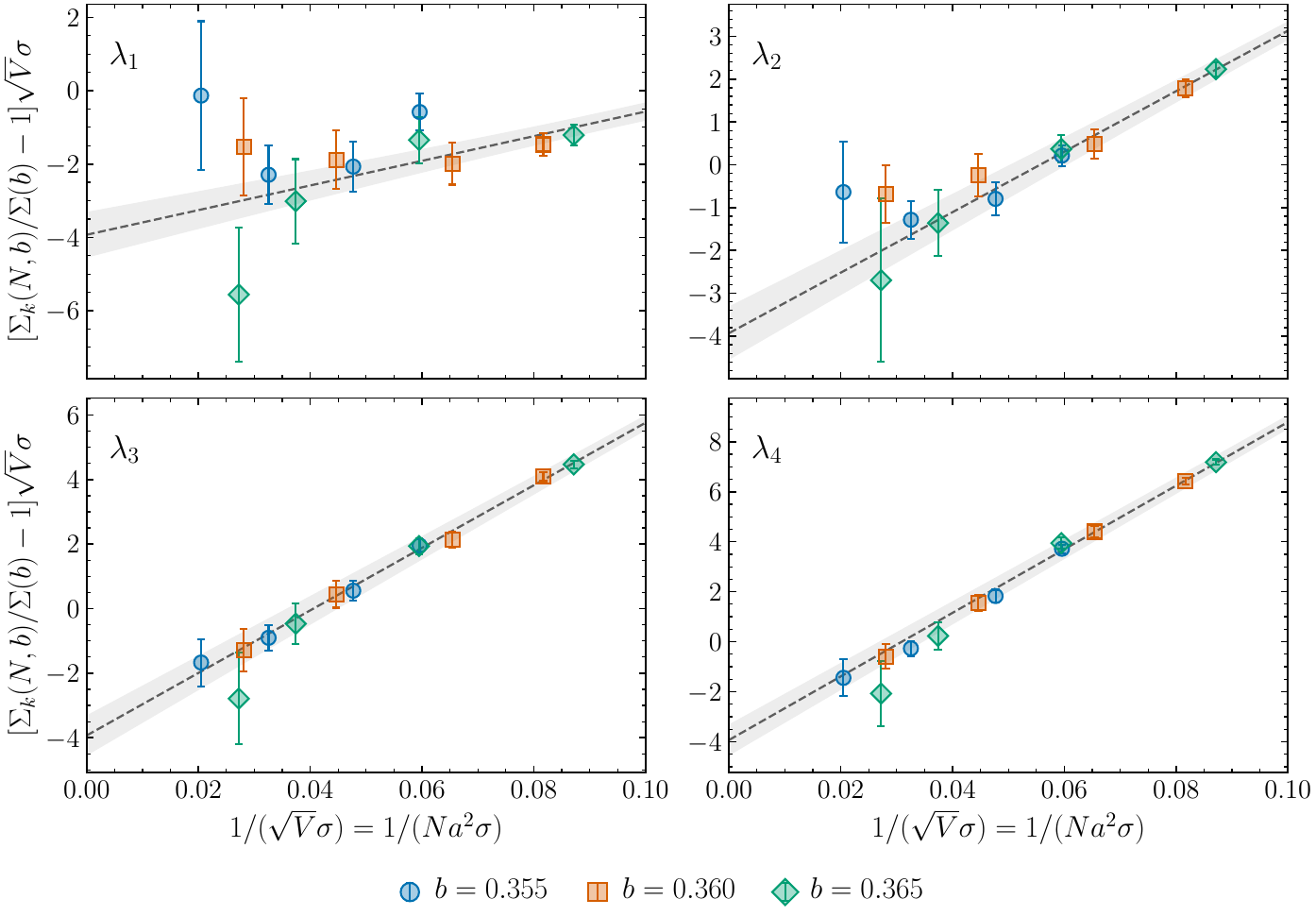}
 \caption{Same data and global fit as in Fig.~\ref{fig:condensate-infinite-volume}, shown in the linear form of Eq.~\eqref{eq:results-RMT-linear}, with one panel for each eigenvalue. Colors and markers distinguish the bare coupling. The dashed lines and grey bands show the global fit and its uncertainty.}
 \label{fig:condensate-infinite-volume-linear}
\end{figure}

The $V\to\infty$ extrapolation is carried out adopting the same best fit ansatz of Ref.~\cite{Bonanno:2025bla}, based on $\epsilon$-regime $\chi$PT~\cite{Gasser:1986vb,Gasser:1987ah}. This chiral regime is obtained when $m_\pi\to0$ and $\ell \to \infty$ with $m_\pi \ell^2 = \mathrm{fixed}$, and it is in this regime that RMT predictions are expected to hold~\cite{Verbaarschot:2000dy,Basile:2007ki}. The effective theory prediction in the $\epsilon$-regime is that finite-volume effects are polynomial in $1/\sqrt{V}$. We thus perform a global best fit of the four eigenvalue determinations at the three bare couplings assuming the following functional form:
\beq
 \frac{\Sigma_k(N,b)}{\Sigma(b)}
 =1+c_1\frac{1}{\sqrt{V}\sigma}
   +c_{2,k}\frac{1}{V\sigma^2}, \qquad V = a^4N^2
 \label{eq:results-RMT}
\eeq
Here $\Sigma(b)$ denotes the common infinite-volume extrapolations of the different spectral determinations of the condensate, corresponding to one independent fit parameter for each $b$. Instead, the fit parameter $c_1$ and the four fit parameters $c_{2,k}$ are assumed to be independent of the coupling. The global $V\to\infty$ extrapolation is shown in Fig.~\ref{fig:condensate-infinite-volume}. The global fit gives a reduced chi-squared of $\chi^2/{\rm dof}=38.97/40$ ($p$-value $=52\%$). Since the finite-volume corrections in Eq.~\eqref{eq:results-RMT} are small compared to the condensate itself, deviations of the data from the fit are better appreciated by rewriting it in the linear form
\beq
\left[\frac{\Sigma_k(N,b)}{\Sigma(b)}-1\right]\sqrt{V}\sigma = c_1 + c_{2,k}\,\frac{1}{\sqrt{V}\sigma}.
\label{eq:results-RMT-linear}
\eeq
The data and the global fit are shown in this form in Fig.~\ref{fig:condensate-infinite-volume-linear}, separately for each eigenvalue. All determinations are well described by a common intercept $c_1=-3.9(6)$ and by slopes $c_{2,k}$ increasing with $k$, i.e., the finite-volume corrections are larger for the higher eigenvalues.
The resulting infinite-volume bare condensates are reported in Tab.~\ref{tab:bare-condensate}.
Combining the scale setting determinations, the chirally extrapolated renormalization factors $\ZS^{\chir}(b)$ determined in Sec.~\ref{sec:res-ZS}, and the infinite-volume large-$N$ bare condensates, one finally obtains the renormalized condensate in physical units, cf.~Eq.~\eqref{eq:overlap-renormalization}.

\begin{table}[t]
    \centering
    \begin{tabular}{c c c}
        \hline\hline
        $b$ & $a^3\Sigma(b)/N$ & $(8t_1)^{3/2}\Sigma_\R(b)/N$\\
        \hline
        $0.355$ & $0.000939(17)$ & $0.0240(14)$\\
        $0.360$ & $0.000586(10)$ & $0.0247(14)$\\
        $0.365$ & $0.0003832(66)$  & $0.0249(15)$\\
        \hline\hline
    \end{tabular}
    \caption{Infinite-volume bare chiral condensate obtained from the global best fit of the determinations based on the first four overlap Dirac eigenvalues and RMT predictions expressed in lattice units. We also report the values of the renormalized one in physical units.}
    \label{tab:bare-condensate}
\end{table}

\begin{figure}[!t]
 \centering
\includegraphics[width=0.75\textwidth]
 {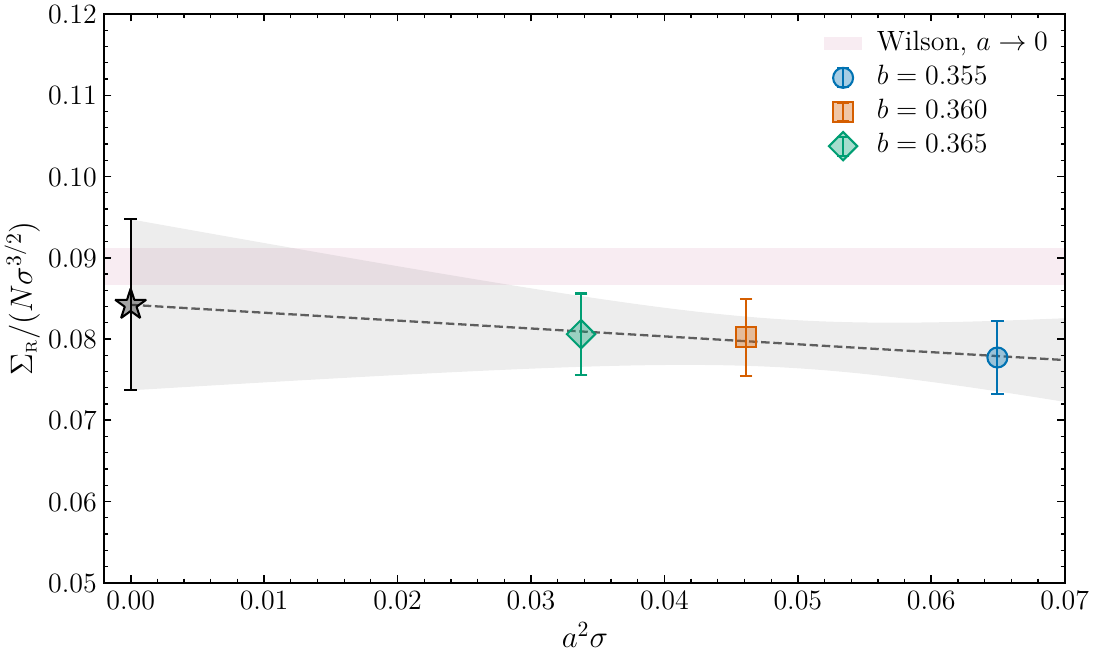}
\caption{Continuum extrapolation of the renormalized overlap condensate in string-tension units. The star denotes the continuum result, the grey band the uncertainty on the continuum fit, and the pink band the Wilson continuum result of Ref.~\cite{Bonanno:2025hzr}.}
\label{fig:condensate-continuum}
\end{figure}

The values of $\Sigma_\R$ obtained at each lattice spacing are collected in Tab.~\ref{tab:bare-condensate}. The extrapolation to the continuum limit, shown in Fig.~\ref{fig:condensate-continuum}, is achieved assuming the leading
$O(a^2)$ artifacts expected for overlap fermions:
\beq
\frac{\Sigma_\R(b)}{N}
 =\frac{\Sigma_\R}{N}
 +C_{\scriptscriptstyle{\Sigma}}a^2 + \mathcal{O}(a^4).
 \label{eq:condensate-continuum-fit}
\eeq
In the end, we obtain ($\chi^2/{\rm dof}=0.02/1$, see Fig.~\ref{fig:condensate-continuum}):
\beq
(8t_1)^{3/2}\frac{\Sigma_\R}{N} &=& 0.0260(32),\\
 \frac{\Sigma_\R}{N\sqrt{\sigma^3}} &=& 0.084(10).
 \label{eq:final-renormalized-condensate}
\eeq
The overlap condensate, Eq.~\eqref{eq:final-renormalized-condensate}, is perfectly compatible with the continuum TEK Wilson result $0.0889(23)$ obtained (for the same renormalization scheme and scale) in Ref.~\cite{Bonanno:2025hzr}.

\subsection{Pion decay constant and the NLO coupling \texorpdfstring{$\bar\ell_4$}{ell4}}

\begin{figure}[!t]
 \centering
 \includegraphics[width=.8\textwidth]
 {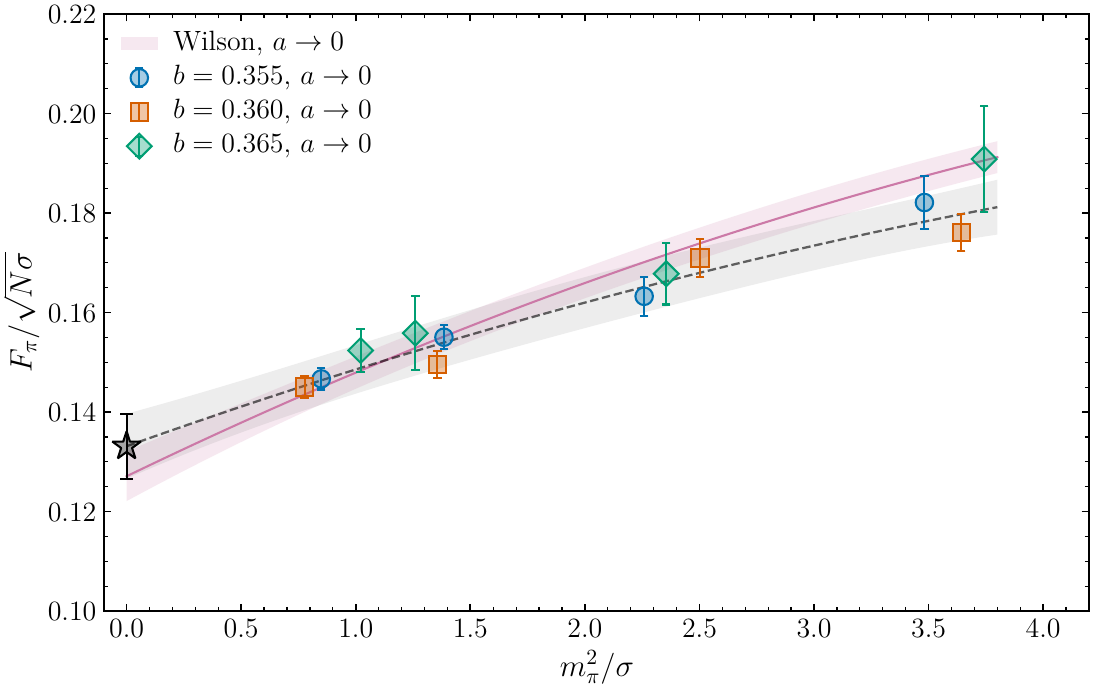}
 \caption{Joint chiral-continuum extrapolation of the renormalized overlap pion decay constant. The fitted $O(a^2)$ artifact has been subtracted from  the data. The grey band is the uncertainty on the continuum curve, while the star denotes the chiral-continuum limit of $F_\pi$. The pink band shows the continuum result obtained with Wilson fermions.}
 \label{fig:Fpi-continuum}
\end{figure}

We determined the pion decay constant directly from the axial current matrix element and studied its pion-mass dependence in order to obtain the LECs $F/\sqrt{N}$ and $\bar{\ell}_4/N$. Using the axial renormalization factors determined in Sec.~\ref{sec:res-quark-mass}, our determinations of the $\pi$ decay constant were renormalized as
\beq
F_{\pi}=\ZA F_{\pi}^{\scriptscriptstyle{(\rm lat)}}.
 \label{eq:Fpi-renormalization}
\eeq
Scale setting is again performed in terms of $\sqrt{\sigma}$, but passing through $\sqrt{8t_1}$, as done for the chiral condensate. After renormalization we performed a combined chiral-continuum extrapolation assuming the following fit function:
\beq
 \frac{1}{\sqrt{\sigma}}\frac{F_{\pi}(a,m_\pi)}{\sqrt{N}}
 = \frac{F}{\sqrt{N \sigma}}\left[1 + \frac{1}{16 \pi^2} \frac{1}{F^2/(N\sigma)}\frac{\bar{\ell}_4}{N} \frac{m_\pi^2}{\sigma} + k_{\scriptscriptstyle{F}}\frac{m_\pi^4}{\sigma^2}\right]
 +C_{\scriptscriptstyle{F}} a^2\sigma.
 \label{eq:results-Fpi-model}
\eeq
The best fit, yielding a reduced chi-squared of $\chi^2/{\rm dof}=6.33/8$, is shown in Fig.~\ref{fig:Fpi-continuum}, where we display the data after subtracting the fitted $O(a^2)$ lattice artifacts, together with the continuum curve and its uncertainty band. The final results:
\beq
 \label{eq:results-Fpi-chiral}
\frac{F}{\sqrt{N\sigma}} &=& 0.1331(65),\\
 \label{eq:results-l4}
\frac{\bar\ell_4}{N} &=& 0.35(11),
\eeq
perfectly agree with the Wilson determinations $0.1271(50)$ and $0.446(55)$ of Ref.~\cite{Bonanno:2025hzr}, respectively. Using instead the axial renormalization factors obtained from the quadratic fits of the PCAC mass discussed in Sec.~\ref{sec:res-quark-mass}, we obtain $F/\sqrt{N\sigma}=0.1440(86)$ and $\bar\ell_4/N=0.40(15)$ ($\chi^2/{\rm dof}=7.29/8$), i.e., a shift of about $1.5$ standard deviations in $F$, still compatible with the Wilson result.

\subsection{Rho meson mass}

\begin{figure}[!t]
 \centering
 \includegraphics[width=.8\textwidth]
   {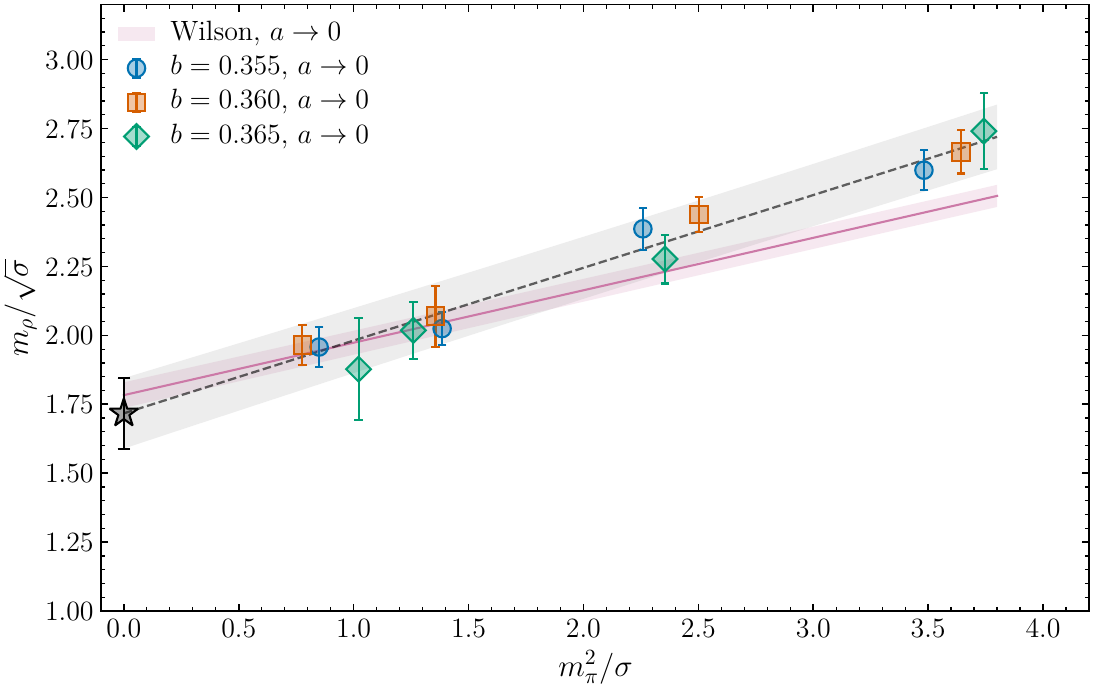}
 \caption{The $\rho$ meson mass as a function of the squared $\pi$ mass. The fitted $O(a^2)$ artifacts have been subtracted from the lattice data. The dashed line and the grey band show the continuum relation and its uncertainties. The star marks the chiral-continuum limit. The pink band shows the continuum result obtained with Wilson fermions.}
 \label{fig:rho-continuum}
\end{figure}

We finally present the results for the pion-mass dependence of the $\rho$ meson mass. We performed a joint fit to all our determinations at the three lattice spacings according to the following fit function:
\beq
\frac{m_\rho(a,m_\pi)}{\sqrt\sigma}
 =\frac{m_\rho^{\chir}}{\sqrt\sigma}
 +(C_\rho\sqrt{\sigma})\,\frac{m_\pi^2}{\sigma}
 +K_\rho a^2 \sigma,
 \label{eq:results-rho-model}
\eeq
where $m_\rho^{\chir}$ denotes the rho mass in the continuum and chiral limits. Again, the masses were first expressed in units of the gradient flow scale $t_1$, and then converted to string-tension units. The global fit results are
\beq
 \frac{m_\rho^{\chir}}{\sqrt\sigma}=1.72(13),\qquad
 C_\rho\sqrt{\sigma}=0.264(25),\qquad
 K_\rho=-1.9(2.1),
 \label{eq:rho-fit-results}
\eeq
with $\chi^2/{\rm dof}=4.47/9$. The best fit is depicted in Fig.~\ref{fig:rho-continuum}, along with the $\rho$ mass data after subtracting the fitted $O(a^2)$ artifacts. The dashed curve and its uncertainty band represent the continuum curve, while the star denotes the chiral-limit result.

\begin{figure}[!t]
 \centering
 \includegraphics[width=.8\textwidth]{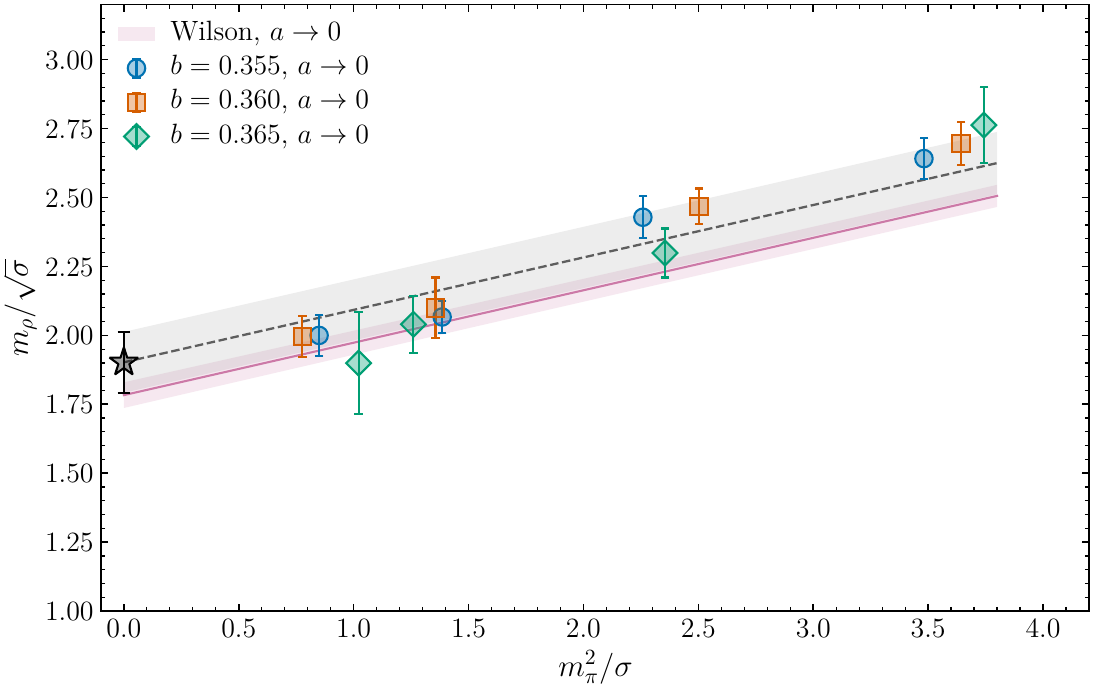}
 \caption{Same as Fig.~\ref{fig:rho-continuum}, with the slope $C_\rho$ fixed to the Wilson continuum value. The $O(a^2)$ artifacts obtained from this fit have been subtracted from the lattice data; the dashed line and the grey band show the corresponding continuum relation, whose uncertainty includes that of the Wilson slope.}
 \label{fig:rho-fixed-slope}
\end{figure}

The chiral limit is in very good agreement with the TEK Wilson result of Ref.~\cite{Bonanno:2025hzr}, $m_\rho^{\chir}/\sqrt{\sigma}=1.783(47)$. The slope turns out to be slightly larger than the Wilson one, $C_\rho\sqrt{\sigma} = 0.1902(65)$, but nonetheless compatible within about $2.5$ standard deviations. We also repeated the fit of Eq.~\eqref{eq:results-rho-model} with the slope fixed to the Wilson continuum value, $C_\rho\sqrt{\sigma}=0.1902(65)$. This constrained fit still provides a good description of our data, with $\chi^2/{\rm dof}=14.4/10$ ($p$-value $=16\%$), and yields $m_\rho^{\chir}/\sqrt{\sigma}=1.90(11)$, compatible with the Wilson result within one standard deviation. The resulting continuum curve is shown in Fig.~\ref{fig:rho-fixed-slope}.
 
\section{Conclusions}\label{sec:conclusions}

In this work we have presented a lattice investigation of chiral dynamics in large-$N$ QCD with chiral quarks, extending the study of Ref.~\cite{Bonanno:2025bla}. As in
our previous work, we exploited large-$N$ twisted volume reduction, which allowed us to reach values of $N$ as large as $N=1156$, and we discretized the valence quarks with the truncated overlap
operator, for which violations of chiral symmetry are kept under control by a suitable choice of the fifth-dimension extent and of the smearing of the gauge fields entering the Wilson kernel. We considered three values of the lattice spacing and, for each of them, four values of the quark mass, performing joint chiral-continuum extrapolations of all considered observables. Using only bare overlap data, we verified that the pion becomes massless when the bare quark mass vanishes, confirming the absence of additive renormalizations in the quark mass.

\begin{table}[!t]
\centering
\renewcommand{\arraystretch}{1.25}
\begin{tabular}{lcc}
\hline\hline
 & Overlap (this work) & Wilson\\
\hline
$\Sigma_\R/(N\sigma^{3/2})$ & $0.084(10)$ & $0.0889(23)$\\
$F/\sqrt{N\sigma}$          & $0.1331(65)$ & $0.1271(50)$\\
$\bar\ell_4/N$              & $0.35(11)$   & $0.446(55)$\\
$m_\rho^{\chir}/\sqrt{\sigma}$ & $1.72(13)$ & $1.783(47)$\\
$C_\rho\sqrt{\sigma}$       & $0.264(25)$  & $0.1902(65)$\\
\hline\hline
\end{tabular}
\caption{Summary of our final continuum- and chiral-extrapolated large-$N$ results obtained with overlap quarks, compared with the TEK determinations of Ref.~\cite{Bonanno:2025hzr} obtained with Wilson quarks. Here, $C_\rho$ denotes the slope of the $\rho$ mass as a function of $m_\pi^2$, cf.~Eq.~\eqref{eq:results-rho-model}. Renormalized quantities are given in the $\overline{\rm MS}$ scheme at $\mu\sqrt{8t_0}=4.30$, corresponding to $\mu=2$ $\mathrm{GeV}$ for $\sqrt{t_0}=0.15$ $\mathrm{fm}$. The same scale-setting convention is adopted for both discretizations.}
\label{tab:final-results}
\end{table}

We collect in Tab.~\ref{tab:final-results} the final results of our investigation. The overlap determinations of the chiral condensate and of the LECs describing the quark mass dependence of the pion decay constant and of the $\rho$ mass agree, within errors, with the continuum TEK determinations obtained in Ref.~\cite{Bonanno:2025hzr} with Wilson
fermions. Although overlap results have larger errors than the Wilson ones due to the larger computational cost, they are still sufficiently precise to allow a meaningful comparison. Beyond its intrinsic theoretical and phenomenological interest, this agreement constitutes an important universality test of our theoretical and numerical framework for large-$N$ volume reduction.

The present study opens the way to further applications of chiral lattice fermions in the context of twisted volume reduction for the study of large-$N$ gauge theories. A natural next step would be the use of the overlap operator in large-$N$ $\mathcal{N}=1$ SUSY (supersymmetric) Yang--Mills, recently studied in the TEK framework with Wilson fermions~\cite{Butti:2022sgy,Bonanno:2024bqg,Bonanno:2024onr}. While the gluino is a massless adjoint Majorana fermion, a lattice formulation of the SUSY theory with a Wilson gluino necessarily requires adding a non-vanishing mass, and then taking the chiral limit. This, in turn, requires defining the chiral point by tuning the bare gluino mass to the point where the mass of the unphysical adjoint-pion vanishes. Using a chirally-preserving discretization would be theoretically cleaner, as the lattice chiral symmetry would protect the gluino mass from additive renormalizations, allowing one to simulate directly in the chiral limit (as done, e.g., in Ref.~\cite{Piemonte:2020wkm}). Clearly, the additional difficulty in the SUSY case is that the gluino is not quenched, and simulations with a dynamical overlap gluino would be much more expensive than the Wilson case. A mixed-action setup (valence overlap gluino and sea Wilson gluino) could be a possible intermediate future step.

\section*{Acknowledgements}
This work is partially supported by the Spanish Research Agency (Agencia Estatal de Investigaci\'on) through the grant IFT Centro de Excelencia Severo Ochoa CEX2025-001574-S, funded by MICIU/AEI/10.13039/501100011033, and, partially, by the grant PID2024-160152NB-I00, funded by MCIN/AEI/10.13039/501100011033. The research presented in this publication falls within the Severo Ochoa research line Quantum Matter and Fields. It is also partially supported under HADRON-2030 funded by the European Union's Horizon Europe research and innovation programme under Grant Agreement No.~101292522. It is also partially funded by the European Commission – NextGenerationEU, through Momentum CSIC Programme: Develop Your Digital Talent. We acknowledge support from the SFT Scientific Initiative of INFN. This work was partially supported by the Simons Foundation grant SFI-MPS-QCD-00014867-08 (Simons Collaboration on Confinement and QCD Strings). Numerical calculations have been performed on the \texttt{Finisterrae~III} cluster at CESGA (Centro de Supercomputaci\'on de Galicia), on the \texttt{Drago} cluster at CSIC (Consejo Superior de Investigaciones Científicas) and on the Hydra and Ciclope clusters at IFT. We acknowledge HPC support by Emilio Ambite, staff hired under the Generation D initiative, promoted by Red.es, an organization attached to the Ministry for Digital Transformation and the Civil Service, for the attraction and retention of talent through grants and training contracts, financed by the Recovery, Transformation and Resilience Plan through the European Union's Next Generation funds. We have also used computational resources of Oakbridge-CX, at the University of Tokyo through the HPCI System Research Project (Project ID: hp230021 and hp220011), of Cygnus at Center for Computational Sciences, University of Tsukuba, of SQUID at D3 Center of Osaka University through the RCNP joint use program.

\appendix
\section*{Appendix}

\section{Raw numerical data}\label{app:data}

In this appendix we collect, in Tabs.~\ref{tab:rawdata-chir_cond}, \ref{tab:rawdata-ratios} and~\ref{tab:rawdata-mesons}, the raw lattice data analyzed in the present study.

\begin{table}[!t]
\centering
\small
\renewcommand{\arraystretch}{1.15}
\begin{tabular}{cccccc}
\hline\hline
$b$ & $N$ & $10^4\times a^3\Sigma_1/N$ & $10^4\times a^3\Sigma_2/N$ & $10^4\times a^3\Sigma_3/N$ & $10^4\times a^3\Sigma_4/N$\\
\hline
\multirow{4}{*}{$0.355$} & $289$ & $9.07(28)$ & $9.51(14)$ & $10.49(11)$ & $11.471(94)$ \\
 & $361$ & $8.46(30)$ & $9.04(17)$ & $9.64(14)$ & $10.21(11)$ \\
 & $529$ & $8.69(24)$ & $9.00(13)$ & $9.12(12)$ & $9.310(91)$ \\
 & $841$ & $9.37(39)$ & $9.27(23)$ & $9.07(14)$ & $9.11(14)$ \\
\hline
\multirow{4}{*}{$0.360$} & $289$ & $5.16(15)$ & $6.72(10)$ & $7.823(69)$ & $8.940(64)$ \\
 & $361$ & $5.10(22)$ & $6.05(13)$ & $6.681(96)$ & $7.550(88)$ \\
 & $529$ & $5.37(21)$ & $5.80(13)$ & $5.98(11)$ & $6.268(83)$ \\
 & $841$ & $5.61(22)$ & $5.75(11)$ & $5.65(11)$ & $5.764(81)$ \\
\hline
\multirow{4}{*}{$0.365$} & $361$ & $3.428(95)$ & $4.579(56)$ & $5.326(38)$ & $6.230(37)$ \\
 & $529$ & $3.53(14)$ & $3.917(74)$ & $4.274(57)$ & $4.731(51)$ \\
 & $841$ & $3.40(16)$ & $3.64(11)$ & $3.766(91)$ & $3.866(77)$ \\
 & $1156$ & $3.25(19)$ & $3.55(20)$ & $3.54(15)$ & $3.62(13)$ \\
\hline\hline
\end{tabular}
    \caption{Bare chiral condensate in lattice units extracted from the $k^{\rm th}$ eigenvalue of the massless overlap Dirac operator, $k=1,\dots,4$, using Eq.~\eqref{eq:condensate-from-means-results}.}
\label{tab:rawdata-chir_cond}
\end{table}

\begin{table}[!t]
\centering
\small
\renewcommand{\arraystretch}{1.15}
\begin{tabular}{cccccccc}
\hline\hline
$b$ & $N$ & $R_{2,1}$ & $R_{3,1}$ & $R_{4,1}$ & $R_{3,2}$ & $R_{4,2}$ & $R_{4,3}$\\
\hline
\multirow{4}{*}{$0.355$} & $289$ & $0.952(27)$ & $0.866(26)$ & $0.789(24)$ & $0.9081(96)$ & $0.8302(98)$ & $0.9106(52)$\\
 & $361$ & $0.939(27)$ & $0.877(29)$ & $0.831(27)$ & $0.937(12)$ & $0.885(13)$ & $0.9410(68)$\\
 & $529$ & $0.966(23)$ & $0.954(24)$ & $0.932(23)$ & $0.989(11)$ & $0.968(11)$ & $0.9744(69)$\\
 & $841$ & $1.012(33)$ & $1.033(34)$ & $1.026(37)$ & $1.022(15)$ & $1.018(17)$ & $0.9912(93)$\\
\hline
\multirow{4}{*}{$0.360$} & $289$ & $0.770(18)$ & $0.661(18)$ & $0.577(16)$ & $0.8582(87)$ & $0.7511(86)$ & $0.8715(46)$\\
 & $361$ & $0.845(29)$ & $0.764(29)$ & $0.676(27)$ & $0.906(13)$ & $0.802(12)$ & $0.8815(62)$\\
 & $529$ & $0.929(32)$ & $0.899(32)$ & $0.857(31)$ & $0.969(15)$ & $0.924(16)$ & $0.9499(86)$\\
 & $841$ & $0.978(36)$ & $0.993(38)$ & $0.975(39)$ & $1.018(15)$ & $0.999(17)$ & $0.977(10)$\\
\hline
\multirow{4}{*}{$0.365$} & $361$ & $0.749(16)$ & $0.645(16)$ & $0.551(14)$ & $0.8606(85)$ & $0.7352(75)$ & $0.8513(35)$\\
 & $529$ & $0.902(29)$ & $0.825(25)$ & $0.745(24)$ & $0.916(11)$ & $0.827(10)$ & $0.8992(59)$\\
 & $841$ & $0.937(34)$ & $0.906(35)$ & $0.881(34)$ & $0.966(14)$ & $0.943(16)$ & $0.9703(85)$\\
 & $1156$ & $0.917(32)$ & $0.918(35)$ & $0.899(37)$ & $1.003(25)$ & $0.983(30)$ & $0.976(12)$\\
 \hline\hline
\end{tabular}
\caption{Ratios $R_{k_1,k_2}$ of Eq.~\eqref{eq:ratio_rmt}, normalized to the RMT predictions, for all ensembles. Agreement with RMT corresponds to $R_{k_1,k_2}=1$.}
\label{tab:rawdata-ratios}
\end{table}

\begin{table}[!t]
\centering
\small
\renewcommand{\arraystretch}{1.15}
\begin{tabular}{cccccccc}
\hline\hline
$b$ & $\hat m$ & $N$ & $am_{\PCAC}$ & $am_\pi$ & $m_\pi\ell$ & $aF_\pi^{\scriptscriptstyle{(\rm lat)}}/\sqrt{N}$ & $am_\rho$\\
\hline
\multirow{4}{*}{$0.355$} & $0.088$ & $361$ & $0.0964(17)$ & $0.476(10)$ & $9.04$ & $0.0467(15)$ & $0.632(19)$ \\
 & $0.063$ & $361$ & $0.06750(98)$ & $0.3830(53)$ & $7.28$ & $0.0414(11)$ & $0.577(19)$ \\
 & $0.042$ & $361$ & $0.04429(39)$ & $0.2999(39)$ & $5.70$ & $0.03905(65)$ & $0.485(15)$ \\
 & $0.025$ & $361$ & $0.02661(28)$ & $0.2349(36)$ & $4.46$ & $0.03670(55)$ & $0.468(19)$ \\
\hline
\multirow{6}{*}{$0.360$} & $0.077$ & $361$ & $0.0828(18)$ & $0.4099(71)$ & $7.79$ & $0.03831(81)$ & $0.554(17)$ \\
 & $0.055$ & $361$ & $0.0598(12)$ & $0.3397(57)$ & $6.45$ & $0.03710(84)$ & $0.505(14)$ \\
 & $0.034$ & $361$ & $0.03525(41)$ & $0.2588(34)$ & $4.92$ & $0.03241(53)$ & $0.454(15)$ \\
 & $0.020$ & $361$ & $0.02106(55)$ & $0.1940(82)$ & $3.69$ & $0.02926(55)$ & $0.424(28)$ \\
 \cline{2-8}
  & $0.034$ & $529$ & $0.03476(57)$ & $0.2501(40)$ & $5.75$ & $0.03211(58)$ & $0.426(24)$ \\
 & $0.020$ & $529$ & $0.02074(25)$ & $0.1893(27)$ & $4.35$ & $0.03107(42)$ & $0.404(16)$ \\
\hline
\multirow{4}{*}{$0.365$} & $0.064$ & $529$ & $0.0696(21)$ & $0.355(16)$ & $8.17$ & $0.0349(20)$ & $0.492(25)$ \\
 & $0.046$ & $529$ & $0.04675(92)$ & $0.2819(83)$ & $6.48$ & $0.0305(11)$ & $0.407(16)$ \\
 & $0.025$ & $529$ & $0.02474(39)$ & $0.2062(38)$ & $4.74$ & $0.0282(14)$ & $0.359(19)$ \\
 & $0.020$ & $529$ & $0.0201(10)$ & $0.186(10)$ & $4.27$ & $0.02753(74)$ & $0.333(34)$ \\
\hline\hline
\end{tabular}
\caption{Meson observables in lattice units obtained at non-vanishing quark mass. The lattice pion decay constant $F_\pi^{\scriptscriptstyle{(\rm lat)}}=F_\pi/\ZA$, cf.~Eq.~\eqref{eq:Fpi-renormalization}. For $b=0.360$ we also report the $N=361$ data at $\hat m=0.020, 0.034$ for completeness; they are not used in the analysis, where the corresponding $N=529$ ensembles are used instead.}
\label{tab:rawdata-mesons}
\end{table}
\section{Normalization of the bare quark mass}\label{app:mass}

In this appendix we show that the bare quark mass is given by $am=M(2-M)\hat m$, cf.~Eq.~\eqref{eq:overlap_normalization}, up to $O(a^2)$ corrections (see also~\cite{Capitani:1999uz,Capitani:1999ay}). We consider the overlap operator of Eq.~\eqref{eq:overlap} and the subtracted propagator of Eq.~\eqref{eq:setup-subtracted-propagator}, with the truncated overlap operator replaced by its $\Nfive\to\infty$ limit $D$ (at $\Nfive=24$, violations of the Ginsparg--Wilson relation are negligible, see Sec.~\ref{sec:setup}).

Writing $D=(1-\hat m)\left[D_0+\hat m/(1-\hat m)\right]$, the mass appears to enter as $\hat m/(1-\hat m)$. This factor, however, cancels in the propagator. Since $\mathbb{I}-D=(1-\hat m)(\mathbb{I}-D_0)$, one has $\widetilde S=D^{-1}(\mathbb{I}-D_0)$, and therefore, exactly and for any gauge field,
\beq\label{eq:app-mass-exact}
\widetilde S=\left(K_0+\hat m\right)^{-1},\qquad K_0\equiv D_0\left(\mathbb{I}-D_0\right)^{-1}.
\eeq
As a consequence of the Ginsparg--Wilson relation, $K_0$ is anti-Hermitian and anticommutes with $\gamma_5$, so the quark mass enters the propagator only through $\hat m$.

The same cancellation can be seen directly from the eigenvalues of $D_0$. On the circle of Eq.~\eqref{eq:overlap_spectrum} they can be written as $\hat\lambda=\ii y/(1+\ii y)$, where $\ii y$ is the corresponding eigenvalue of $K_0$. The pole condition $(1-\hat m)\hat\lambda+\hat m=0$ then becomes
\beq
(1-\hat m)\,\ii y+\hat m\,(1+\ii y)=\ii y+\hat m=0,
\eeq
i.e., the factor $(1-\hat m)$ is compensated by the real part of $\hat\lambda$, and the pole is located at $\ii y=-\hat m$.

Finally, for free fields $K_0$ is odd in the momentum,
\beq
K_0=\ii a\slashed{p}/M(2-M)\left[1+O(a^2p^2)\right],
\eeq
which gives
\beq
\widetilde S=M(2-M)\left(\ii a\slashed{p}+am\right)^{-1}\left[1+O(a^2)\right],
\eeq
where $am=M(2-M)\,\hat m$.

\bibliographystyle{JHEP}
\bibliography{biblio}

@article{Zhu:2026zvb,
    author = "Zhu, Isaac and Romero-L{\'o}pez, Fernando and Shanahan, Phiala E.",
    title = "{The structure of the pion at large $N_c$}",
    eprint = "2611.12771",
    archivePrefix = "arXiv",
    primaryClass = "hep-lat",
    reportNumber = "MIT-CTP/6146",
    month = "11",
    year = "2026"
}

@article{Bonanno:2026gay,
    author = "Bonanno, Claudio and others",
    title = "{Large-N $\pi^+\pi^+$ scattering from twisted Eguchi--Kawai reduction}",
    eprint = "2610.35176",
    archivePrefix = "arXiv",
    primaryClass = "hep-lat",
    month = "10",
    year = "2026"
}

@article{JLQCD:2007ppn,
    author = "Fukaya, H. and others",
    collaboration = "JLQCD",
    title = "{Two-flavor lattice QCD simulation in the epsilon-regime with exact chiral symmetry}",
    eprint = "hep-lat/0702003",
    archivePrefix = "arXiv",
    reportNumber = "RIKEN-TH-91, UTHEP-538, KEK-CP-191, YITP-07-4, NTUTH-07-505A",
    doi = "10.1103/PhysRevLett.98.172001",
    journal = "Phys. Rev. Lett.",
    volume = "98",
    pages = "172001",
    year = "2007"
}

@article{TWQCD:2007eit,
    author = "Fukaya, H. and Aoki, S. and Chiu, T. W. and Hashimoto, S. and Kaneko, T. and Matsufuru, H. and Noaki, J. and Ogawa, K. and Onogi, T. and Yamada, N.",
    collaboration = "TWQCD",
    title = "{Two-flavor lattice QCD in the epsilon-regime and chiral Random Matrix Theory}",
    eprint = "0705.3322",
    archivePrefix = "arXiv",
    primaryClass = "hep-lat",
    reportNumber = "RIKEN-TH-98, UTHEP-545, KEK-CP-194, YITP-07-27, NTUTH-07-505D",
    doi = "10.1103/PhysRevD.76.054503",
    journal = "Phys. Rev. D",
    volume = "76",
    pages = "054503",
    year = "2007"
}

@article{Hernandez:1999cu,
    author = "Hernandez, Pilar and Jansen, Karl and Lellouch, Laurent",
    title = "{Finite size scaling of the quark condensate in quenched lattice QCD}",
    eprint = "hep-lat/9907022",
    archivePrefix = "arXiv",
    reportNumber = "CERN-TH-99-197, CPT-99-PE-3856, FTUV-99-49, IFIC-99-51",
    doi = "10.1016/S0370-2693(99)01244-7",
    journal = "Phys. Lett. B",
    volume = "469",
    pages = "198--204",
    year = "1999"
}

@article{Donini:2016lwz,
    author = "Donini, A. and Hern{\'a}ndez, P. and Pena, C. and Romero-L{\'o}pez, F.",
    title = "{Nonleptonic kaon decays at large $N_c$}",
    eprint = "1607.03262",
    archivePrefix = "arXiv",
    primaryClass = "hep-ph",
    reportNumber = "IFIC-16-39, IFT-UAM-CSIC-16-063, FTUAM-16-26",
    doi = "10.1103/PhysRevD.94.114511",
    journal = "Phys. Rev. D",
    volume = "94",
    number = "11",
    pages = "114511",
    year = "2016"
}

@article{tHooft:1986nc,
    author = "'t Hooft, Gerard",
    title = "{How Instantons Solve the U(1) Problem}",
    doi = "10.1016/0370-1573(86)90117-1",
    journal = "Phys. Rept.",
    volume = "142",
    pages = "357--387",
    year = "1986"
}

@article{Murakami:2021kdk,
    author = "Murakami, Yuko and Sekiguchi, Motoo and Wada, Hiroaki and Wakayama, Masayuki",
    title = "{Properties of the five dimensions for the truncated overlap fermions}",
    doi = "10.1088/2399-6528/ac1b71",
    journal = "J. Phys. Comm.",
    volume = "5",
    number = "8",
    pages = "085009",
    year = "2021"
}

@article{Donini:2020qfu,
    author = "Donini, Andrea and Hern{\'a}ndez, Pilar and Pena, Carlos and Romero-L{\'o}pez, Fernando",
    title = "{Dissecting the $\Delta I= 1/2$ rule at large $N_c$}",
    eprint = "2003.10293",
    archivePrefix = "arXiv",
    primaryClass = "hep-lat",
    doi = "10.1140/epjc/s10052-020-8192-3",
    journal = "Eur. Phys. J. C",
    volume = "80",
    number = "7",
    pages = "638",
    year = "2020"
}

@article{Bocchia:2026lty,
    author = "Bocchia, Sebastiano and Guerrieri, Andrea",
    title = "{Large $N$ hadron scattering bootstrap: massive pions and glueballs}",
    eprint = "2610.31578",
    archivePrefix = "arXiv",
    primaryClass = "hep-th",
    month = "10",
    year = "2026"
}

@article{Piemonte:2020wkm,
    author = "Piemonte, Stefano and Bergner, Georg and L{\'o}pez, Camilo",
    title = "{Monte Carlo simulations of overlap Majorana fermions}",
    eprint = "2005.02236",
    archivePrefix = "arXiv",
    primaryClass = "hep-lat",
    doi = "10.1103/PhysRevD.102.014503",
    journal = "Phys. Rev. D",
    volume = "102",
    number = "1",
    pages = "014503",
    year = "2020"
}

@article{Eguchi:1982ta,
    author = "Eguchi, Tohru and Nakayama, Ryuichi",
    title = "{Simplification of Quenching Procedure for Large $N$ Spin Models}",
    reportNumber = "UT-391-TOKYO",
    doi = "10.1016/0370-2693(83)91168-1",
    journal = "Phys. Lett. B",
    volume = "122",
    pages = "59--62",
    year = "1983"
}

@article{Ambjorn:1999ts,
    author = "Ambjorn, Jan and Makeenko, Y. M. and Nishimura, J. and Szabo, R. J.",
    title = "{Finite N matrix models of noncommutative gauge theory}",
    eprint = "hep-th/9911041",
    archivePrefix = "arXiv",
    reportNumber = "NBI-HE-99-44, ITEP-TH-58-99",
    doi = "10.1088/1126-6708/1999/11/029",
    journal = "JHEP",
    volume = "11",
    pages = "029",
    year = "1999"
}

@article{Ambjorn:2000nb,
    author = "Ambjorn, Jan and Makeenko, Y. M. and Nishimura, J. and Szabo, R. J.",
    title = "{Nonperturbative dynamics of noncommutative gauge theory}",
    eprint = "hep-th/0002158",
    archivePrefix = "arXiv",
    reportNumber = "NBI-HE-00-08, ITEP-TH-10-00",
    doi = "10.1016/S0370-2693(00)00391-9",
    journal = "Phys. Lett. B",
    volume = "480",
    pages = "399--408",
    year = "2000"
}

@article{Jenkins:1995vb,
    author = "Jenkins, Elizabeth Ellen and Manohar, Aneesh V. and Wise, Mark B.",
    title = "{Chiral perturbation theory for vector mesons}",
    eprint = "hep-ph/9506356",
    archivePrefix = "arXiv",
    reportNumber = "UCSD-PTH-95-08, CALT-68-2000",
    doi = "10.1103/PhysRevLett.75.2272",
    journal = "Phys. Rev. Lett.",
    volume = "75",
    pages = "2272--2275",
    year = "1995"
}

@article{Booth:1996hk,
    author = "Booth, Michael and Chiladze, George and Falk, Adam F.",
    title = "{Quenched chiral perturbation theory for vector mesons}",
    eprint = "hep-ph/9610532",
    archivePrefix = "arXiv",
    reportNumber = "JHU-TIPAC-96016",
    doi = "10.1103/PhysRevD.55.3092",
    journal = "Phys. Rev. D",
    volume = "55",
    pages = "3092--3100",
    year = "1997"
}

@article{Yaffe:2026ouz,
    author = "Yaffe, Laurence G.",
    title = "{Resurrecting the coherent state variational algorithm for large N gauge theories}",
    eprint = "2601.20150",
    archivePrefix = "arXiv",
    primaryClass = "hep-lat",
    doi = "10.1007/JHEP05(2026)091",
    journal = "JHEP",
    volume = "05",
    pages = "091",
    year = "2026"
}

@article{Narayanan:2006rf,
    author = "Narayanan, R. and Neuberger, H.",
    title = "{Infinite N phase transitions in continuum Wilson loop operators}",
    eprint = "hep-th/0601210",
    archivePrefix = "arXiv",
    doi = "10.1088/1126-6708/2006/03/064",
    journal = "JHEP",
    volume = "03",
    pages = "064",
    year = "2006"
}

@article{Karthik:2022fdb,
    author = "Karthik, Nikhil and Narayanan, Rajamani",
    title = "{Parton physics of the large-Nc mesons}",
    eprint = "2205.02252",
    archivePrefix = "arXiv",
    primaryClass = "hep-lat",
    reportNumber = "JLAB-THY-22-3613",
    doi = "10.1103/PhysRevD.106.014503",
    journal = "Phys. Rev. D",
    volume = "106",
    number = "1",
    pages = "014503",
    year = "2022"
}

@article{Hietanen:2009tu,
    author = "Hietanen, A. and Narayanan, R. and Patel, R. and Prays, C.",
    title = "{The vector meson mass in the large N limit of QCD}",
    eprint = "0901.3752",
    archivePrefix = "arXiv",
    primaryClass = "hep-lat",
    doi = "10.1016/j.physletb.2009.02.054",
    journal = "Phys. Lett. B",
    volume = "674",
    pages = "80--82",
    year = "2009"
}

@article{Gulpers:2015bba,
    author = {G\"ulpers, Vera and von Hippel, Georg and Wittig, Hartmut},
    title = "{The scalar radius of the pion from Lattice QCD in the continuum limit}",
    eprint = "1507.01749",
    archivePrefix = "arXiv",
    primaryClass = "hep-lat",
    reportNumber = "MITP-15-050",
    doi = "10.1140/epja/i2015-15158-0",
    journal = "Eur. Phys. J. A",
    volume = "51",
    number = "12",
    pages = "158",
    year = "2015"
}

@article{vonHippel:2025uhr,
    author = "von Hippel, Georg and Ottnad, Konstantin",
    title = "{Low-Energy Constants of Chiral Perturbation Theory from Pion Scalar Form Factors in~Nf=2+1-Flavor Lattice QCD with Controlled Errors}",
    eprint = "2503.20689",
    archivePrefix = "arXiv",
    primaryClass = "hep-lat",
    reportNumber = "MITP-25-025",
    doi = "10.1103/f4x5-frx1",
    journal = "Phys. Rev. Lett.",
    volume = "135",
    number = "7",
    pages = "071904",
    year = "2025"
}

@article{Ottnad:2025hkw,
    author = "Ottnad, Konstantin and von Hippel, Georg",
    title = "{Scalar size of the pion from lattice QCD}",
    eprint = "2503.20687",
    archivePrefix = "arXiv",
    primaryClass = "hep-lat",
    reportNumber = "MITP-25-024",
    doi = "10.1103/qh7q-9nyy",
    journal = "Phys. Rev. D",
    volume = "112",
    number = "3",
    pages = "034504",
    year = "2025"
}

@article{Feng:2009ij,
    author = "Feng, Xu and Jansen, Karl and Renner, Dru B.",
    title = "{The $\pi^+ \pi^+$ scattering length from maximally twisted mass lattice QCD}",
    eprint = "0909.3255",
    archivePrefix = "arXiv",
    primaryClass = "hep-lat",
    reportNumber = "DESY-09-141, SFB-CPP-09-82, MS-TP-09-17",
    doi = "10.1016/j.physletb.2010.01.018",
    journal = "Phys. Lett. B",
    volume = "684",
    pages = "268--274",
    year = "2010"
}

@article{ETM:2015bzg,
    author = "Helmes, C. and Jost, C. and Knippschild, B. and Liu, C. and Liu, J. and Liu, L. and Urbach, C. and Ueding, M. and Wang, Z. and Werner, M.",
    collaboration = "ETM",
    title = "{Hadron-hadron interactions from N$_{f}$ = 2 + 1 + 1 lattice QCD: isospin-2 \ensuremath{\pi}\ensuremath{\pi} scattering length}",
    eprint = "1506.00408",
    archivePrefix = "arXiv",
    primaryClass = "hep-lat",
    doi = "10.1007/JHEP09(2015)109",
    journal = "JHEP",
    volume = "09",
    pages = "109",
    year = "2015"
}

@article{ETM:2009ztk,
    author = "Baron, Remi and others",
    collaboration = "ETM",
    title = "{Light Meson Physics from Maximally Twisted Mass Lattice QCD}",
    eprint = "0911.5061",
    archivePrefix = "arXiv",
    primaryClass = "hep-lat",
    reportNumber = "HISKP-TH-09-37, DESY-09-187, SFB-CPP-09-115, MS-TP-09-27, IFT-UAM-CSIC-09-58, ROM2F-2009-24",
    doi = "10.1007/JHEP08(2010)097",
    journal = "JHEP",
    volume = "08",
    pages = "097",
    year = "2010"
}

@article{Brandt:2013dua,
    author = {Brandt, Bastian B. and J\"uttner, Andreas and Wittig, Hartmut},
    title = "{The pion vector form factor from lattice QCD and NNLO chiral perturbation theory}",
    eprint = "1306.2916",
    archivePrefix = "arXiv",
    primaryClass = "hep-lat",
    reportNumber = "HIM-2013-03, MITP-13-029, SHEP",
    doi = "10.1007/JHEP11(2013)034",
    journal = "JHEP",
    volume = "11",
    pages = "034",
    year = "2013"
}

@article{Engel:2014eea,
    author = "Engel, Georg P. and Giusti, Leonardo and Lottini, Stefano and Sommer, Rainer",
    title = "{Spectral density of the Dirac operator in two-flavor QCD}",
    eprint = "1411.6386",
    archivePrefix = "arXiv",
    primaryClass = "hep-lat",
    doi = "10.1103/PhysRevD.91.054505",
    journal = "Phys. Rev. D",
    volume = "91",
    number = "5",
    pages = "054505",
    year = "2015"
}

@article{Bazavov:2010yq,
    author = "Bazavov, A. and others",
    editor = "Rossi, Giancarlo",
    title = "{Staggered chiral perturbation theory in the two-flavor case and SU(2) analysis of the MILC data}",
    eprint = "1011.1792",
    archivePrefix = "arXiv",
    primaryClass = "hep-lat",
    journal = "PoS",
    volume = "LATTICE2010",
    pages = "083",
    year = "2010"
}

@article{Borsanyi:2012zv,
    author = "Borsanyi, Szabolcs and Durr, Stephan and Fodor, Zoltan and Krieg, Stefan and Schafer, Andreas and Scholz, Enno E and Szabo, Kalman K",
    title = "{SU(2) chiral perturbation theory low-energy constants from 2+1 flavor staggered lattice simulations}",
    eprint = "1205.0788",
    archivePrefix = "arXiv",
    primaryClass = "hep-lat",
    doi = "10.1103/PhysRevD.88.014513",
    journal = "Phys. Rev. D",
    volume = "88",
    pages = "014513",
    year = "2013"
}

@article{Budapest-Marseille-Wuppertal:2013vij,
    author = {D\"urr, Stephan and others},
    collaboration = "Budapest-Marseille-Wuppertal",
    title = "{Lattice QCD at the physical point meets SU(2) chiral perturbation theory}",
    eprint = "1310.3626",
    archivePrefix = "arXiv",
    primaryClass = "hep-lat",
    reportNumber = "CPT-P005-2013, WUB-13-14",
    doi = "10.1103/PhysRevD.90.114504",
    journal = "Phys. Rev. D",
    volume = "90",
    number = "11",
    pages = "114504",
    year = "2014"
}

@article{Boyle:2015exm,
    author = "Boyle, P. A. and others",
    title = "{Low energy constants of SU(2) partially quenched chiral perturbation theory from N$_f$=2+1 domain wall QCD}",
    eprint = "1511.01950",
    archivePrefix = "arXiv",
    primaryClass = "hep-lat",
    doi = "10.1103/PhysRevD.93.054502",
    journal = "Phys. Rev. D",
    volume = "93",
    number = "5",
    pages = "054502",
    year = "2016"
}

@article{Cossu:2016eqs,
    author = "Cossu, Guido and Fukaya, Hidenori and Hashimoto, Shoji and Kaneko, Takashi and Noaki, Jun-Ichi",
    title = "{Stochastic calculation of the Dirac spectrum on the lattice and a determination of chiral condensate in 2+1-flavor QCD}",
    eprint = "1607.01099",
    archivePrefix = "arXiv",
    primaryClass = "hep-lat",
    reportNumber = "KEK-CP-346, OU-HET-898",
    doi = "10.1093/ptep/ptw129",
    journal = "PTEP",
    volume = "2016",
    number = "9",
    pages = "093B06",
    year = "2016"
}

@article{Aoki:2017paw,
    author = "Aoki, S. and Cossu, G. and Fukaya, H. and Hashimoto, S. and Kaneko, T.",
    collaboration = "JLQCD",
    title = {{Topological susceptibility of QCD with dynamical M\"obius domain-wall fermions}},
    eprint = "1705.10906",
    archivePrefix = "arXiv",
    primaryClass = "hep-lat",
    reportNumber = "OU-HET-937, KEK-CP-359, YITP-17-51",
    doi = "10.1093/ptep/pty041",
    journal = "PTEP",
    volume = "2018",
    number = "4",
    pages = "043B07",
    year = "2018"
}

@article{Cichy:2013gja,
    author = "Cichy, Krzysztof and Garcia-Ramos, Elena and Jansen, Karl",
    title = "{Chiral condensate from the twisted mass Dirac operator spectrum}",
    eprint = "1303.1954",
    archivePrefix = "arXiv",
    primaryClass = "hep-lat",
    reportNumber = "DESY-13-038, HU-EP-13-08, SFB-CPP-13-17",
    doi = "10.1007/JHEP10(2013)175",
    journal = "JHEP",
    volume = "10",
    pages = "175",
    year = "2013"
}

@article{Alexandrou:2017bzk,
    author = "Alexandrou, Constantia and Athenodorou, Andreas and Cichy, Krzysztof and Constantinou, Martha and Horkel, Derek P. and Jansen, Karl and Koutsou, Giannis and Larkin, Conor",
    title = "{Topological susceptibility from twisted mass fermions using spectral projectors and the gradient flow}",
    eprint = "1709.06596",
    archivePrefix = "arXiv",
    primaryClass = "hep-lat",
    reportNumber = "DESY-17-139",
    doi = "10.1103/PhysRevD.97.074503",
    journal = "Phys. Rev. D",
    volume = "97",
    number = "7",
    pages = "074503",
    year = "2018"
}

@article{Frezzotti:2008dr,
    author = "Frezzotti, R. and Lubicz, V. and Simula, S.",
    collaboration = "ETM",
    title = "{Electromagnetic form factor of the pion from twisted-mass lattice QCD at N(f) = 2}",
    eprint = "0812.4042",
    archivePrefix = "arXiv",
    primaryClass = "hep-lat",
    reportNumber = "RM3-TH-08-15, ROM2F-2008-24",
    doi = "10.1103/PhysRevD.79.074506",
    journal = "Phys. Rev. D",
    volume = "79",
    pages = "074506",
    year = "2009"
}

@article{MILC:2010hzw,
    author = "Bazavov, A. and others",
    editor = "Rossi, Giancarlo",
    collaboration = "MILC",
    title = "{Results for light pseudoscalar mesons}",
    eprint = "1012.0868",
    archivePrefix = "arXiv",
    primaryClass = "hep-lat",
    doi = "10.22323/1.105.0074",
    journal = "PoS",
    volume = "LATTICE2010",
    pages = "074",
    year = "2010"
}

@article{Beane:2011zm,
    author = "Beane, S. R. and Detmold, W. and Junnarkar, P. M. and Luu, T. C. and Orginos, K. and Parreno, A. and Savage, M. J. and Torok, A. and Walker-Loud, A.",
    title = "{SU(2) Low-Energy Constants from Mixed-Action Lattice QCD}",
    eprint = "1108.1380",
    archivePrefix = "arXiv",
    primaryClass = "hep-lat",
    reportNumber = "JLAB-THY-11-1380, NT-LBNL-11-013, NT-UW-11-14, UCB-NPAT-11-008, UNH-11-4",
    doi = "10.1103/PhysRevD.86.094509",
    journal = "Phys. Rev. D",
    volume = "86",
    pages = "094509",
    year = "2012"
}

@article{Bonanno:2025cmh,
    author = "Bonanno, Claudio and others",
    title = "{Lattice determination of the QCD low-energy constant {\ensuremath{\ell}}7}",
    eprint = "2512.09880",
    archivePrefix = "arXiv",
    primaryClass = "hep-lat",
    doi = "10.1103/8sy5-f78d",
    journal = "Phys. Rev. D",
    volume = "113",
    number = "7",
    pages = "074510",
    year = "2026"
}

@article{Bonanno:2023xkg,
    author = "Bonanno, Claudio and D'Angelo, Francesco and D'Elia, Massimo",
    title = "{The chiral condensate of N$_{f}$ = 2 + 1 QCD from the spectrum of the staggered Dirac operator}",
    eprint = "2308.01303",
    archivePrefix = "arXiv",
    primaryClass = "hep-lat",
    doi = "10.1007/JHEP11(2023)013",
    journal = "JHEP",
    volume = "11",
    pages = "013",
    year = "2023"
}

@article{ETM:2010cqp,
    author = "Baron, R. and others",
    editor = "Rossi, Giancarlo",
    collaboration = "ETM",
    title = "{Light hadrons from Nf=2+1+1 dynamical twisted mass fermions}",
    eprint = "1101.0518",
    archivePrefix = "arXiv",
    primaryClass = "hep-lat",
    reportNumber = "HU-EP-10-75, SFB-CPP-10-110, DESY-10-205, HU-EP-10/75, SFB/CPP-10-110, MS-TP-10-22, DESY 10-205",
    doi = "10.22323/1.105.0123",
    journal = "PoS",
    volume = "LATTICE2010",
    pages = "123",
    year = "2010"
}

@article{Cheung:2026lpv,
    author = "Cheung, Clifford and Jeong, Jaehoon and Ko, Pyungwon and Pomarol, Alex and Remmen, Grant N. and Sciotti, Francesco",
    title = "{Multipositivity Constrains the Chiral Lagrangian}",
    eprint = "2605.21582",
    archivePrefix = "arXiv",
    primaryClass = "hep-th",
    reportNumber = "CALT-TH 2026-019, KIAS-Q26009",
    month = "5",
    year = "2026"
}

@article{Gasser:1983yg,
    author = "Gasser, J. and Leutwyler, H.",
    title = "{Chiral Perturbation Theory to One Loop}",
    reportNumber = "CERN-TH-3689",
    doi = "10.1016/0003-4916(84)90242-2",
    journal = "Annals Phys.",
    volume = "158",
    pages = "142",
    year = "1984"
}

@article{Colangelo:2001df,
    author = "Colangelo, G. and Gasser, J. and Leutwyler, H.",
    title = "{$\pi \pi$ scattering}",
    eprint = "hep-ph/0103088",
    archivePrefix = "arXiv",
    reportNumber = "ZU-TH-3-01, BUTP-01-1",
    doi = "10.1016/S0550-3213(01)00147-X",
    journal = "Nucl. Phys. B",
    volume = "603",
    pages = "125--179",
    year = "2001"
}

@article{Kaiser:2000gs,
    author = "Kaiser, Roland and Leutwyler, H.",
    title = "{Large $N_c$ in chiral perturbation theory}",
    eprint = "hep-ph/0007101",
    archivePrefix = "arXiv",
    reportNumber = "BUTP-00-19",
    doi = "10.1007/s100520000499",
    journal = "Eur. Phys. J. C",
    volume = "17",
    pages = "623--649",
    year = "2000"
}

@article{Bocchia:2026kew,
    author = "Bocchia, Sebastiano and Vichi, Alessandro",
    title = "{Primal Bootstrap for Pion Scattering at Large-Nc}",
    eprint = "2606.14676",
    archivePrefix = "arXiv",
    primaryClass = "hep-th",
    month = "6",
    year = "2026"
}

@article{Li:2023qzs,
    author = "Li, Yue-Zhou",
    title = "{Effective field theory bootstrap, large-$N$ $\chi$PT and holographic QCD}",
    eprint = "2310.09698",
    archivePrefix = "arXiv",
    primaryClass = "hep-th",
    doi = "10.1007/JHEP01(2024)072",
    journal = "JHEP",
    volume = "01",
    pages = "072",
    year = "2024"
}

@article{Guerrieri:2024jkn,
    author = {Guerrieri, Andrea and H\"aring, Kelian and Su, Ning},
    title = "{From data to the analytic S-matrix: A Bootstrap fit of the pion scattering amplitude}",
    eprint = "2410.23333",
    archivePrefix = "arXiv",
    primaryClass = "hep-th",
    reportNumber = "CERN-TH-2024-177, CALT-TH-2024-038",
    month = "10",
    year = "2024"
}

@article{Albert:2022oes,
    author = "Albert, Jan and Rastelli, Leonardo",
    title = "{Bootstrapping pions at large $N$}",
    eprint = "2203.11950",
    archivePrefix = "arXiv",
    primaryClass = "hep-th",
    reportNumber = "YITP-SB-2022-07",
    doi = "10.1007/JHEP08(2022)151",
    journal = "JHEP",
    volume = "08",
    pages = "151",
    year = "2022"
}

@article{Fernandez:2022kzi,
    author = "Fernandez, Clara and Pomarol, Alex and Riva, Francesco and Sciotti, Francesco",
    title = "{Cornering large-$N_{c}$ QCD with positivity bounds}",
    eprint = "2211.12488",
    archivePrefix = "arXiv",
    primaryClass = "hep-th",
    doi = "10.1007/JHEP06(2023)094",
    journal = "JHEP",
    volume = "06",
    pages = "094",
    year = "2023"
}

@article{Albert:2023jtd,
    author = "Albert, Jan and Rastelli, Leonardo",
    title = "{Bootstrapping pions at large $N$. Part II. Background gauge fields and the chiral anomaly}",
    eprint = "2307.01246",
    archivePrefix = "arXiv",
    primaryClass = "hep-th",
    reportNumber = "YITP-SB-2023-15",
    doi = "10.1007/JHEP09(2024)039",
    journal = "JHEP",
    volume = "09",
    pages = "039",
    year = "2024"
}

@article{Albert:2023seb,
    author = "Albert, Jan and Henriksson, Johan and Rastelli, Leonardo and Vichi, Alessandro",
    title = "{Bootstrapping mesons at large $N$: Regge trajectory from spin-two maximization}",
    eprint = "2312.15013",
    archivePrefix = "arXiv",
    primaryClass = "hep-th",
    reportNumber = "YITP-SB-2023-41",
    doi = "10.1007/JHEP09(2024)172",
    journal = "JHEP",
    volume = "09",
    pages = "172",
    year = "2024"
}

@article{Albert:2026xyz,
    author = "Albert, Jan and Kosva, Dilara and Rastelli, Leonardo",
    title = "{Bootstrapping Pion Form Factors at Large $N$}",
    eprint = "2606.19420",
    archivePrefix = "arXiv",
    primaryClass = "hep-th",
    month = "6",
    year = "2026"
}

@article{Ma:2023vgc,
    author = "Ma, Teng and Pomarol, Alex and Sciotti, Francesco",
    title = "{Bootstrapping the chiral anomaly at large N$_{c}$}",
    eprint = "2307.04729",
    archivePrefix = "arXiv",
    primaryClass = "hep-th",
    doi = "10.1007/JHEP11(2023)176",
    journal = "JHEP",
    volume = "11",
    pages = "176",
    year = "2023"
}

@article{Cirigliano:2003yq,
    author = "Cirigliano, V. and Ecker, G. and Neufeld, H. and Pich, A.",
    title = "{Meson resonances, large $N_c$ and chiral symmetry}",
    eprint = "hep-ph/0305311",
    archivePrefix = "arXiv",
    reportNumber = "IFIC-03-23, UWTHPH-2003-8",
    doi = "10.1088/1126-6708/2003/06/012",
    journal = "JHEP",
    volume = "06",
    pages = "012",
    year = "2003"
}

@article{Weinberg:2013cfa,
    author = "Weinberg, Steven",
    title = "{Tetraquark Mesons in Large $N$ Quantum Chromodynamics}",
    eprint = "1303.0342",
    archivePrefix = "arXiv",
    primaryClass = "hep-ph",
    reportNumber = "UTTG-06-13",
    doi = "10.1103/PhysRevLett.110.261601",
    journal = "Phys. Rev. Lett.",
    volume = "110",
    pages = "261601",
    year = "2013"
}

@article{Cohen:2014vta,
    author = "Cohen, Thomas and Llanes-Estrada, Felipe J. and Pelaez, J. R. and Ruiz de Elvira, J.",
    title = "{Nonordinary light meson couplings and the $1/N_c$ expansion}",
    eprint = "1405.4831",
    archivePrefix = "arXiv",
    primaryClass = "hep-ph",
    doi = "10.1103/PhysRevD.90.036003",
    journal = "Phys. Rev. D",
    volume = "90",
    number = "3",
    pages = "036003",
    year = "2014"
}

@article{Pelaez:2015qba,
    author = "Pelaez, J. R.",
    title = "{From controversy to precision on the sigma meson: a review on the status of the non-ordinary $f_0(500)$ resonance}",
    eprint = "1510.00653",
    archivePrefix = "arXiv",
    primaryClass = "hep-ph",
    doi = "10.1016/j.physrep.2016.09.001",
    journal = "Phys. Rept.",
    volume = "658",
    pages = "1",
    year = "2016"
}

@article{Bonanno:2025bla,
    author = "Bonanno, Claudio and Garc{\'\i}a P{\'e}rez, Margarita and Gonz{\'a}lez-Arroyo, Antonio and Ishikawa, Ken-Ichi and Okawa, Masanori and Panfalone, Dario",
    title = "{Universal features of chiral symmetry breaking in large-N QCD}",
    eprint = "2510.25644",
    archivePrefix = "arXiv",
    primaryClass = "hep-lat",
    reportNumber = "HUPD-2507, IFT-UAM/CSIC-25-114",
    doi = "10.1007/JHEP04(2026)015",
    journal = "JHEP",
    volume = "04",
    pages = "015",
    year = "2026"
}

@article{Gonzalez-Arroyo:1983zev,
    author = "Gonz{\'a}lez-Arroyo, Antonio and Korthals Altes, C. P.",
    title = "{Reduced Model for Large $N$ Continuum Field
Theories}",
    reportNumber = "FTUAM-83-16",
    doi = "10.1016/0370-2693(83)90526-9",
    journal = "Phys. Lett. B",
    volume = "131",
    pages = "396--398",
    year = "1983"
}

@article{Ambjorn:2000cs,
    author = "Ambjorn, Jan and Makeenko, Y. M. and Nishimura, J. and Szabo, R. J.",
    title = "{Lattice gauge fields and discrete noncommutative Yang-Mills theory}",
    eprint = "hep-th/0004147",
    archivePrefix = "arXiv",
    reportNumber = "NBI-HE-00-21, ITEP-TH-11-00",
    doi = "10.1088/1126-6708/2000/05/023",
    journal = "JHEP",
    volume = "05",
    pages = "023",
    year = "2000"
}

@article{Douglas:2001ba,
    author = "Douglas, Michael R. and Nekrasov, Nikita A.",
    title = "{Noncommutative field theory}",
	  eprint = "hep-th/0106048",
	  archivePrefix = "arXiv",
	  reportNumber = "ITEP-TH-31-01, IHES-P-01-27, RUNHETC-2001-18",
	  doi = "10.1103/RevModPhys.73.977",
	  journal = "Rev. Mod. Phys.",
	  volume = "73",
    pages = "977--1029",
    year = "2001"
}

@article{Edwards:1999ra,
    author = "Edwards, Robert G. and Heller, Urs M. and Kiskis, Joe E. and Narayanan, Rajamani",
    title = "{Quark spectra, topology and random matrix theory}",
    eprint = "hep-th/9902117",
    archivePrefix = "arXiv",
    reportNumber = "FSU-SCRI-99-08, UCD-1999-3",
    doi = "10.1103/PhysRevLett.82.4188",
    journal = "Phys. Rev. Lett.",
    volume = "82",
    pages = "4188--4191",
    year = "1999"
}

@article{Damgaard:1999tk,
    author = "Damgaard, Poul H. and Edwards, Robert G. and Heller, Urs M. and Narayanan, Rajamani",
    title = "{Universal scaling of the chiral condensate in finite volume gauge theories}",
    eprint = "hep-lat/9907016",
    archivePrefix = "arXiv",
    reportNumber = "NBI-HE-99-23, FSU-SCRI-99-41",
    doi = "10.1103/PhysRevD.61.094503",
    journal = "Phys. Rev. D",
    volume = "61",
    pages = "094503",
    year = "2000"
}

@article{Bietenholz:2003mi,
    author = "Bietenholz, Wolfgang and Jansen, K. and Shcheredin, S.",
    title = "{Spectral properties of the overlap Dirac operator in QCD}",
    eprint = "hep-lat/0306022",
    archivePrefix = "arXiv",
    reportNumber = "HU-EP-03-33, DESY-03-070, SFB-CPP-03-10",
    doi = "10.1088/1126-6708/2003/07/033",
    journal = "JHEP",
    volume = "07",
    pages = "033",
    year = "2003"
}

@article{Nishigaki:1998is,
    author = "Nishigaki, Shinsuke M. and Damgaard, Poul H. and Wettig, Tilo",
    title = "{Smallest Dirac eigenvalue distribution from random matrix theory}",
    eprint = "hep-th/9803007",
    archivePrefix = "arXiv",
    reportNumber = "NSF-ITP-98-021, NBI-HE-98-03, TUM-T39-98-7",
    doi = "10.1103/PhysRevD.58.087704",
    journal = "Phys. Rev. D",
    volume = "58",
    pages = "087704",
    year = "1998"
}

@article{Neuberger:1997fp,
    author = "Neuberger, Herbert",
    title = "{Exactly massless quarks on the lattice}",
    eprint = "hep-lat/9707022",
    archivePrefix = "arXiv",
    reportNumber = "RU-97-63",
    doi = "10.1016/S0370-2693(97)01368-3",
    journal = "Phys. Lett. B",
    volume = "417",
    pages = "141--144",
    year = "1998"
}

@article{Kaplan:1992bt,
    author = "Kaplan, David B.",
    title = "{A Method for simulating chiral fermions on the lattice}",
    eprint = "hep-lat/9206013",
    archivePrefix = "arXiv",
    reportNumber = "UCSD-PTH-92-16",
    doi = "10.1016/0370-2693(92)91112-M",
    journal = "Phys. Lett. B",
    volume = "288",
    pages = "342--347",
    year = "1992"
}

@article{Neuberger:1998wv,
    author = "Neuberger, Herbert",
    title = "{More about exactly massless quarks on the lattice}",
    eprint = "hep-lat/9801031",
    archivePrefix = "arXiv",
    reportNumber = "RU-98-03",
    doi = "10.1016/S0370-2693(98)00355-4",
    journal = "Phys. Lett. B",
    volume = "427",
    pages = "353--355",
    year = "1998"
}

@article{Edwards:2000qv,
    author = "Edwards, Robert G. and Heller, Urs M.",
    title = "{Domain wall fermions with exact chiral symmetry}",
    eprint = "hep-lat/0005002",
    archivePrefix = "arXiv",
    reportNumber = "JLAB-THY-00-14",
    doi = "10.1103/PhysRevD.63.094505",
    journal = "Phys. Rev. D",
    volume = "63",
    pages = "094505",
    year = "2001"
}

@article{Borici:1999zw,
    author = "Borici, A.",
    editor = "Campostrini, M. and Caracciolo, S. and Cosmai, L. and Di Giacomo, A. and Rossi, P. and Rapuano, F.",
    title = "{Truncated overlap fermions}",
    eprint = "hep-lat/9909057",
    archivePrefix = "arXiv",
    doi = "10.1016/S0920-5632(00)91802-4",
    journal = "Nucl. Phys. B Proc. Suppl.",
    volume = "83",
    pages = "771--773",
    year = "2000"
}

@article{Damgaard:2000ah,
    author = "Damgaard, Poul H. and Nishigaki, Shinsuke M.",
    title = "{Distribution of the k-th smallest Dirac operator eigenvalue}",
    eprint = "hep-th/0006111",
    archivePrefix = "arXiv",
    reportNumber = "TIT-HEP-446",
    doi = "10.1103/PhysRevD.63.045012",
    journal = "Phys. Rev. D",
    volume = "63",
    pages = "045012",
    year = "2001"
}

@article{Brower:2005qw,
    author = "Brower, R. C. and Neff, H. and Orginos, K.",
    editor = "Alexandrou, Constantia and Panagopoulos, Haralambos and Schierholz, Gerrit",
    title = "{Mobius fermions}",
    eprint = "hep-lat/0511031",
    archivePrefix = "arXiv",
    doi = "10.1016/j.nuclphysbps.2006.01.047",
    journal = "Nucl. Phys. B Proc. Suppl.",
    volume = "153",
    pages = "191--198",
    year = "2006"
}

@article{Shamir:1993zy,
    author = "Shamir, Yigal",
    title = "{Chiral fermions from lattice boundaries}",
    eprint = "hep-lat/9303005",
    archivePrefix = "arXiv",
    reportNumber = "WIS-93-20-PH",
    doi = "10.1016/0550-3213(93)90162-I",
    journal = "Nucl. Phys. B",
    volume = "406",
    pages = "90--106",
    year = "1993"
}

@article{Furman:1994ky,
    author = "Furman, Vadim and Shamir, Yigal",
    title = "{Axial symmetries in lattice QCD with Kaplan fermions}",
    eprint = "hep-lat/9405004",
    archivePrefix = "arXiv",
    reportNumber = "WIS-94-19-PH, TAUP-2148-94",
    doi = "10.1016/0550-3213(95)00031-M",
    journal = "Nucl. Phys. B",
    volume = "439",
    pages = "54--78",
    year = "1995"
}

@article{Kikukawa:2000ac,
    author = "Kikukawa, Yoshio and Noguchi, Tatsuya",
    editor = "Campostrini, M. and Caracciolo, S. and Cosmai, L. and Di Giacomo, A. and Rossi, P. and Rapuano, F.",
    title = "{Low-energy effective action of domain wall fermion and the Ginsparg-Wilson relation}",
    eprint = "hep-lat/9902022",
    archivePrefix = "arXiv",
    reportNumber = "KUNS-1560",
    doi = "10.1016/S0920-5632(00)91758-4",
    journal = "Nucl. Phys. B Proc. Suppl.",
    volume = "83",
    pages = "630--632",
    year = "2000"
}

@article{Hasenfratz:2008ce,
    author = "Hasenfratz, Anna and Hoffmann, Roland and Schaefer, Stefan",
    title = "{Low energy chiral constants from epsilon-regime simulations with improved Wilson fermions}",
    eprint = "0806.4586",
    archivePrefix = "arXiv",
    primaryClass = "hep-lat",
    reportNumber = "HU-EP-08-23, SFB-CPP-08-35",
    doi = "10.1103/PhysRevD.78.054511",
    journal = "Phys. Rev. D",
    volume = "78",
    pages = "054511",
    year = "2008"
}

@article{Gasser:1987ah,
    author = "Gasser, J. and Leutwyler, H.",
    title = "{Thermodynamics of Chiral Symmetry}",
    reportNumber = "BUTP-87/2-BERN",
    doi = "10.1016/0370-2693(87)91652-2",
    journal = "Phys. Lett. B",
    volume = "188",
    pages = "477--481",
    year = "1987"
}

@article{Basile:2007ki,
    author = "Basile, Francesco and Akemann, Gernot",
    title = "{Equivalence of QCD in the epsilon-regime and chiral random matrix theory with or without chemical potential}",
    eprint = "0710.0376",
    archivePrefix = "arXiv",
    primaryClass = "hep-th",
    reportNumber = "CPT-P33-2007",
    doi = "10.1088/1126-6708/2007/12/043",
    journal = "JHEP",
    volume = "12",
    pages = "043",
    year = "2007"
}

@article{Ishikawa:2013rxa,
    author = "Ishikawa, Ken-Ichi",
    title = "{Testing reweighting method for truncated Overlap fermions}",
    eprint = "1310.8022",
    archivePrefix = "arXiv",
    primaryClass = "hep-lat",
    doi = "10.22323/1.187.0026",
    journal = "PoS",
    volume = "LATTICE2013",
    pages = "026",
    year = 2014
}

@article{DelDebbio:2007wk,
    author = "Del Debbio, Luigi and Lucini, Biagio and Patella, Agostino and Pica, Claudio",
    title = "{Quenched mesonic spectrum at large N}",
    eprint = "0712.3036",
    archivePrefix = "arXiv",
    primaryClass = "hep-th",
    reportNumber = "BNL-NT-07-56, IFUP-TH-2007-36, NI07089",
    doi = "10.1088/1126-6708/2008/03/062",
    journal = "JHEP",
    volume = "03",
    pages = "062",
    year = "2008"
}

@article{Shuryak:1992pi,
    author = "Shuryak, Edward V. and Verbaarschot, J. J. M.",
    title = "{Random matrix theory and spectral sum rules for the Dirac operator in QCD}",
    eprint = "hep-th/9212088",
    archivePrefix = "arXiv",
    reportNumber = "SUNY-NTG-92-45",
    doi = "10.1016/0375-9474(93)90098-I",
    journal = "Nucl. Phys. A",
    volume = "560",
    pages = "306--320",
    year = "1993"
}

@article{Verbaarschot:1993pm,
    author = "Verbaarschot, J. J. M. and Zahed, I.",
    title = "{Spectral density of the QCD Dirac operator near zero virtuality}",
    eprint = "hep-th/9303012",
    archivePrefix = "arXiv",
    reportNumber = "SUNY-NTG-93-2",
    doi = "10.1103/PhysRevLett.70.3852",
    journal = "Phys. Rev. Lett.",
    volume = "70",
    pages = "3852--3855",
    year = "1993"
}

@article{Bernardoni:2008ei,
    author = "Bernardoni, F. and Damgaard, P. H. and Fukaya, H. and Hern\'andez, P.",
    title = "{Finite Volume Scaling of Pseudo Nambu-Goldstone Bosons in QCD}",
    eprint = "0808.1986",
    archivePrefix = "arXiv",
    primaryClass = "hep-lat",
    reportNumber = "IFIC-08-38, FTUV-08-0815",
    doi = "10.1088/1126-6708/2008/10/008",
    journal = "JHEP",
    volume = "10",
    pages = "008",
    year = "2008"
}

@article{Bernardoni:2010nf,
    author = "Bernardoni, Fabio and Hern\'andez, Pilar and Garron, Nicolas and Necco, Silvia and Pena, Carlos",
    title = "{Probing the chiral regime of $N_{f}$= 2 QCD with mixed actions}",
    eprint = "1008.1870",
    archivePrefix = "arXiv",
    primaryClass = "hep-lat",
    reportNumber = "IFIC-10-27, FTUV-10-0810, FTUAM-10-10, IFT-UAM-CSIC-10-47, CERN-PH-TH-2010-178, EDINBURGH-2010-22",
    doi = "10.1103/PhysRevD.83.054503",
    journal = "Phys. Rev. D",
    volume = "83",
    pages = "054503",
    year = "2011"
}

@article{Wennekers:2005wa,
    author = "Wennekers, Jan and Wittig, Hartmut",
    title = "{On the renormalized scalar density in quenched QCD}",
    eprint = "hep-lat/0507026",
    archivePrefix = "arXiv",
    reportNumber = "DESY-05-129",
    doi = "10.1088/1126-6708/2005/09/059",
    journal = "JHEP",
    volume = "09",
    pages = "059",
    year = "2005"
}

@article{Deuzeman:2011dh,
    author = "Deuzeman, Albert and Wenger, Urs and Wuilloud, Jair",
    title = "{Spectral properties of the Wilson Dirac operator in the $\epsilon$-regime}",
    eprint = "1110.4002",
    archivePrefix = "arXiv",
    primaryClass = "hep-lat",
    doi = "10.1007/JHEP12(2011)109",
    journal = "JHEP",
    volume = "12",
    pages = "109",
    year = "2011"
}

@article{Catillo:2017qbz,
    author = "Catillo, M. and Glozman, L. Ya.",
    title = "{Distribution law of the Dirac eigenmodes in QCD}",
    eprint = "1709.01886",
    archivePrefix = "arXiv",
    primaryClass = "hep-lat",
    doi = "10.1142/S0217751X18500549",
    journal = "Int. J. Mod. Phys. A",
    volume = "33",
    number = "10",
    pages = "1850054",
    year = "2018"
}

@article{Verbaarschot:2000dy,
    author = "Verbaarschot, J. J. M. and Wettig, T.",
    title = "{Random matrix theory and chiral symmetry in QCD}",
    eprint = "hep-ph/0003017",
    archivePrefix = "arXiv",
    reportNumber = "SUNY-NTG-00-5",
    doi = "10.1146/annurev.nucl.50.1.343",
    journal = "Ann. Rev. Nucl. Part. Sci.",
    volume = "50",
    pages = "343--410",
    year = "2000"
}

@article{Giusti:2003gf,
    author = "Giusti, Leonardo and L{\"u}scher, Martin and Weisz, Peter and Wittig, Hartmut",
    title = "{Lattice QCD in the epsilon regime and random matrix theory}",
    eprint = "hep-lat/0309189",
    archivePrefix = "arXiv",
    reportNumber = "CERN-TH-2003-213, CPT-2003-PE-4569, DESY-03-135, MPP-2003-74",
    doi = "10.1088/1126-6708/2003/11/023",
    journal = "JHEP",
    volume = "11",
    pages = "023",
    year = "2003"
}

@article{Leutwyler:1992yt,
    author = "Leutwyler, H. and Smilga, Andrei V.",
    title = "{Spectrum of Dirac operator and role of winding number in QCD}",
    reportNumber = "BUTP-92-10",
    doi = "10.1103/PhysRevD.46.5607",
    journal = "Phys. Rev. D",
    volume = "46",
    pages = "5607--5632",
    year = "1992"
}

@article{Verbaarschot:1994qf,
    author = "Verbaarschot, Jacobus J. M.",
    title = "{The Spectrum of the QCD Dirac operator and chiral random matrix theory: The Threefold way}",
    eprint = "hep-th/9401059",
    archivePrefix = "arXiv",
    reportNumber = "SUNY-NTG-94-1",
    doi = "10.1103/PhysRevLett.72.2531",
    journal = "Phys. Rev. Lett.",
    volume = "72",
    pages = "2531--2533",
    year = "1994"
}

@article{Aoki:2001su,
    author = "Aoki, S. and Taniguchi, Y.",
    title = "{Chiral properties of domain wall fermions from eigenvalues of four-dimensional Wilson-Dirac operator}",
    eprint = "hep-lat/0109022",
    archivePrefix = "arXiv",
    reportNumber = "UTHEP-447, UTCCP-P-112",
    doi = "10.1103/PhysRevD.65.074502",
    journal = "Phys. Rev. D",
    volume = "65",
    pages = "074502",
    year = "2002"
}

@article{Baeza-Ballesteros:2022azb,
    author = "Baeza-Ballesteros, Jorge and Hern\'andez, Pilar and Romero-L\'opez, Fernando",
    title = "{A lattice study of \ensuremath{\pi}\ensuremath{\pi} scattering at large N$_{c}$}",
    eprint = "2202.02291",
    archivePrefix = "arXiv",
    primaryClass = "hep-lat",
    reportNumber = "MIT-CTP/5395",
    doi = "10.1007/JHEP06(2022)049",
    journal = "JHEP",
    volume = "06",
    pages = "049",
    year = "2022"
}

@article{Baeza-Ballesteros:2025iee,
    author = "Baeza-Ballesteros, Jorge and Hern{\'a}ndez, Pilar and Romero-L{\'o}pez, Fernando",
    title = "{The $\pi\pi$ scattering amplitude at large $N_{c}$}",
    eprint = "2503.13978",
    archivePrefix = "arXiv",
    primaryClass = "hep-lat",
    reportNumber = "DESY-25-040, MIT-CTP/5850",
    doi = "10.1007/JHEP08(2025)110",
    journal = "JHEP",
    volume = "08",
    pages = "110",
    year = "2025"
}

@article{Butti:2023hfp,
    author = "Butti, Pietro and Gonz\'alez-Arroyo, Antonio",
    title = "{Testing (asymptotic) scaling in Yang-Mills theories in the large-$N_c$ limit}",
    eprint = "2311.18696",
    archivePrefix = "arXiv",
    primaryClass = "hep-lat",
    reportNumber = "IFT-UAM/CSIC-23-58",
    doi = "10.22323/1.453.0381",
    journal = "PoS",
    volume = "LATTICE2023",
    pages = "381",
    year = "2024"
}

@article{Bribian:2019ybc,
    author = "Bribi\'an, Eduardo I. and Garc\'ia P\'erez, Margarita",
    title = "{The twisted gradient flow coupling at one loop}",
    eprint = "1903.08029",
    archivePrefix = "arXiv",
    primaryClass = "hep-lat",
    reportNumber = "IFT-UAM/CSIC-19-26",
    doi = "10.1007/JHEP03(2019)200",
    journal = "JHEP",
    volume = "03",
    pages = "200",
    year = "2019"
}

@article{Bonanno:2023ypf,
    author = "Bonanno, Claudio and Butti, Pietro and Garc\'\i{}a P\'erez, Margarita and Gonz\'alez-Arroyo, Antonio and Ishikawa, Ken-Ichi and Okawa, Masanori",
    title = "{The large-$N$ limit of the chiral condensate from twisted reduced models}",
    eprint = "2309.15540",
    archivePrefix = "arXiv",
    primaryClass = "hep-lat",
    reportNumber = "IFT-UAM/CSIC-23-118, HUPD-2303",
    doi = "10.1007/JHEP12(2023)034",
    journal = "JHEP",
    volume = "12",
    pages = "034",
    year = "2023"
}

@article{Bonanno:2025hzr,
    author = "Bonanno, Claudio and Garc{\'\i}a P{\'e}rez, Margarita and Gonz{\'a}lez-Arroyo, Antonio and Ishikawa, Ken-Ichi and Okawa, Masanori",
    title = "{Non-perturbative determination of meson masses and low-energy constants in large-N QCD}",
    eprint = "2508.05446",
    archivePrefix = "arXiv",
    primaryClass = "hep-lat",
    reportNumber = "HUPD-2505, IFT-UAM/CSIC-25-83",
    doi = "10.1007/JHEP12(2025)096",
    journal = "JHEP",
    volume = "12",
    pages = "096",
    year = "2025"
}

@article{Bonanno:2024bqg,
    author = "Bonanno, Claudio and Butti, Pietro and Garc\'\i{}a P\'erez, Margarita and Gonz\'alez-Arroyo, Antonio and Ishikawa, Ken-Ichi and Okawa, Masanori",
    title = "{Nonperturbative determination of the $\mathcal{N}=1$ supersymmetric Yang-Mills gluino condensate at large $N$}",
    eprint = "2406.08995",
    archivePrefix = "arXiv",
    primaryClass = "hep-th",
    reportNumber = "IFT-UAM/CSIC-24-90, HUPD-2404",
    doi = "10.1103/PhysRevD.110.074507",
    journal = "Phys. Rev. D",
    volume = "110",
    number = "7",
    pages = "074507",
    year = "2024"
}

@article{Bonanno:2024onr,
    author = "Bonanno, Claudio and Garc\'\i{}a P\'erez, Margarita and Gonz\'alez-Arroyo, Antonio and Ishikawa, Ken-Ichi and Okawa, Masanori",
    title = "{The mass of the gluino-glue bound state in large-$N$ $\mathcal{N} = 1$ Supersymmetric Yang-Mills theory}",
    eprint = "2412.02348",
    archivePrefix = "arXiv",
    primaryClass = "hep-lat",
    reportNumber = "HUPD-2410, IFT-UAM/CSIC-24-167",
    doi = "10.1007/JHEP03(2025)174",
    journal = "JHEP",
    volume = "03",
    pages = "174",
    year = "2025"
}

@article{Aldazabal:1983ec,
author = "Aldazabal, G. and Parga, N. and Okawa, M. and Gonz\'alez-Arroyo, Antonio",
title = "{Large $N$ Reduced Models and Stochastic Quantization}",
reportNumber = "FTUAM/83-8",
doi = "10.1016/0370-2693(83)90735-9",
journal = "Phys. Lett. B",
volume = "129",
pages = "90--94",
year = "1983"
}

@article{DeGrand:2023hzz,
    author = "DeGrand, Thomas A. and Wickenden, Evan",
    title = "{Lattice study of the chiral properties of large-$N_c$ QCD}",
    eprint = "2309.12270",
    archivePrefix = "arXiv",
    primaryClass = "hep-lat",
    reportNumber = "FERMILAB-PUB-23-479-V",
    doi = "10.1103/PhysRevD.108.094516",
    journal = "Phys. Rev. D",
    volume = "108",
    number = "9",
    pages = "094516",
    year = "2023"
}

@article{DeGrand:2016pur,
    author = "DeGrand, Thomas and Liu, Yuzhi",
    title = "{Lattice study of large $N_c$ QCD}",
    eprint = "1606.01277",
    archivePrefix = "arXiv",
    primaryClass = "hep-lat",
    reportNumber = "COLO-HEP-590",
    doi = "10.1103/PhysRevD.94.034506",
    journal = "Phys. Rev. D",
    volume = "94",
    number = "3",
    pages = "034506",
    year = "2016",
    note = "[Erratum: Phys.Rev.D 95, 019902 (2017)]"
}

@inproceedings{Gonzalez-Arroyo:1997ugn,
    author = "Gonz\'alez-Arroyo, Antonio",
    title = "{Yang-Mills fields on the four-dimensional torus. Part 1.: Classical theory}",
    booktitle = "{Advanced Summer School on Nonperturbative Quantum Field Physics}",
    eprint = "hep-th/9807108",
    archivePrefix = "arXiv",
    reportNumber = "FTUAM-97-18",
    pages = "57--91",
    year = "1997"
}

@article{Azeyanagi:2007su,
    author = "Azeyanagi, Tatsuo and Hanada, Masanori and Hirata, Tomoyoshi and Ishikawa, Tomomi",
    title = "{Phase structure of twisted Eguchi-Kawai model}",
    eprint = "0711.1925",
    archivePrefix = "arXiv",
    primaryClass = "hep-lat",
    reportNumber = "KUNS-2068, RIKEN-TH-97, RBRC-682, WIS-19-07-NOV-DPP",
    doi = "10.1088/1126-6708/2008/01/025",
    journal = "JHEP",
    volume = "01",
    pages = "025",
    year = "2008"
}

@article{GarciaPerez:2018fkj,
    author = "Garc\'\i{}a P\'erez, Margarita and Gonz\'alez-Arroyo, Antonio and Koren, Mateusz and Okawa, Masanori",
    title = "{The spectrum of 2+1 dimensional Yang-Mills theory on a twisted spatial torus}",
    eprint = "1807.03481",
    archivePrefix = "arXiv",
    primaryClass = "hep-th",
    reportNumber = "IFT-UAM-CSIC-18-71, FTUAM-18-19, DESY-18-108, HUPD-1804",
    doi = "10.1007/JHEP07(2018)169",
    journal = "JHEP",
    volume = "07",
    pages = "169",
    year = "2018"
}

@article{Ishikawa:2003,
    author = "Ishikawa, T. and Okawa, M.",
    title = "{$Z_N^D$ symmetry breaking on the numerical simulation of twisted Eguchi-Kawai model}",
    journal = "talk given at the \emph{Annual Meeting of the Physical Society of Japan}, March 28–31, Sendai, Japan",
    year = "2003"
}

@article{Chamizo:2016msz,
    author = "Chamizo, Fernando and Gonz\'alez-Arroyo, Antonio",
    title = "{Tachyonic instabilities in 2  +  1 dimensional Yang\textendash{}Mills theory and its connection to number theory}",
    eprint = "1610.07972",
    archivePrefix = "arXiv",
    primaryClass = "hep-th",
    reportNumber = "IFT-UAM-CSIC-16-088, FTUAM-16-34",
    doi = "10.1088/1751-8121/aa7346",
    journal = "J. Phys. A",
    volume = "50",
    number = "26",
    pages = "265401",
    year = "2017"
}

@article{Perez:2017jyq,
    author = "Garc\'\i{}a P\'erez, Margarita  and Gonz\'alez-Arroyo, Antonio and Okawa, Masanori",
    title = "{Perturbative contributions to Wilson loops in twisted lattice boxes and reduced models}",
    eprint = "1708.00841",
    archivePrefix = "arXiv",
    primaryClass = "hep-lat",
    reportNumber = "IFT-UAM-CSIC-17-067, FTUAM-17-12, HUPD-1707",
    doi = "10.1007/JHEP10(2017)150",
    journal = "JHEP",
    volume = "10",
    pages = "150",
    year = "2017"
}

@article{Bietenholz:2006cz,
    author = "Bietenholz, Wolfgang and Nishimura, Jun and Susaki, Yoshiaki and Volkholz, Jan",
    title = "{A Non-perturbative study of 4-D U(1) non-commutative gauge theory: The Fate of one-loop instability}",
    eprint = "hep-th/0608072",
    archivePrefix = "arXiv",
    reportNumber = "HU-EP-06-16, KEK-TH-1079, UTEP-517",
    doi = "10.1088/1126-6708/2006/10/042",
    journal = "JHEP",
    volume = "10",
    pages = "042",
    year = "2006"
}

@article{Teper:2006sp,
    author = "Teper, Michael and Vairinhos, Helvio",
    title = "{Symmetry breaking in twisted Eguchi-Kawai models}",
    eprint = "hep-th/0612097",
    archivePrefix = "arXiv",
    doi = "10.1016/j.physletb.2007.06.037",
    journal = "Phys. Lett. B",
    volume = "652",
    pages = "359--369",
    year = "2007"
}

@article{FlavourLatticeAveragingGroupFLAG:2021npn,
    author = "Aoki, Y. and others",
    collaboration = "Flavour Lattice Averaging Group (FLAG)",
    title = "{FLAG Review 2021}",
    eprint = "2111.09849",
    archivePrefix = "arXiv",
    primaryClass = "hep-lat",
    reportNumber = "CERN-TH-2021-191, JLAB-THY-21-3528, FERMILAB-PUB-21-620-SCD-T",
    doi = "10.1140/epjc/s10052-022-10536-1",
    journal = "Eur. Phys. J. C",
    volume = "82",
    number = "10",
    pages = "869",
    year = "2022"
}

@article{PhysRevLett.48.1063,
  title = {Reduction of Dynamical Degrees of Freedom in the Large-${N}$ Gauge Theory},
  author = {Eguchi, Tohru and Kawai, Hikaru},
  journal = {Phys. Rev. Lett.},
  volume = {48},
  issue = {16},
  pages = {1063--1066},
  numpages = {0},
  year = {1982},
  month = {Apr},
  publisher = {American Physical Society},
  doi = {10.1103/PhysRevLett.48.1063},
  url = {https://link.aps.org/doi/10.1103/PhysRevLett.48.1063}
}

@article{Gross:1982at,
    author = "Gross, David J. and Kitazawa, Yoshihisa",
    title = "{A Quenched Momentum Prescription for Large ${N}$ Theories}",
    reportNumber = "Print-82-0477 (PRINCETON)",
    doi = "10.1016/0550-3213(82)90278-4",
    journal = "Nucl. Phys. B",
    volume = "206",
    pages = "440--472",
    year = "1982"
}

@article{Neuberger:2020wpx,
    author = "Neuberger, Herbert",
    title = "{Quenched Eguchi-Kawai model revisited}",
    eprint = "2009.09539",
    archivePrefix = "arXiv",
    primaryClass = "hep-lat",
    doi = "10.1103/PhysRevD.102.094503",
    journal = "Phys. Rev. D",
    volume = "102",
    number = "9",
    pages = "094503",
    year = "2020"
}

@article{BHANOT198247,
title = {The quenched {E}guchi-{K}awai model},
author = {Bhanot, Gyan and Heller, Urs M. and Neuberger, Herbert},
journal = {Physics Letters B},
volume = {113},
number = {1},
pages = {47-50},
year = {1982},
issn = {0370-2693},
doi = {https://doi.org/10.1016/0370-2693(82)90106-X},
url = {https://www.sciencedirect.com/science/article/pii/037026938290106X}
}

@article{Gonzalez-Arroyo:1982hwr,
    author = "Gonzalez-Arroyo, Antonio and Okawa, M.",
    title = "{A Twisted Model for Large $N$ Lattice Gauge Theory}",
    reportNumber = "BNL-31689",
    doi = "10.1016/0370-2693(83)90647-0",
    journal = "Phys. Lett. B",
    volume = "120",
    pages = "174--178",
    year = "1983"
}

@article{Capitani:1999ay,
    author = "Capitani, S. and Gockeler, M. and Horsley, R. and Rakow, Paul E. L. and Schierholz, G.",
    editor = "Campostrini, M. and Caracciolo, S. and Cosmai, L. and Di Giacomo, A. and Rossi, P. and Rapuano, F.",
    title = "{On-shell and off-shell improvement for Ginsparg-Wilson fermions}",
    eprint = "hep-lat/9909167",
    archivePrefix = "arXiv",
    reportNumber = "TPR-99-18, HUB-EP-99-52",
    doi = "10.1016/S0920-5632(00)91837-1",
    journal = "Nucl. Phys. B Proc. Suppl.",
    volume = "83",
    pages = "893--895",
    year = "2000"
}

@article{Capitani:1999uz,
    author = "Capitani, S. and Gockeler, M. and Horsley, R. and Rakow, Paul E. L. and Schierholz, G.",
    title = "{Operator improvement for Ginsparg-Wilson fermions}",
    eprint = "hep-lat/9908029",
    archivePrefix = "arXiv",
    reportNumber = "DESY-99-115, TPR-99-15A, HUB-EP-99-41, HLRZ-99-36",
    doi = "10.1016/S0370-2693(99)01142-9",
    journal = "Phys. Lett. B",
    volume = "468",
    pages = "150--160",
    year = "1999"
}

@article{Gasser:1986vb,
    author = "Gasser, J. and Leutwyler, H.",
    title = "{Light Quarks at Low Temperatures}",
    reportNumber = "BUTP-86/19-BERN",
    doi = "10.1016/0370-2693(87)90492-8",
    journal = "Phys. Lett. B",
    volume = "184",
    pages = "83--88",
    year = "1987"
}

@article{Cordoba:2026wsy,
    author = "Cordoba, Rafael and He, Yifei and Kruczenski, Martin and Ping, Derek and Mohanty, Jignesh and Alegria, Emil",
    title = "{Gauge Theory Bootstrap with varying $N_c$}",
    eprint = "2611.16824",
    archivePrefix = "arXiv",
    primaryClass = "hep-th",
    month = "11",
    year = "2026"
}

@article{PhysRevD.27.2397,
  title = {Twisted {E}guchi-{K}awai model: A reduced model for large-${N}$ lattice gauge theory},
  author = {Gonz\'alez-Arroyo, A. and Okawa, M.},
  journal = {Phys. Rev. D},
  volume = {27},
  issue = {10},
  pages = {2397--2411},
  numpages = {0},
  year = {1983},
  month = {May},
  publisher = {American Physical Society},
  doi = {10.1103/PhysRevD.27.2397},
  url = {https://link.aps.org/doi/10.1103/PhysRevD.27.2397}
}

@article{Kovtun:2007py,
author = "Kovtun, Pavel and {\"U}nsal, Mithat and Yaffe, Laurence G.",
title = "{Volume independence in large $N_c$ QCD-like gauge theories}",
eprint = "hep-th/0702021",
archivePrefix = "arXiv",
reportNumber = "SLAC-PUB-12331, NSF-KITP-07-07",
doi = "10.1088/1126-6708/2007/06/019",
journal = "JHEP",
volume = "06",
pages = "019",
year = "2007"
}

@article{Gonzalez-Arroyo:1983cyv,
    author = "Gonz\'alez-Arroyo, Antonio and Okawa, M.",
    title = "{String Tension for Large $N$ Gauge Theory}",
    reportNumber = "BNL-32988",
    doi = "10.1016/0370-2693(83)90818-3",
    journal = "Phys. Lett. B",
    volume = "133",
    pages = "415--418",
    year = "1983"
}

@article{Gonzalez-Arroyo:2014dua,
    author = "Gonz\'alez-Arroyo, Antonio and Okawa, Masanori",
    title = "{Testing volume independence of SU$(N)$ pure gauge theories at large $N$}",
    eprint = "1410.6405",
    archivePrefix = "arXiv",
    primaryClass = "hep-lat",
    reportNumber = "IFT-UAM-CSIC-14-102, FTUAM-14-39, HUPD-1406",
    doi = "10.1007/JHEP12(2014)106",
    journal = "JHEP",
    volume = "12",
    pages = "106",
    year = "2014"
}

@article{Bringoltz:2011by,
    author = "Bringoltz, Barak and Koren, Mateusz and Sharpe, Stephen R.",
    title = "{Large-$N$ reduction in QCD with two adjoint Dirac fermions}",
    eprint = "1106.5538",
    archivePrefix = "arXiv",
    primaryClass = "hep-lat",
    doi = "10.1103/PhysRevD.85.094504",
    journal = "Phys. Rev. D",
    volume = "85",
    pages = "094504",
    year = "2012"
}

@article{Kiskis:2002gr,
author = "Kiskis, J. and Narayanan, R. and Neuberger, H.",
title = "{Proposal for the numerical solution of planar QCD}",
eprint = "hep-lat/0203005",
archivePrefix = "arXiv",
reportNumber = "UCD2002-3, RU-02-01, UCD2002-3-RU-02-01",
doi = "10.1103/PhysRevD.66.025019",
journal = "Phys. Rev. D",
volume = "66",
pages = "025019",
year = "2002"
}

@article{Kiskis:2009rf,
    author = "Kiskis, Joe and Narayanan, Rajamani",
    title = "{Computation of the string tension in four-dimensional Yang-Mills theory using large N reduction}",
    eprint = "0908.1451",
    archivePrefix = "arXiv",
    primaryClass = "hep-lat",
    reportNumber = "UCD-0905",
    doi = "10.1016/j.physletb.2009.10.043",
    journal = "Phys. Lett. B",
    volume = "681",
    pages = "372--375",
    year = "2009"
}

@article{Lohmayer:2013spa,
    author = "Lohmayer, Robert and Narayanan, Rajamani",
    title = "{Weak-coupling analysis of the single-site large-${N}$ gauge theory coupled to adjoint fermions}",
    eprint = "1305.1279",
    archivePrefix = "arXiv",
    primaryClass = "hep-lat",
    doi = "10.1103/PhysRevD.87.125024",
    journal = "Phys. Rev. D",
    volume = "87",
    number = "12",
    pages = "125024",
    year = "2013"
}

@article{Hietanen:2012ma,
    author = "Hietanen, A. and Narayanan, R.",
    title = "{Numerical evidence for non-analytic behavior in the beta function of large N SU(N) gauge theory coupled to an adjoint Dirac fermion}",
    eprint = "1204.0331",
    archivePrefix = "arXiv",
    primaryClass = "hep-lat",
    reportNumber = "CP3-ORIGINS-2012-009, DIAS-2012-10",
    doi = "10.1103/PhysRevD.86.085002",
    journal = "Phys. Rev. D",
    volume = "86",
    pages = "085002",
    year = "2012"
}

@article{Hietanen:2010fx,
    author = "Hietanen, A. and Narayanan, R.",
    title = "{Large-$N$ reduction of SU($N$) Yang\textendash{}Mills theory with massive adjoint overlap fermions}",
    eprint = "1011.2150",
    archivePrefix = "arXiv",
    primaryClass = "hep-lat",
    doi = "10.1016/j.physletb.2011.03.009",
    journal = "Phys. Lett. B",
    volume = "698",
    pages = "171--174",
    year = "2011"
}

@article{Hietanen:2009ex,
    author = "Hietanen, A. and Narayanan, R.",
    title = "{The large N limit of four dimensional Yang-Mills field coupled to adjoint fermions on a single site lattice}",
    eprint = "0911.2449",
    archivePrefix = "arXiv",
    primaryClass = "hep-lat",
    doi = "10.1007/JHEP01(2010)079",
    journal = "JHEP",
    volume = "01",
    pages = "079",
    year = "2010"
}

@article{Narayanan:2003fc,
    author = "Narayanan, R. and Neuberger, H.",
    title = "{Large N reduction in continuum}",
    eprint = "hep-lat/0303023",
    archivePrefix = "arXiv",
    doi = "10.1103/PhysRevLett.91.081601",
    journal = "Phys. Rev. Lett.",
    volume = "91",
    pages = "081601",
    year = "2003"
}

@article{Narayanan:2004cp,
    author = "Narayanan, R. and Neuberger, H.",
    title = "{Chiral symmetry breaking at large $N_c$}",
    eprint = "hep-lat/0405025",
    archivePrefix = "arXiv",
    doi = "10.1016/j.nuclphysb.2004.07.002",
    journal = "Nucl. Phys. B",
    volume = "696",
    pages = "107--140",
    year = "2004"
}

@article{Unsal:2008ch,
    author = "{\"U}nsal, Mithat and Yaffe, Laurence G.",
    title = "{Center-stabilized Yang-Mills theory: Confinement and large $N$ volume independence}",
    eprint = "0803.0344",
    archivePrefix = "arXiv",
    primaryClass = "hep-th",
    reportNumber = "SLAC-PUB-13144, NSF-KITP-08-16",
    doi = "10.1103/PhysRevD.78.065035",
    journal = "Phys. Rev. D",
    volume = "78",
    pages = "065035",
    year = "2008"
}

@article{GarciaPerez:2015rda,
    author = "Garc\'\i{}a P\'erez, Margarita and Gonz\'alez-Arroyo, Antonio and Keegan, Liam and Okawa, Masanori",
    title = "{Mass anomalous dimension of Adjoint QCD at large N from twisted volume reduction}",
    eprint = "1506.06536",
    archivePrefix = "arXiv",
    primaryClass = "hep-lat",
    reportNumber = "CERN-PH-TH-2015-138, IFT-UAM-CSIC-15-060, FTUAM-15-17, HUPD-1501",
    doi = "10.1007/JHEP08(2015)034",
    journal = "JHEP",
    volume = "08",
    pages = "034",
    year = "2015"
}

@article{GarciaPerez:2014azn,
    author = "Garc\'\i{}a P\'erez, Margarita and Gonz\'alez-Arroyo, Antonio and Keegan, Liam and Okawa, Masanori",
    title = "{The $SU(\infty)$ twisted gradient flow running coupling}",
    eprint = "1412.0941",
    archivePrefix = "arXiv",
    primaryClass = "hep-lat",
    reportNumber = "CERN-PH-TH-2014-233, IFT-UAM-CSIC-14-125, FTUAM-14-50, HUPD-1411",
    doi = "10.1007/JHEP01(2015)038",
    journal = "JHEP",
    volume = "01",
    pages = "038",
    year = "2015"
}

@article{Gonzalez-Arroyo:2015bya,
    author = "Gonz\'alez-Arroyo, Antonio and Okawa, Masanori",
    title = "{Large N meson masses from a matrix model}",
    eprint = "1510.05428",
    archivePrefix = "arXiv",
    primaryClass = "hep-lat",
    reportNumber = "IFT-UAM-CSIC-15-106, FTUAM-15-30, HUPD-1506",
    doi = "10.1016/j.physletb.2016.02.001",
    journal = "Phys. Lett. B",
    volume = "755",
    pages = "132--137",
    year = "2016"
}

@article{Hernandez:2020tbc,
    author = "Hern\'andez, Pilar and Romero-L\'opez, Fernando",
    title = "{The large $N_c$ limit of QCD on the lattice}",
    eprint = "2012.03331",
    archivePrefix = "arXiv",
    primaryClass = "hep-lat",
    doi = "10.1140/epja/s10050-021-00374-2",
    journal = "Eur. Phys. J. A",
    volume = "57",
    number = "2",
    pages = "52",
    year = "2021"
}

@article{Butti:2022sgy,
    author = "Butti, Pietro and Garc\'\i{}a P\'erez, Margarita and Gonz\'alez-Arroyo, Antonio and Ishikawa, Ken-Ichi and Okawa, Masanori",
    title = "{Scale setting for large-N SUSY Yang-Mills on the lattice}",
    eprint = "2205.03166",
    archivePrefix = "arXiv",
    primaryClass = "hep-lat",
    reportNumber = "IFT-UAM/CSIC-22-39, HUPD-2206",
    doi = "10.1007/JHEP07(2022)074",
    journal = "JHEP",
    volume = "07",
    pages = "074",
    year = "2022"
}

@article{Gonzalez-Arroyo:2005dgf,
    author = "Gonz\'alez-Arroyo, Antonio and Narayanan, R. and Neuberger, H.",
    title = "{Large N reduction on a twisted torus}",
    eprint = "hep-lat/0509074",
    archivePrefix = "arXiv",
    doi = "10.1016/j.physletb.2005.10.006",
    journal = "Phys. Lett. B",
    volume = "631",
    pages = "133--140",
    year = "2005"
}

@article{Gonzalez-Arroyo:2013bta,
    author = "Gonz\'alez-Arroyo, Antonio and Okawa, Masanori",
    title = "{Twisted space-time reduced model of large N QCD with two adjoint Wilson fermions}",
    eprint = "1305.6253",
    archivePrefix = "arXiv",
    primaryClass = "hep-lat",
    reportNumber = "HUPD-1303, IFT-UAM-CSIC-13-050, FTUAM-13-8",
    doi = "10.1103/PhysRevD.88.014514",
    journal = "Phys. Rev. D",
    volume = "88",
    pages = "014514",
    year = "2013"
}

@article{Gonzalez-Arroyo:2010omx,
    author = "Gonz\'alez-Arroyo, Antonio and Okawa, Masanori",
    title = "{Large $N$ reduction with the Twisted Eguchi-Kawai model}",
    eprint = "1005.1981",
    archivePrefix = "arXiv",
    primaryClass = "hep-th",
    reportNumber = "IFT-UAM-CSIC-10-32, FTUAM-2010-9, HUPD-1002",
    doi = "10.1007/JHEP07(2010)043",
    journal = "JHEP",
    volume = "07",
    pages = "043",
    year = "2010"
}

@article{Gonzalez-Arroyo:2012euf,
    author = "Gonz\'alez-Arroyo, Antonio and Okawa, Masanori",
    title = "{The string tension from smeared Wilson loops at large N}",
    eprint = "1206.0049",
    archivePrefix = "arXiv",
    primaryClass = "hep-th",
    reportNumber = "IFT-UAM-CSIC-12-50, FTUAM-2012-92, HUPD-1201",
    doi = "10.1016/j.physletb.2012.12.027",
    journal = "Phys. Lett. B",
    volume = "718",
    pages = "1524--1528",
    year = "2013"
}

@article{Bali:2013kia,
    author = "Bali, Gunnar S. and Bursa, Francis and Castagnini, Luca and Collins, Sara and Del Debbio, Luigi and Lucini, Biagio and Panero, Marco",
    title = "{Mesons in large-N QCD}",
    eprint = "1304.4437",
    archivePrefix = "arXiv",
    primaryClass = "hep-lat",
    reportNumber = "HIP-2013-01-TH",
    doi = "10.1007/JHEP06(2013)071",
    journal = "JHEP",
    volume = "06",
    pages = "071",
    year = "2013"
}

@article{Castagnini:2015ejr,
    author = "Castagnini, Luca",
    title = "{Meson spectroscopy in Large-$N$ QCD}",
    year = "2015",
    journal = "\href{https://inspirehep.net/literature/1411974}{\texttt{\emph{[inspirehep/1411974]}}}"
}

@article{Perez:2020vbn,
    author = "Garc\'\i{}a P\'erez, Margarita  and Gonz\'alez-Arroyo, Antonio and Okawa, Masanori",
    title = "{Meson spectrum in the large $N$ limit}",
    eprint = "2011.13061",
    archivePrefix = "arXiv",
    primaryClass = "hep-lat",
    reportNumber = "IFT-UAM/CSIC-20-110, HUPD-2003, FTUAM-20-13",
    doi = "10.1007/JHEP04(2021)230",
    journal = "JHEP",
    volume = "04",
    pages = "230",
    year = "2021"
}

@article{Perez:2015ssa,
    author = "Garc\'\i{}a P\'erez, Margarita  and Gonz\'alez-Arroyo, Antonio and Keegan, Liam and Okawa, Masanori and Ramos, Alberto",
    title = "{A comparison of updating algorithms for large N reduced models}",
    eprint = "1505.05784",
    archivePrefix = "arXiv",
    primaryClass = "hep-lat",
    reportNumber = "CERN-PH-TH-2015-030, IFT-UAM-CSIC-15-032, FTUAM-15-10, HUPD-1502",
    doi = "10.1007/JHEP06(2015)193",
    journal = "JHEP",
    volume = "06",
    pages = "193",
    year = "2015"
}

@article{Hernandez:2019qed,
    author = "Hern\'andez, Pilar and Pena, Carlos and Romero-L\'opez, Fernando",
    title = "{Large $N_c$ scaling of meson masses and decay constants}",
    eprint = "1907.11511",
    archivePrefix = "arXiv",
    primaryClass = "hep-lat",
    doi = "10.1140/epjc/s10052-019-7395-y",
    journal = "Eur. Phys. J. C",
    volume = "79",
    number = "10",
    pages = "865",
    year = "2019"
}

@article{Lucini:2012gg,
    author = "Lucini, Biagio and Panero, Marco",
    title = "{SU($N$) gauge theories at large $N$}",
    eprint = "1210.4997",
    archivePrefix = "arXiv",
    primaryClass = "hep-th",
    reportNumber = "HIP-2012-24-TH, NSF-KITP-12-190",
    doi = "10.1016/j.physrep.2013.01.001",
    journal = "Phys. Rept.",
    volume = "526",
    pages = "93--163",
    year = "2013"
}

@article{Gasser:1984gg,
    author = "Gasser, J. and Leutwyler, H.",
    title = "{Chiral Perturbation Theory: Expansions in the Mass of the Strange Quark}",
    reportNumber = "CERN-TH-3798",
    doi = "10.1016/0550-3213(85)90492-4",
    journal = "Nucl. Phys. B",
    volume = "250",
    pages = "465--516",
    year = "1985"
}

@article{GonzalezArroyo:2012bh,
    author = "Gonz\'alez-Arroyo, Antonio and Okawa, Masanori",
    editor = "Leinweber, Derek and Kamleh, Waseem and Mahbub, Selim and Matevosyan, Hrayr and Thomas, Anthony and Williams, Anthony G. and Young, Ross and Zanotti, James",
    title = "{The string tension for Large N gauge theory from smeared Wilson loops}",
    eprint = "1212.3835",
    archivePrefix = "arXiv",
    primaryClass = "hep-lat",
    reportNumber = "HUPD-1206, IFT-UAM-CSIC-12-119, FTUAM-12-118",
    doi = "10.22323/1.164.0221",
    journal = "PoS",
    volume = "LATTICE2012",
    pages = "221",
    year = "2012"
}

@article{Morningstar:2003gk,
    author = "Morningstar, Colin and Peardon, Mike J.",
    title = "{Analytic smearing of SU(3) link variables in lattice QCD}",
    eprint = "hep-lat/0311018",
    archivePrefix = "arXiv",
    doi = "10.1103/PhysRevD.69.054501",
    journal = "Phys. Rev. D",
    volume = "69",
    pages = "054501",
    year = "2004"
}

@article{Ce:2016awn,
      author         = "C\`e, Marco and Garc\'ia Vera, Miguel and Giusti, Leonardo
                        and Schaefer, Stefan",
      title          = "{The topological susceptibility in the large-$N$ limit of
                        SU($N$) Yang-Mills theory}",
      journal        = "Phys. Lett. B",
      volume         = "762",
      year           = "2016",
      pages          = "232-236",
      doi            = "10.1016/j.physletb.2016.09.029",
      eprint         = "1607.05939",
      archivePrefix  = "arXiv",
      primaryClass   = "hep-lat",
      reportNumber   = "DESY-16-131",
      SLACcitation   = "%%CITATION = ARXIV:1607.05939;%%"
}

@article{Luscher:2010iy,
      author         = "L{\"u}scher, Martin",
      title          = "{Properties and uses of the Wilson flow in lattice QCD}",
      journal        = "JHEP",
      volume         = "08",
      year           = "2010",
      pages          = "071",
      doi            = "10.1007/JHEP08(2010)071, 10.1007/JHEP03(2014)092",
      note           = "[Erratum: JHEP03,092(2014)]",
      eprint         = "1006.4518",
      archivePrefix  = "arXiv",
      primaryClass   = "hep-lat",
      reportNumber   = "CERN-PH-TH-2010-143",
      SLACcitation   = "%%CITATION = ARXIV:1006.4518;%%"
}

@article{Giusti:2008vb,
    author = "Giusti, Leonardo and L{\"u}scher, Martin",
    title = "{Chiral symmetry breaking and the Banks-Casher relation in lattice QCD with Wilson quarks}",
    eprint = "0812.3638",
    archivePrefix = "arXiv",
    primaryClass = "hep-lat",
    reportNumber = "CERN-PH-TH-2008-239",
    doi = "10.1088/1126-6708/2009/03/013",
    journal = "JHEP",
    volume = "03",
    pages = "013",
    year = "2009"
}

@article{DelDebbio:2004ns,
      author         = "Del Debbio, Luigi and Giusti, Leonardo and Pica, Claudio",
      title          = "{Topological susceptibility in the $SU(3)$ gauge theory}",
      journal        = "Phys. Rev. Lett.",
      volume         = "94",
      year           = "2005",
      pages          = "032003",
      doi            = "10.1103/PhysRevLett.94.032003",
      eprint         = "hep-th/0407052",
      archivePrefix  = "arXiv",
      primaryClass   = "hep-th",
      reportNumber   = "CERN-PH-TH-2004-117, CPT-2004-P-041",
      SLACcitation   = "%%CITATION = HEP-TH/0407052;%%"
}

@article{Luscher:1998pqa,
      author         = "L{\"u}scher, Martin",
      title          = "{Exact chiral symmetry on the lattice and the
                        Ginsparg-Wilson relation}",
      journal        = "Phys. Lett. B",
      volume         = "428",
      year           = "1998",
      pages          = "342-345",
      doi            = "10.1016/S0370-2693(98)00423-7",
      eprint         = "hep-lat/9802011",
      archivePrefix  = "arXiv",
      primaryClass   = "hep-lat",
      reportNumber   = "DESY-98-014",
      SLACcitation   = "%%CITATION = HEP-LAT/9802011;%%"
}

@article{Veneziano:1979ec,
    author = "Veneziano, G.",
    title = "{$U(1)$ Without Instantons}",
    reportNumber = "CERN-TH-2651",
    doi = "10.1016/0550-3213(79)90332-8",
    journal = "Nucl. Phys. B",
    volume = "159",
    pages = "213--224",
    year = "1979"
}

@article{DiVecchia:1980yfw,
    author = "Di Vecchia, P. and Veneziano, G.",
    title = "{Chiral Dynamics in the Large $N$ Limit}",
    reportNumber = "CERN-TH-2814",
    doi = "10.1016/0550-3213(80)90370-3",
    journal = "Nucl. Phys. B",
    volume = "171",
    pages = "253--272",
    year = "1980"
}

@article{Witten:1980sp,
    author = "Witten, Edward",
    title = "{Large $N$ Chiral Dynamics}",
    reportNumber = "HUTP-80/A005",
    doi = "10.1016/0003-4916(80)90325-5",
    journal = "Annals Phys.",
    volume = "128",
    pages = "363",
    year = "1980"
}

@article{tHooft:1973alw,
    author = "'t Hooft, Gerard",
    editor = "Taylor, J.C.",
    title = "{A Planar Diagram Theory for Strong Interactions}",
    reportNumber = "CERN-TH-1786",
    doi = "10.1016/0550-3213(74)90154-0",
    journal = "Nucl. Phys. B",
    volume = "72",
    pages = "461",
    year = "1974"
}

@article{Witten:1979vv,
    author = "Witten, Edward",
    title = "{Current Algebra Theorems for the $U(1)$ Goldstone Boson}",
    reportNumber = "HUTP-79/A014",
    doi = "10.1016/0550-3213(79)90031-2",
    journal = "Nucl. Phys. B",
    volume = "156",
    pages = "269--283",
    year = "1979"
}

@article{tHooft:1976rip,
    author = "'t Hooft, Gerard",
    editor = "Shifman, Mikhail A.",
    title = "{Symmetry Breaking Through Bell--Jackiw Anomalies}",
    reportNumber = "PRINT-76-0254 (HARVARD)",
    doi = "10.1103/PhysRevLett.37.8",
    journal = "Phys. Rev. Lett.",
    volume = "37",
    pages = "8--11",
    year = "1976"
}

\end{document}